\documentclass[a4paper,11pt]{article}
\pdfoutput=1
\usepackage{jheppub}
\usepackage[T1]{fontenc}
\usepackage{bm,amsmath,amssymb,bbm} 
\usepackage{graphicx}
\usepackage[most]{tcolorbox}
\tcbuselibrary{skins}

\DeclareMathOperator\diag{diag}
\DeclareMathOperator\Tr{Tr}
\DeclareMathOperator\rme{\mathrm{e}}
\newcommand{\nn}{\nonumber}

\newcommand{\der}{\partial}
\renewcommand{\bar}[1]{\overline{#1}}

\newcommand{\bep}{\begin{pmatrix}} 
\newcommand{\eep}{\end{pmatrix}}

\newcommand{\SU}{\text{SU}}

\renewcommand{\O}{\text{O}}
\newcommand{\U}{\text{U}}
\newcommand{\1}{\mathbbm{1}}

\renewcommand{\epsilon}{\varepsilon}
\newcommand{\rmd}{\mathrm{d}}
\newcommand{\be}{\begin{eqnarray}}
\newcommand{\ee}{\end{eqnarray}}

\def\wt#1{\widetilde{#1}}
\def\wh#1{\widehat{#1}}
\def\ba#1\ea{\begin{align}#1\end{align}}
\def\mkakko#1{\left( #1 \right)}
\def\ckakko#1{\left\{ #1 \right\}}
\def\kkakko#1{\left[ #1 \right]}
\newcommand{\kk}{\kappa}
\newcommand{\eref}[1]{(\ref{#1})}

\title{Cascade of phase transitions in a planar Dirac material}

\author[a]{Takuya Kanazawa,}
\author[b]{Mario Kieburg}
\author[c]{and Jacobus J.M. Verbaarschot}

\affiliation[a]{Research and Development Group, Hitachi, Ltd.,\\
Kokubunji, Tokyo 185-8601, Japan}
\affiliation[b]{School of Mathematics and Statistics, University of Melbourne,\\
Parkville, Melbourne VIC 3010, Australia}
\affiliation[c]{Department of Physics and Astronomy, Stony Brook University,\\
Stony Brook, NY 11794, U.S.A.}

\emailAdd{tkanazawa@nt.phys.s.u-tokyo.ac.jp}
\emailAdd{m.kieburg@unimelb.edu.au}
\emailAdd{jacobus.verbaarschot@stonybrook.edu}

\abstract{We investigate a model of interacting Dirac fermions
in $2+1$ dimensions  with $M$
  flavors and $N$ colors having the  U($M$)$\times$SU($N$) symmetry.
  In the large-$N$ limit, we find that the U($M$) symmetry is spontaneously broken in a variety of ways. In the vacuum,
  when the parity-breaking flavor-singlet mass is varied, the ground state undergoes a sequence of $M$ first-order phase transitions,
  experiencing $M+1$ phases characterized by symmetry breaking U($M$)$\to$\mbox{U($M-k$)}$\times$U($k$)
  with $k\in\{0,1,2,\cdots,M\}$, bearing a close resemblance to the vacuum structure of three-dimensional QCD.
  At finite temperature and chemical potential, a rich phase diagram with first and second-order phase
  transitions and tricritical points is observed. Also exotic phases with spontaneous symmetry breaking of the form as U(3)$\to$U(1)$^3$,
  U(4)$\to$U(2)$\times$U(1)$^2$, and U(5)$\to$U(2)$^2\times$U(1) exist. For a large flavor-singlet mass, the increase of the chemical
  potential $\mu$ brings about $M$ consecutive first-order transitions that separate the low-$\mu$ phase diagram with
  vanishing fermion density from the high-$\mu$ region with a high fermion density.}

\begin{document} 
\maketitle
\flushbottom

\section{Introduction}

Dirac fermions play a central role in physics -- not only in elementary particle physics but also in condensed matter physics
\cite{Vafek:2013mpa,Wehling:2014cla,Hasan:2017hwf,Armitage:2017cjs}.  Interactions of Dirac fermions are essential
in determining the ground state of various physical systems, and models with quartic interactions have been studied for decades
in a variety of fields. For instance, in nuclear and hadron physics, the Nambu--Jona-Lasinio (NJL) model \cite{Nambu:1961tp,Nambu:1961fr}
is famous as a phenomenological effective theory of QCD \cite{Klevansky:1992qe,Hatsuda:1994pi}. Lower-dimensional
four-fermion models such as the Gross-Neveu model in $1+1$ dimensions \cite{Gross:1974jv} have also played a pivotal role in
advancing our understanding of phenomena like dynamical symmetry breaking, asymptotic freedom and dimensional transmutation.
Recently there are renewed interests in Dirac fermions in $2+1$ dimensions. They appear in some condensed matter systems
\cite{Fu:2007uya,CastroNeto:2009zz,Tajima_2009,Kobayashi_2009,Lim_2009,Lan:2011qh,10.1093/nsr/nwu080,PhysRevLett.115.126803,Son:2015xqa,Potter:2015cdn,Isobe:2015myw}
and understanding the effects of interactions is therefore imperative. Historically, four-fermion models of Dirac fermions
in $2+1$ dimensions have been thoroughly studied both analytically \cite{PhysRevB.33.3257,PhysRevB.33.3263,Semenoff:1989dm,Rosenstein:1990nm,Hong:1993qk,Gusynin:1994re,Esposito:1998ki,Babaev:1999in,Appelquist:2000mb,Hofling:2002hj,Kneur:2007vm,Braun:2010tt,Klimenko:2012tk,Scherer:2013pda,Cao:2014uva} and by numerical simulations \cite{Hands:1992be,DelDebbio:1997dv,Christofi:2007ye,Chandrasekharan:2013aya,Ayyar:2015lrd,Hands:2016foa,Winstel:2019zfn,Narayanan:2020uqt}, and intriguing features such as superfluidity, Kosterlitz-Thouless transitions, non-Gaussian Ultra-Violet (UV) fixed points and magnetic catalysis have been elucidated. These studies have provided a tractable avenue for understanding nonperturbative aspects of $(2+1)$-dimensional strongly coupled gauge theories, including QED$_3$ and QCD$_3$ as prominent examples. 

Recently QCD$_3$ has experienced a flurry of revived attention \cite{Komargodski:2017keh,Gomis:2017ixy,Armoni:2017jkl,Karthik:2018nzf,Choi:2018tuh,Kanazawa:2019oxu,Argurio:2019tvw,Armoni:2019lgb,Akhond:2019ued}. In \cite{Kanazawa:2019oxu} the present authors have proposed
a new random matrix theory (RMT) which, when random matrix elements are integrated out, reduces to a four-fermion model that
spontaneously breaks symmetries in exactly the same way as does QCD$_3$ with a Chern-Simons term \cite{Komargodski:2017keh}, thus extending the previous work \cite{Verbaarschot:1994ip}. Although RMT is a zero-dimensional theory with no gauge interactions, it provides exact descriptions of the low-lying Dirac spectrum owing to the universality of the microscopic domain \cite{Leutwyler:1992yt,Shuryak:1992pi,Verbaarschot:1993pm,Verbaarschot:1997bf,Verbaarschot:2000dy}.  

In this work, we study thermodynamics and symmetry breaking of an unconventional interacting model of Dirac fermions in $2+1$ dimensions at finite temperature and chemical potential in the large-$N$ limit, where $N$ denotes the number of ``colors.'' Each
fermion comes in $M$ different flavors.
This model can be viewed as a generalization of the RMT proposed in \cite{Kanazawa:2019oxu}. The model has three key ingredients: a repulsive interaction, an attractive interaction, and a flavor-symmetric parity-breaking mass term. Their interplay leads to a surprisingly
rich phase diagram. At zero temperature and zero density, the model exhibits a spontaneous symmetry breaking patterns $\U(M)\to\U(M-k)\times\U(k)$
with various $k$ and experiences a sequence of first-order phase transitions, bearing a close resemblance to three-dimensional QCD \cite{Komargodski:2017keh,Armoni:2019lgb}. The model reduces to a sigma model on a complex Grassmannian at low energy. At nonzero
temperature or chemical potential, there appear even more exotic phases where the symmetry is broken as $\U(3)\to\U(1)\times\U(1)\times\U(1)$, $\U(4)\to\U(2)\times\U(1)\times\U(1)$, and $\U(5)\to\U(2)\times\U(2)\times\U(1)$, to name but a few. All these patterns show up in a single model with a few adjustable parameters. 

The present work is structured as follows. In section~\ref{sc:defm}, the model is defined and the thermodynamic potential is derived.
In section~\ref{sc:vac}, the ground state at zero temperature and density is analyzed. In section~\ref{sc:temp} the effect of nonzero temperature is considered. In section~\ref{sc:mu}, a nonzero chemical potential is introduced, and the fermion number density is calculated.  In section~\ref{sc:fullpd}, phases at nonzero temperature and density are studied. It is shown that the phase structure changes dramatically,
depending on the interaction strength and the flavor-singlet mass. We conclude in section~\ref{sc:conc}, and technical details are worked out in several
appendices.
Throughout this article we will work in the natural units where $\hbar=c=k_{\rm B}=1$ and with Einstein's summation convention where we sum over repeated indices.

\section{\label{sc:defm}Planar four-fermion model}

We consider a system of two-component Dirac fermions $\psi^i_{s\alpha}$ in $2+1$ dimensions. 
Here $\alpha=1,2$ are spinor indices, $i=1,\cdots,N$ are color indices and $s=1,\cdots,M$ are flavor indices. 
The Lagrangian in the Euclidean spacetime is given by
\ba
	\mathcal{L} & = \bar\psi^i_s (\sigma_\nu\der_\nu +\kk-\mu\sigma_3) \psi^i_s + \frac{g_1^2}{N} (\bar\psi_s^i \psi_s^i)^2 - \frac{g_2^2}{N}(\bar\psi^i_s \psi^i_{s'})(\bar\psi^j_{s'} \psi^j_{s}), 
	\label{eq:Ldef}
\ea
where $\sigma_\nu=(\sigma_1, \sigma_2, \sigma_3)$ are the Pauli matrices in   spinor space.
The couplings have dimensions $[g_1]=[g_2]=-1/2$. The Lagrangian $\mathcal{L}$ is invariant under $\U(1)\times\SU(N)\times\SU(M)$ transformations of $\psi$.%
\footnote{A three-dimensional four-fermion model having this symmetry was investigated in \cite{Vshivtsev:1996vs,Vshivtsev:1998fm}. We thank K.~G.~Klimenko for bringing these references to our attention.} The mass term $\kk\bar\psi\psi$ breaks parity symmetry, and $\mu$ is the baryon chemical potential.
The four-fermion interactions of the form \eqref{eq:Ldef} arise in the random matrix model proposed in \cite{Kanazawa:2019oxu} which also gives the sign
of the interaction terms.
We underline that these signs are essential for the results of the present
work.

To rephrase the four-fermion terms in two quadratic ones, we perform the Hubbard-Stratonovich transformation and obtain
\ba
	Z=\int \mathcal{D}(\bar\psi,\psi,\phi,\Phi) \exp\mkakko{-\int_0^{\beta} \hspace{-1mm}\rmd \tau \int \rmd^2x\;\mathcal{\wt{L}}}
\ea
with $\beta=1/T$ the inverse temperature and the Lagrangian
\ba
	\wt{\mathcal{L}} & = \bar\psi^i_s (\sigma_\nu\der_\nu +\kk-\mu\sigma_3
	+ 2ig_1\phi + 2g_2\Phi)_{ss'} \psi^i_{s'} + N (\phi^2 + \Tr\Phi^2),
\ea
where $\phi$ is a scalar field and $\Phi$ is a Hermitian $M\times M$ matrix field, i.e., $\Phi^\dagger=\Phi$. 
Fermions can now be integrated out, yielding
\ba
	\hspace{-2mm}
	Z = \int \mathcal{D}\phi \int \mathcal{D}\Phi~
        {\det}^N
        (\sigma_\nu\der_\nu +\kk-\mu\sigma_3 + 2ig_1\phi + 2g_2\Phi) ~ \exp\ckakko{- N\int {\rmd\tau \rmd^2x} (\phi^2+\Tr\Phi^2) }.
	\label{eq:Z00}
        \ea
Next, we introduce a shifted field $\Phi' \equiv \Phi + \frac{ig_1}{g_2}\phi \1_M + \frac{\kk}{2g_2}\1_M$ to obtain
\ba
Z = & \int \mathcal{D}\phi \int \mathcal{D}\Phi'~
{\det}^N (\sigma_\nu\der_\nu -\mu\sigma_3 + 2g_2\Phi')
	\notag
	\\
	& \times \exp\kkakko{- N \int {\rmd\tau \rmd^2x} \ckakko{\phi^2+\Tr\mkakko{\Phi' - \frac{ig_1}{g_2}\phi \1_M-\frac{\kk}{2g_2}\1_M}^2} }.
\ea
Assuming that the condition $g_2^2 > Mg_1^2$ is fulfilled, the integral over the $\phi$ field can be carried out and leads to the result
\ba
\begin{split}
  Z \propto& \int \mathcal{D}\Phi'~
  {\det}^N (\sigma_\nu\der_\nu - \mu\sigma_3 + 2g_2\Phi')\\
	&\times\exp\kkakko{-N \int {\rmd\tau \rmd^2x} \ckakko{\frac{g_1^2}{g_2^2-Mg_1^2}\mkakko{\Tr \Phi'-m}^2 +\Tr \Phi'^2} }
	\end{split}
	\label{eq:ZnE}
\ea
with
\ba
	m \equiv \frac{g_2\kk}{2g_1^2}\,.
        \ea
        After substituting this into \eqref{eq:Z00} and the shifted field $\Phi' \equiv \Phi + \frac{ig_1}{g_2}\phi \1_M + \frac{\kk}{2g_2}\1_M$
 we arrive at \eqref{eq:ZnE}.
Alternatively, the Gaussian integral over $\phi$ in equation \eqref{eq:Z00} can also be evaluated from the saddle point equation in $\displaystyle\phi$ with the saddle point
$\displaystyle\phi = (ig_1/g_2)\Tr \Phi$. 

In the large-$N$ limit the partition function is dominated by saddle points of the effective potential
\ba
	V_{\rm eff}(\Phi') & = \frac{g_1^2}{g_2^2-Mg_1^2}\mkakko{\Tr \Phi'-m}^2 +\Tr \Phi'^2 - \frac T{L^2}\log \det (\sigma_\nu\der_\nu -\mu\sigma_3 + 2g_2\Phi')\, 
        \ea
where $L$ is the linear extent of the plane. Assuming a constant field $\Phi'(\tau,x_1,x_2) = \Phi'$ we find
\ba
	V_{\rm eff}(\Phi') & = \frac{g_1^2}{g_2^2-Mg_1^2}\mkakko{\Tr \Phi'-m}^2 +\Tr \Phi'^2 
	\notag
	\\
	& \quad 
	- T \int \frac{\rmd^2p}{(2\pi)^2}\sum_{n=-\infty}^{\infty} \mathrm{tr} \log [ip_1\sigma_1+ip_2\sigma_2 + (i\omega_n-\mu) \sigma_3 + 2g_2\Phi']\, ,
\ea
where $\omega_n=(2n+1)\pi T$ and tr is the trace over the spinor and flavor indices. Next, we perform the diagonalization $\Phi'=UEU^\dagger$ with $E=\diag(E_1,\cdots,E_M)$%
\footnote{This change of variables yields a Jacobian  $\prod_{1\leq i<j\leq M}|E_i-E_j|^2$, which does not play a role at leading order of the large-$N$ expansion because $M$ is fixed.} and combine terms with $n\geq 0$ and $n<0$ to get
\ba
	V_{\rm eff}(E) & = \frac{g_1^2}{g_2^2-Mg_1^2}\mkakko{\sum_{k=1}^{M}E_k-m}^2 
	+ \sum_{k=1}^{M}E_k^2 
	\notag
	\\
	& \quad 
	- \frac{T}{2} \sum_{k=1}^{M}\int \frac{\rmd^2p}{(2\pi)^2}\Bigg[
	\sum_{n=-\infty}^{\infty} \log \ckakko{\beta^2\omega_n^2+\beta^2\mkakko{\sqrt{\mathbf{p}^2+4g_2^2E_k^2}+\mu}^2}
	\notag
	\\
	& \quad 
	+ \sum_{n=-\infty}^{\infty} \log \ckakko{\beta^2\omega_n^2+\beta^2\mkakko{\sqrt{\mathbf{p}^2+4g_2^2E_k^2}-\mu}^2}
	\Bigg].
        \ea
        We have included a factor $\beta^2$ in the argument
        of the logarithm which just amounts to an overall
        normalization constant.
Finally, we use the standard formula for summation over Matsubara frequencies \cite{Kapusta:2006pm,wolfram_cosh}
\ba
	\sum_{n=-\infty}^{\infty}\log\mkakko{\frac{\beta^2\omega_n^2+z^2}{\beta^2\omega_n^2}}
	= z + 2 \log (1+\rme^{-z}) - 2\log 2
\ea
to obtain
\ba
	V_{\rm eff}(E) & = \frac{g_1^2}{g_2^2-Mg_1^2}\mkakko{\sum_{k=1}^{M}E_k-m}^2 
	+ \sum_{k=1}^{M}E_k^2 
	- \sum_{k=1}^{M}\int \frac{\rmd^2p}{(2\pi)^2}\Bigg\{
		\sqrt{\mathbf{p}^2+4g_2^2E_k^2}
	\notag
	\\
	& \quad 
		+ T \log \kkakko{1+\rme^{-\beta \mkakko{\sqrt{\mathbf{p}^2+4g_2^2E_k^2}+\mu}}}
		+ T \log \kkakko{1+\rme^{-\beta \mkakko{\sqrt{\mathbf{p}^2+4g_2^2E_k^2}-\mu}}}
	\Bigg\}.
	\label{eq:mainV}
\ea
This is the main result of this section. The momentum integral for the zero-temperature part is UV divergent and we regularize it by a cutoff $\Lambda$.

What happens if we switch off the Gross-Neveu-type interaction by letting $g_1\to 0$? In this limit the potential becomes
\ba
	V_{\rm eff}(E) & = 
	\sum_{k=1}^{M} \Bigg[ E_k^2  - \frac{\kk}{g_2} E_k
	- \int \frac{\rmd^2p}{(2\pi)^2}\Bigg\{
		\sqrt{\mathbf{p}^2+4g_2^2E_k^2}
		+ T \log \kkakko{1+\rme^{-\beta \mkakko{\sqrt{\mathbf{p}^2+4g_2^2E_k^2}+\mu}}}
		\notag
		\\
		& \quad 
		+ T \log \kkakko{1+\rme^{-\beta \mkakko{\sqrt{\mathbf{p}^2+4g_2^2E_k^2}-\mu}}}
	\Bigg\}\Bigg]\,,
	\label{eq:V00}
\ea
where a divergent constant independent of $E$ has been dropped. 
For any $\kk\ne0$ the origin of the $E_k$ is unstable due to the presence of a linear
term and $E_k$ develops a nonzero vacuum expectation value (VEV) $\langle E \rangle\propto \1_M$ at all temperatures. There is no spontaneous breaking of $\U(M)$ symmetry. As will be shown in the following sections, the situation is dramatically different for $g_1\ne 0$; we will see a rich pattern of symmetry breaking taking place.

According to the Coleman-Mermin-Wagner-Hohenberg theorem,
in the absence of long range interactions, continuous
symmetries cannot be broken spontaneously in two-dimensions which includes
2+1 dimensions at nonzero temperature. However, fluctuations that destroy
the condensate, are suppressed for $N\to \infty$ and spontaneously
symmetry breaking is possible also at nonzero temperature. Below we always
analyze the large $N$ limit, but in appendix \ref{ap:comm} we argue that even
at finite $N$ the same phase transitions still may be observed  in particular
if they are of first order.

\section{\label{sc:vac}Vacuum}
\subsection{Numerical results}
We begin our discussion with the vacuum, $T=\mu=0$, to see what the underlying phases are. The momentum integral can be done analytically and yields the effective potential
\ba
	V_{\rm eff}(E) 
	& = \frac{g_1^2}{g_2^2-Mg_1^2}\mkakko{\sum_{k=1}^{M}E_k-m}^2 
	+ \sum_{k=1}^{M} v(E_k)
        \ea
        with $v(E)$ given by
\ba
    v(E_k)= E_k^2+\frac{4}{3\pi}|g_2E_k|^3- \frac{1}{6\pi}\mkakko{\Lambda^2+4g_2^2E_k^2}^{3/2}. 
	\label{eq:Vvac}
\ea
It is convenient to introduce dimensionless variables
\begin{tcolorbox}[ams equation,enhanced,colback=white]
	e_k = \frac{E_k}{\Lambda^{3/2}}\,,\qquad 
	\wt{g}_{1,2} =g_{1,2} \sqrt{\Lambda}\,,\qquad 
	\wt\lambda =  \frac{m} {\Lambda^{3/2}}
	\label{eq:dimless}
\end{tcolorbox}\noindent 
which will be used primarily for the numerical results in 
this paper. For the analytical results
we stick to a different notation, see next subsection,  
to simplify mathematical manipulations.

The dimensionless potential reads
\ba
	\frac {V_{\rm eff}(E)}{\Lambda^3}  
	& = \frac{\wt{g}_1^2}{\wt{g}_2^2-M\wt{g}_1^2}\mkakko
              {\sum_{k=1}^{M}e_k-\wt\lambda}^2 
	+ \sum_{k=1}^{M} \wt{v}(e_k)
\ea
with
\be
        \wt{v}(e) = e^2+\frac{4}{3\pi}|\wt{g}_2 e|^3
          - \frac{1}{6\pi}\mkakko{1+4\wt{g}_2^2e^2}^{3/2}. 
\ee

When $\wt{g}_2^2 > \pi$, so that the potential $\wt{v}(e)$ is confining, $\wt{v}(e)$ has two minima. For large negative $\wt\lambda$, all the $e_k$ are degenerate and negative. With increasing $\wt\lambda$, one of the eigenvalues jumps to a positive value. After increasing $\wt\lambda$ further, another eigenvalue jumps. This continues until all eigenvalues become degenerate and positive. 

\begin{figure}[tb]
	\centering
	\vspace{\baselineskip}
	\includegraphics[width=.6\textwidth]{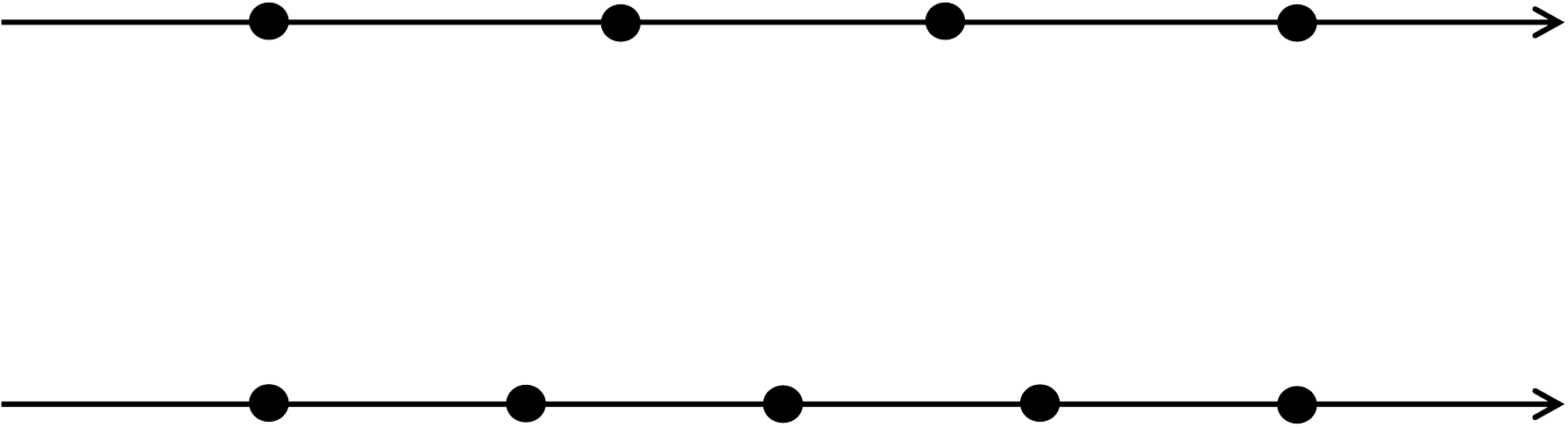}
	\put(-245,75){\footnotesize $\U(4)$}
	\put(-209,51){\footnotesize $\U(3)\times\U(1)$}
	\put(-154,75){\footnotesize $\U(2)\times\U(2)$}
	\put(-97,51){\footnotesize $\U(1)\times\U(3)$}
	\put(-34,75){\footnotesize $\U(4)$}
	\put(7,63){$\lambda$}
	\put(-245,12){\footnotesize $\U(5)$}
	\put(-217,-12){\footnotesize $\U(4)\times\U(1)$}
	\put(-174,12){\footnotesize $\U(3)\times\U(2)$}
	\put(-132,-12){\footnotesize $\U(2)\times\U(3)$}
	\put(-89,12){\footnotesize $\U(1)\times\U(4)$}
	\put(-34,-12){\footnotesize $\U(5)$}
	\put(7,0){$\lambda$}
	\vspace{.6\baselineskip}
	\caption{\label{fg:phases_at_T=mu=0}The phase diagram at $T=\mu=0$ with $|\wt{g}_2|>\sqrt{\pi}$ for $M=4$ (top) and $M=5$ (bottom) at large $N$. The blobs denote first-order phase transitions. Each phase is labeled with its unbroken symmetry group. The phase structure shown here generalizes to higher $M$ in an obvious manner.}
\end{figure}
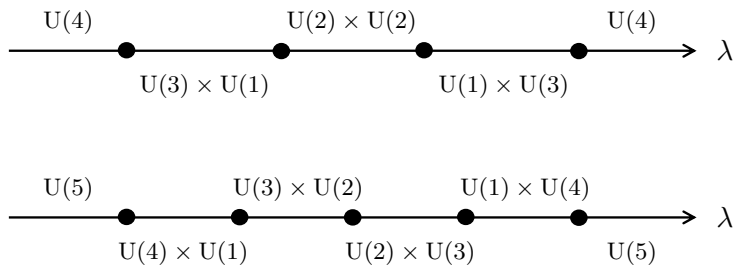

        To understand this phenomenon quantitatively, we performed numerical minimization of the potential. In figure~\ref{fg:T0} we display the  $\wt{\lambda}$ dependence of the
        $e_k$ at the minimum of the potential
$V_{\rm eff}(E)$. The $e_k$ jump $M$ times, marking $M$ sequential first-order phase transitions. Hence, there are in total $M+1$ different vacuum states. 
\begin{figure}[tb]
	\centering
	\includegraphics[width=.4\textwidth]{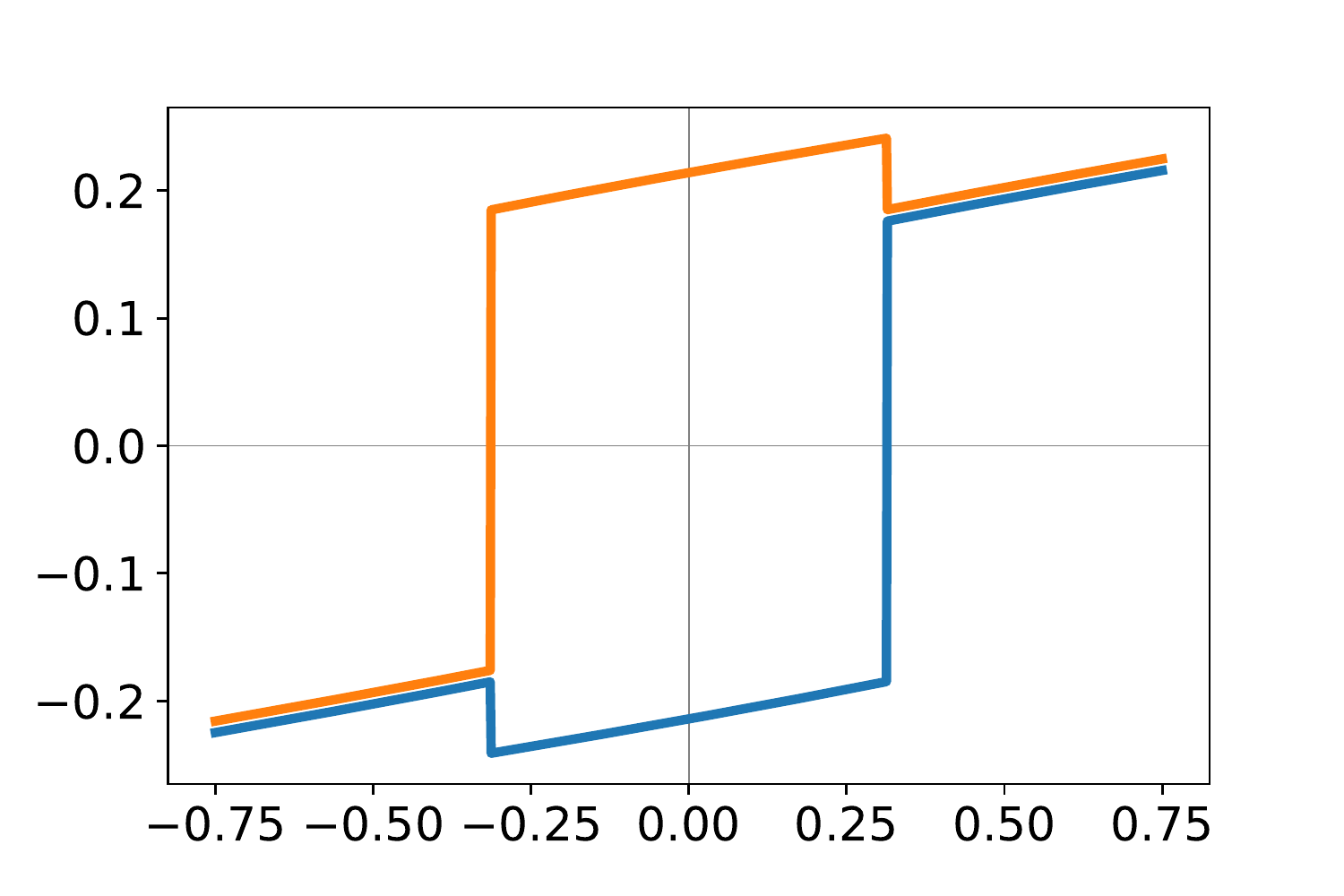}
	\qquad~~
	\includegraphics[width=.4\textwidth]{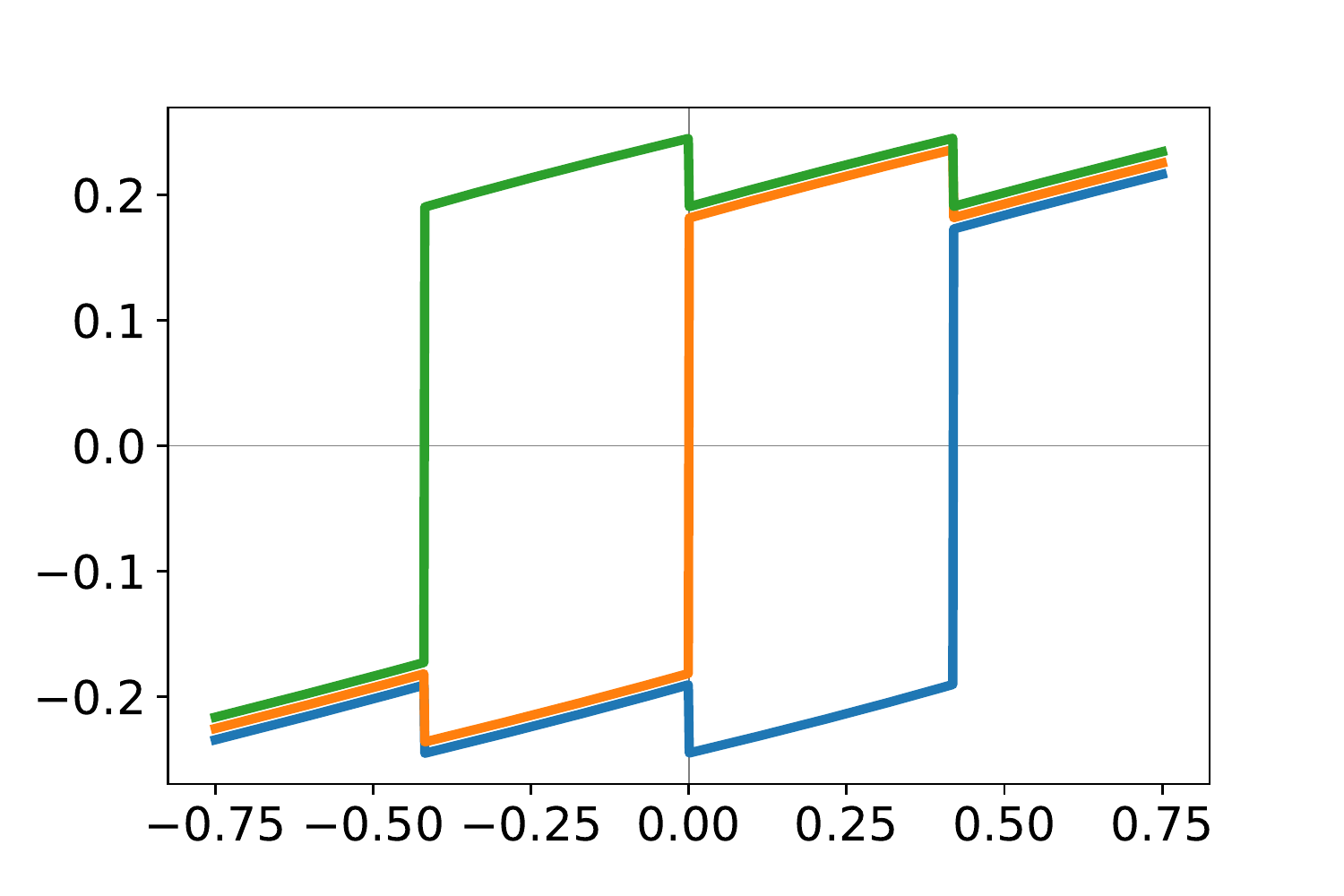}
	\put(-306,113){\large $M=2$}
	\put(-98,113){\large $M=3$}
	\put(-289,-17){\large $\wt{\lambda}$}
	\put(-80,-17){\large $\wt{\lambda}$}
	\put(-195,56){\large $\{e_k\}$}
	\put(-403,56){\large $\{e_k\}$}
	\vspace{\baselineskip}
	\\
	\includegraphics[width=.4\textwidth]{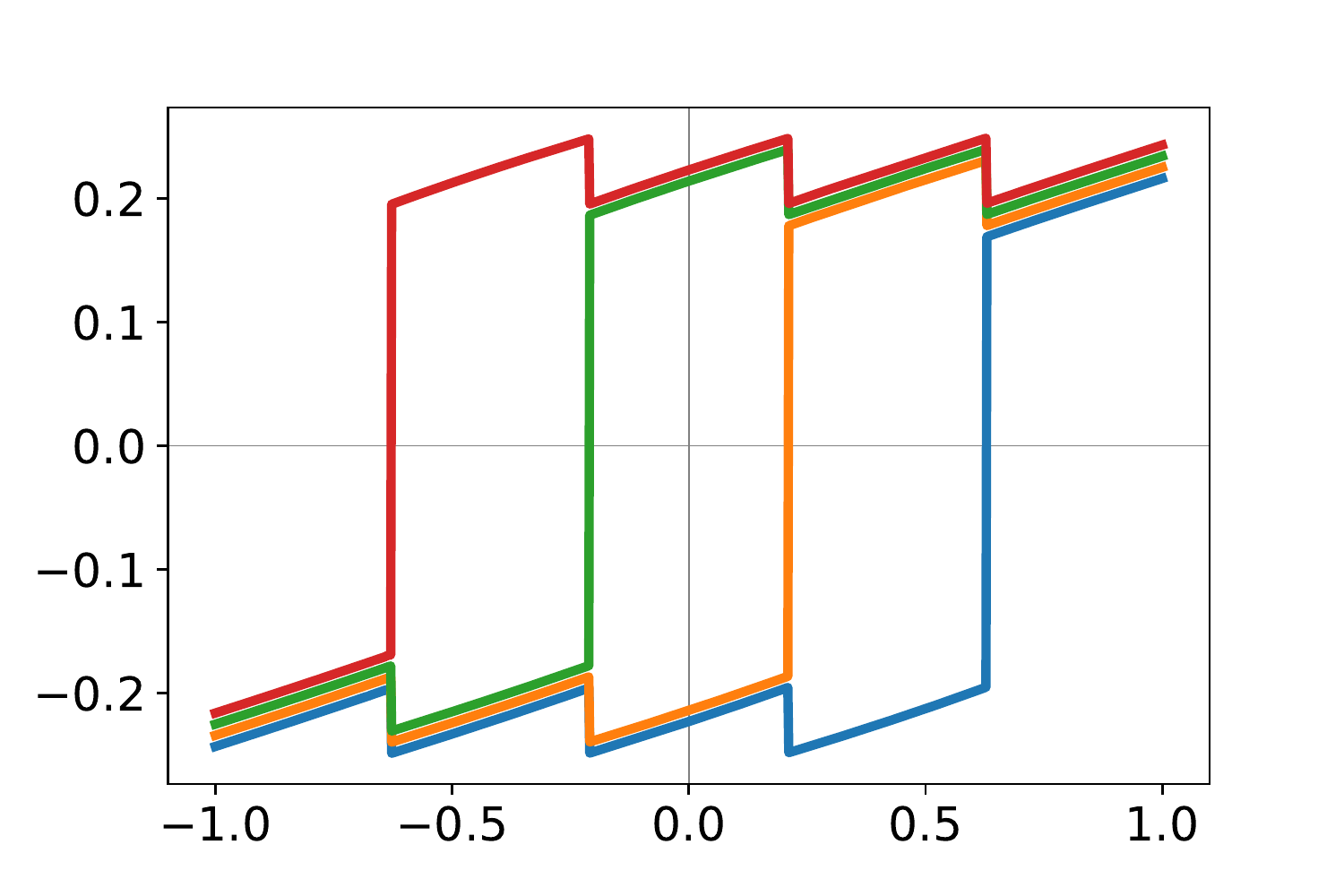}
	\qquad~~
	\includegraphics[width=.4\textwidth]{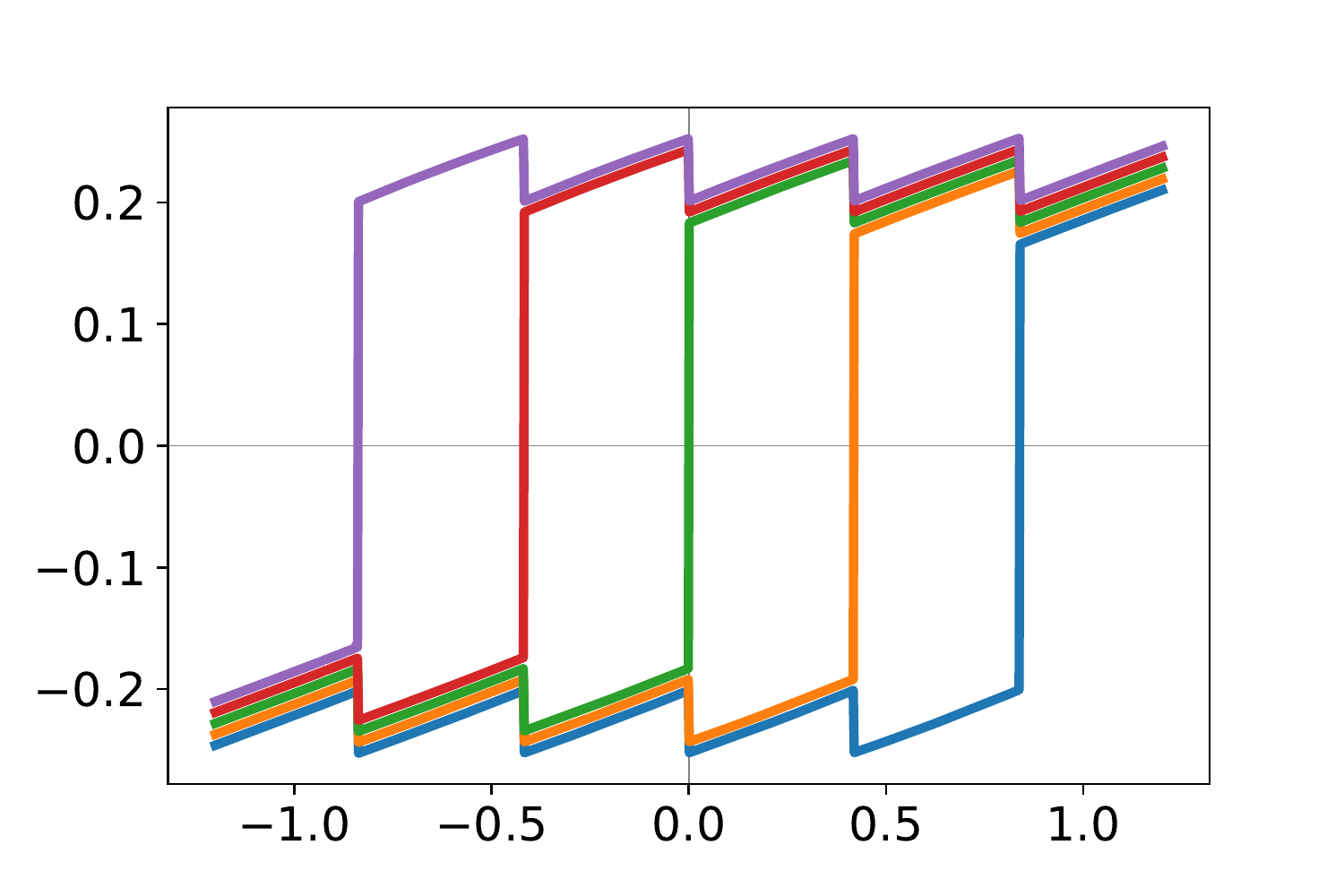}
	\put(-306,113){\large $M=4$}
	\put(-98,113){\large $M=5$}
	\put(-289,-17){\large $\wt{\lambda}$}
	\put(-80,-17){\large $\wt{\lambda}$}
	\put(-195,56){\large $\{e_k\}$}
	\put(-403,56){\large $\{e_k\}$}
	\vspace{-2mm}
	\caption{\label{fg:T0}The $\wt{\lambda}$-dependence of the minimum of $V_{\rm eff}(E)$ for $\wt{g}_1=1$ and $\wt{g}_2=3$ at $T=\mu=0$. Each $e_k$ is represented by a different color. The plots are vertically displaced slightly so that they do not overlap exactly.}
\end{figure}
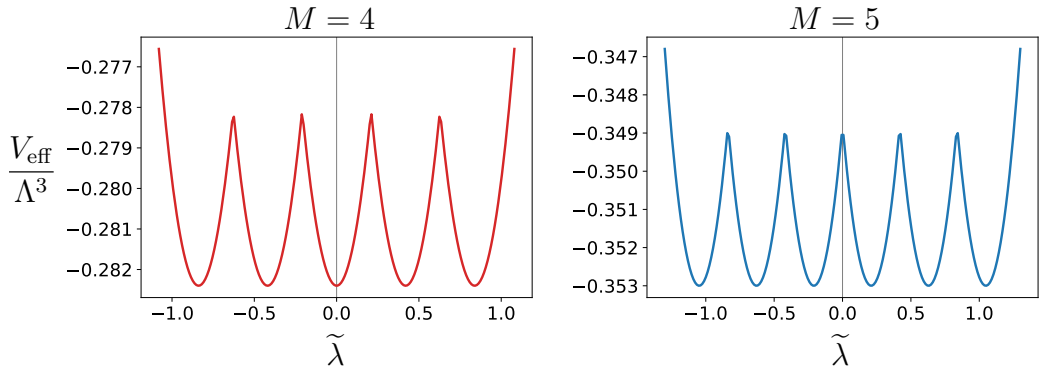
\begin{figure}[tb]
	\centering
	\includegraphics[height=.25\textwidth]{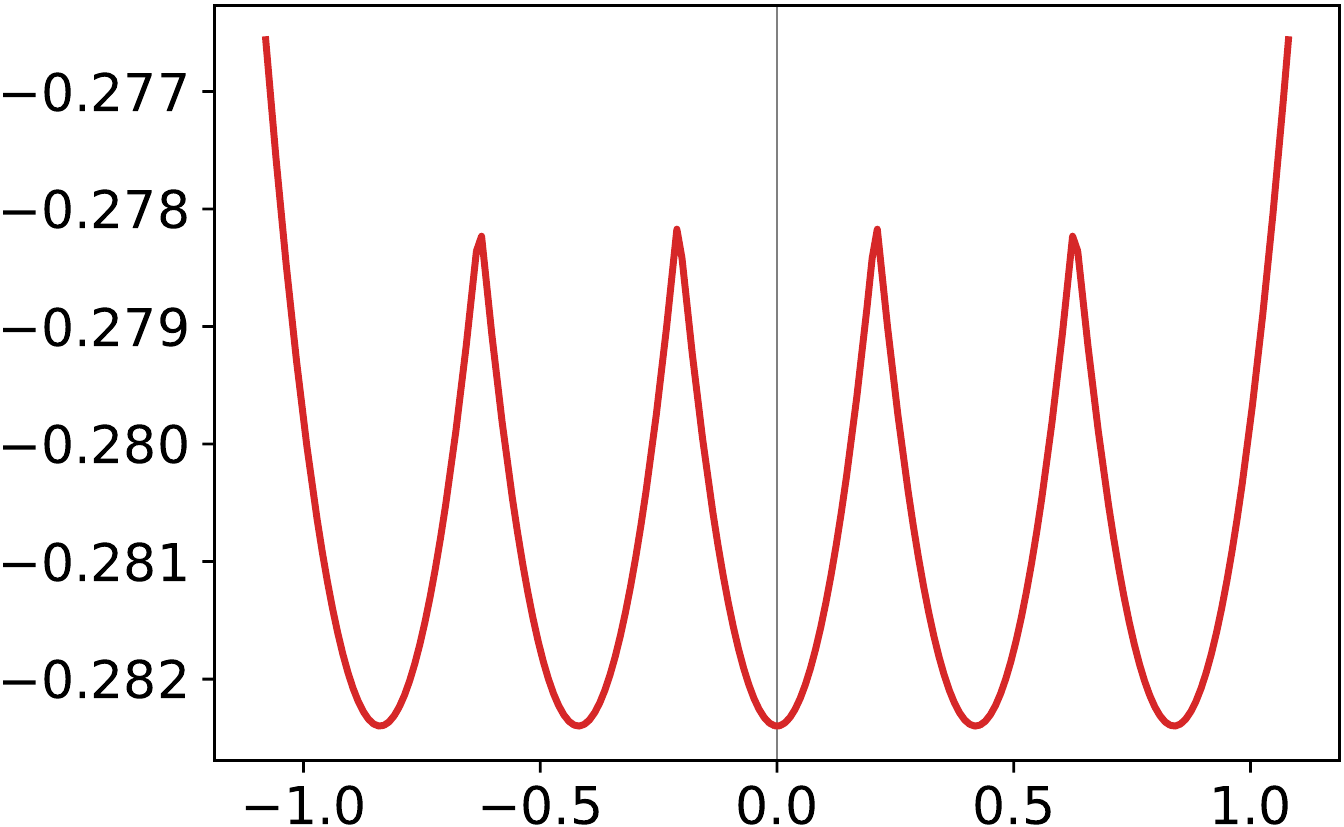}
	\quad 
	\includegraphics[height=.25\textwidth]{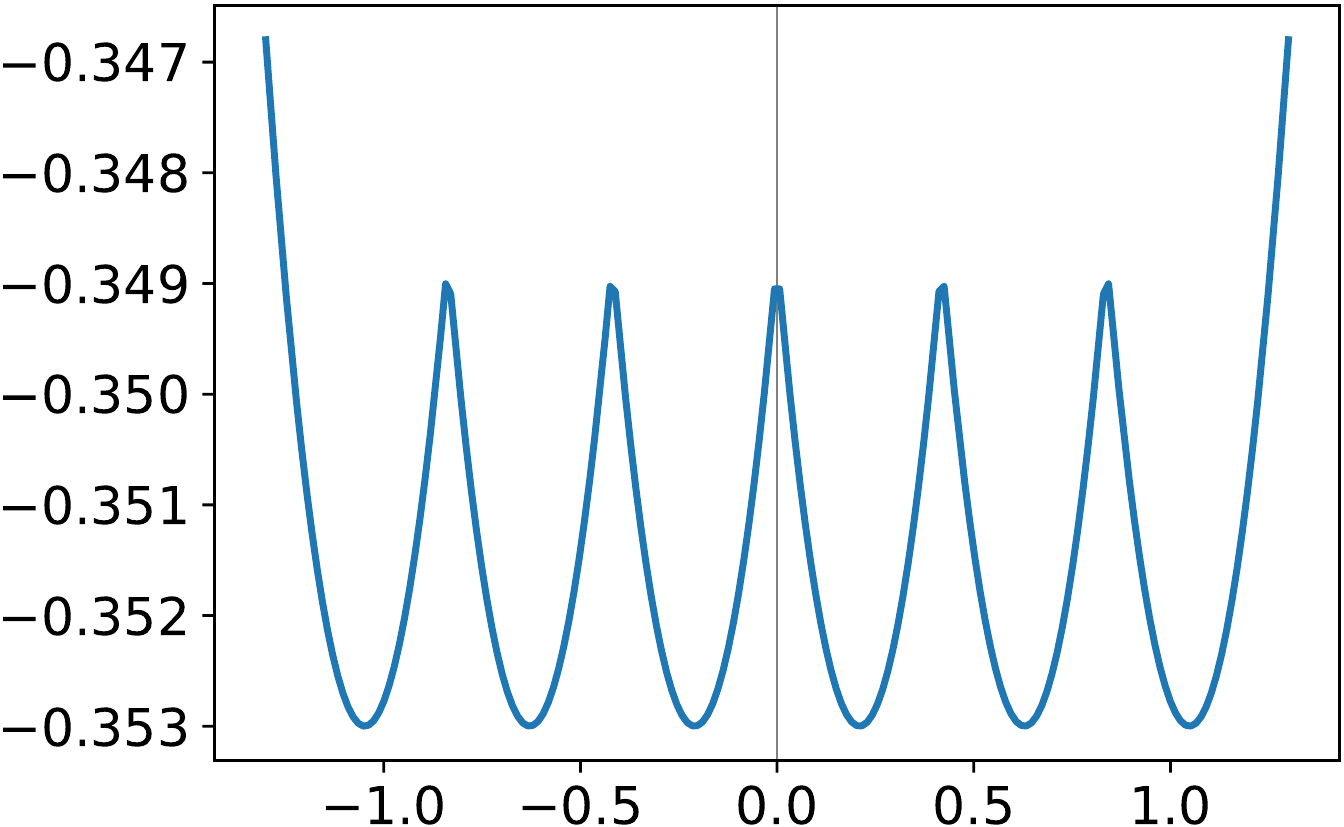}
	\put(-267,-17){\large$\wt\lambda$}
	\put(-78,-17){\large$\wt\lambda$}
	\put(-283,110){\large$M=4$}
	\put(-94,110){\large$M=5$}
	\put(-387,53){\large$\displaystyle\frac{V_{\rm eff}}{\Lambda^3}$}
	\vspace{-3mm}
	\caption{\label{fg:vac_energy}The ground state energy density for $\wt{g}_1=1$ and $\wt{g}_2=3$ as a function of $\wt\lambda$.}
\end{figure}
The ground state energy density shown in figure~\ref{fg:vac_energy} exhibits $M$ sharp kinks associated with the jump of the eigenvalues. In figure~\ref{fg:phases_at_T=mu=0} a schematic phase diagram is drawn for even $M$ and odd $M$, respectively. For large $|\wt{\lambda}|$ the $\U(M)$ symmetry is restored, while at intermediate $\wt\lambda$ the eigenvalues form two clumps, triggering symmetry breaking $\U(M)\to\U(M-k)\times\U(k)$. 

The above numerical findings will be rigorously derived in the next subsection.

 The natural pattern of flavor symmetry breaking for QCD$_3$ is $\U(M)\to \U(M/2)\times \U(M/2)$ for even $M$ and $\U(M)\to \U([M-1]/2)\times \U([M+1]/2)\times \mathbb{Z}_2$ for odd $M$, see~\cite{Pisarski:1984dj,Appelquist:1986qw,Appelquist:1989tc}. However, the sign of the Chern-Simons term can nullify
 this phase so that subleading saddle points become dominant resulting
 in symmetry breaking patterns $\U(M)\to\U(j)\times\U(M-j)$ and a cascade
 of phase transitions \cite{Komargodski:2017keh}.
 Recently it was proposed that QCD$_3$ with a large number of colors would undergo a sequence of first-order transitions when the flavor-singlet mass is varied, in exactly the same fashion as figure~\ref{fg:phases_at_T=mu=0} \cite{Armoni:2019lgb}.
   Although   we may have local minima leading to the symmetry breaking pattern
   $\U(M)\to\U(j)\times\U(M-j-1)\times\U(1)$, saddle points with even
   less symmetry
  are unnatural and unlikely to be global minima of the free energy. It may
  require fine tuning of the parameters if
  they exist. This is indeed the case for the model analyzed in the present work and makes our model a fascinating theoretical laboratory of ideas and methods for QCD$_3$. 
 
\subsection{\label{sc:mk}Analytical considerations}

\newcommand{\f}{{e}}

To simplify the expressions, in this subsection  we use the following variables 
\begin{tcolorbox}[ams equation,enhanced ,colback=white]
	\f_k=2 \wt{g}_2 \frac{E_k}{\Lambda^{3/2}} \,,\quad
	\gamma_1=\frac{3\pi \wt{g}_1^2}
	{2\wt{g}_2^2 (\wt{g}_2^2-M \wt{g}_1^2)}\geq 0\,,
	\quad \gamma_2=\frac{3\pi}{2\wt{g}_2^2}>0\,,\quad
	{\lambda}=2 \wt{g}_2 \wt{\lambda}
	\label{new-units}
\end{tcolorbox}\noindent
 Then, the potential takes the form
\ba
	\wh V_{\rm eff}(e) \equiv 
	\frac{6\pi V_{\rm eff}(E) }{\Lambda^3}
	& = \gamma_1\mkakko{\sum_{k=1}^{M}e_k-\lambda}^2 
	+ \sum_{k=1}^{M} v(e_k),
        & v(e) = {\gamma_2e^2+|e|^3- \mkakko{1+e^2}^{3/2}}
	\label{eq:Vvac.b}
\ea
with three parameters $\gamma_1$, $\gamma_2$ and ${\lambda}$ determining the phases. They essentially correspond to the relative strength of the two quartic interactions, the inverse strength of the interaction proportional to $g_2$ and the mass term, respectively, cf.~\eqref{eq:Ldef}. 

Our primary goal is to find the global minimum of $\wh V_{\rm eff}(e)$.  The extrema are determined by the following saddle point equations ($k=1,\ldots,M$)
\begin{equation}\label{saddpoint-vac}
\partial_{e_k}\wh V_{\rm eff}(e)=2s+g(e_k)=0
\end{equation}
where  
\begin{equation}
s=\gamma_1\left(\sum_{j=1}^M e_j-\lambda\right)\quad {\rm and}\quad  g(e)=\partial_e v\left(e\right)=2\gamma_2 e+3e\mkakko{|e|-\sqrt{1+e^2}}.
\end{equation}
In Appendix~\ref{sc:appA}, we study the solutions and some properties of the corresponding phase diagrams for general $g(e)$ and illustrate it with a simple but non-trivial toy model.

\begin{figure}[t!]
	\centering
	\includegraphics[width=.45\textwidth]{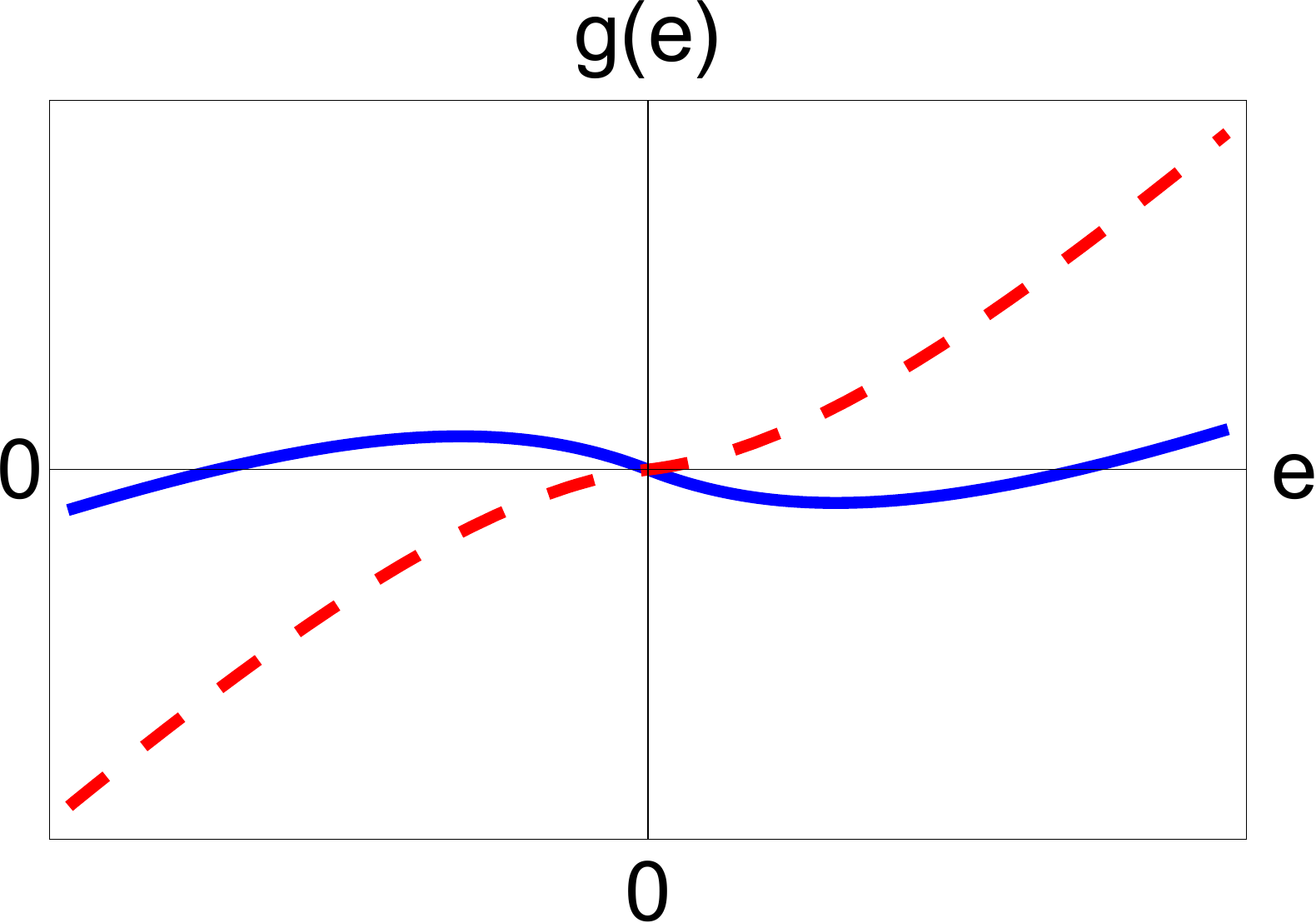}
	\caption{\label{fig:curve} The two forms of the function $g(e)=2\gamma_2 e+3e(|e|-\sqrt{1+e^2})$ for $\gamma_2<3/2$ (blue solid curve) and $\gamma_2\geq 3/2$ (red dashed curve).}
\end{figure}

To solve $\partial_{e_k}\wh V_{\rm eff}(e)=0$ for a fixed $s$, we note that $g(e)$ is a strictly monotonously increasing function if $\gamma_2\geq 3/2$ ($\Lambda<\pi/g_2^2$), see the red dashed curve in figure~\ref{fig:curve}, as can be seen from its derivative
\begin{equation}
\begin{split}
  g'(e)=&\;2\gamma_2+3\left(2|e|-\frac{1+2e^2}{\sqrt{1+e^2}}\right)\\
  =&\;2\gamma_2-3 +3(1+2|e|)\left(1-\frac{1+2e^2}{\sqrt{(1+2e^2)^2+e^2(4|e|^{-1}+1+4|e|)}}\right).
  \end{split}
\end{equation}
Hence, in the regime $\gamma_2\geq 3/2$  we have only a single real solution for $e_k=e^{(0)}$ with unbroken flavor symmetry. Explicit expressions for
$e^{(0)}$ are  derived in Appendix~\ref{app:Comp}. 
Let us emphasize that this part of the phase diagram will be avoided
when the cutoff $\Lambda$ is chosen large enough.

When $\gamma_2<3/2$, there may be two minima $e^{-}(s)<0<e^{+}(s)$
and a maximum $e^{0}(s)$ of the confining potential $v(e)$ for each $\f_k$ when  $s$ is fixed because the derivative $g(e)=v'(e)$ has the form of the blue curve sketched in figure~\ref{fig:curve}.  When $e_{\min}$ is the position of the local minimum of $g= v'(e)$, see \eqref{emin}, and $-e_{\min}$ the position of its local maximum,
this is the case if  $s\in[g(e_{\min})/2,g(-e_{\min})/2]$. When $s$ lies outside this interval, at large values of $|\lambda|$, all $e_k$ are the same and we are in
a phase without flavor symmetry breaking, see Appendix~\ref{app:Comp}
for explicit expressions.

\begin{figure}[t!]
	\centering
	\includegraphics[width=.45\textwidth]{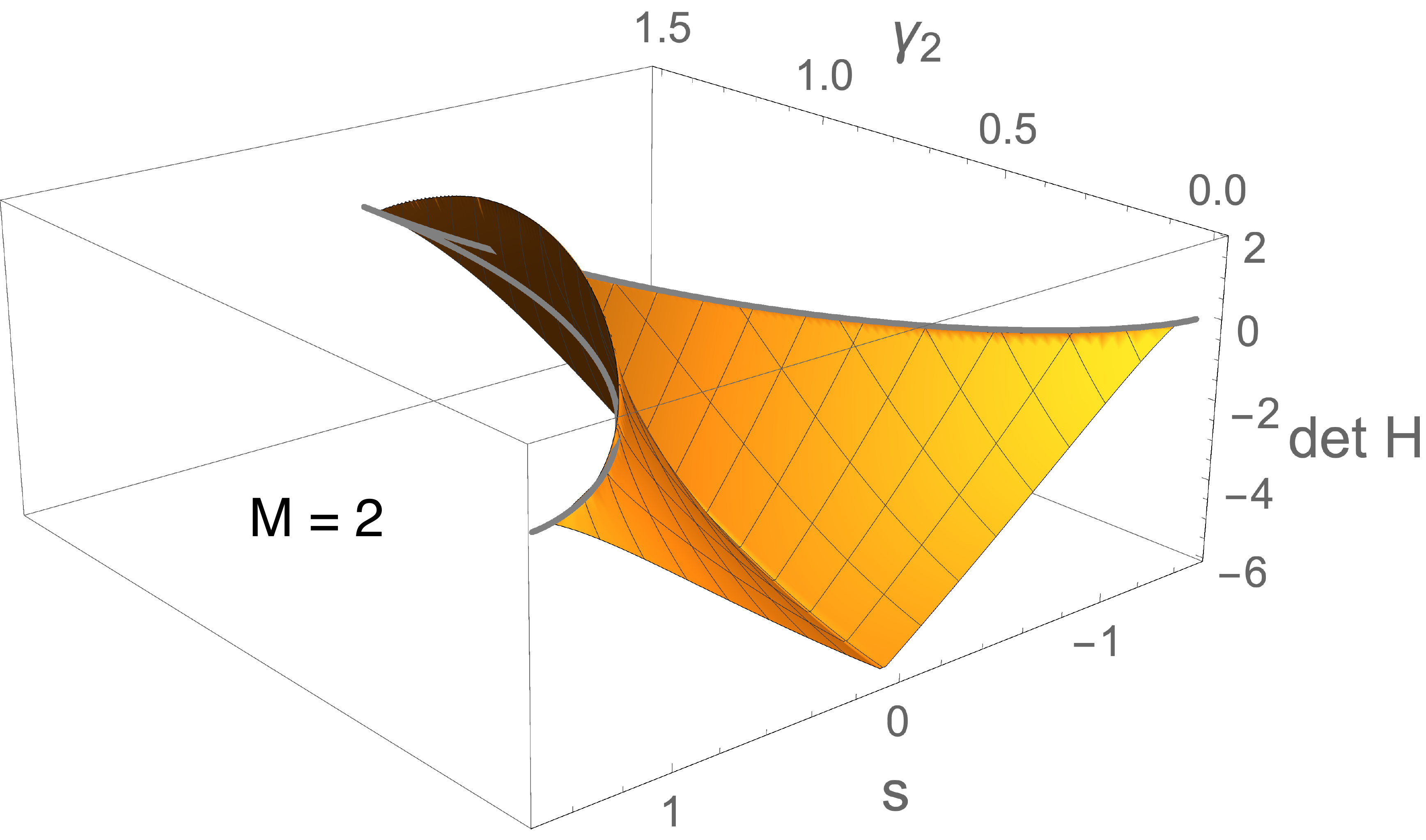}
	\includegraphics[width=.45\textwidth]{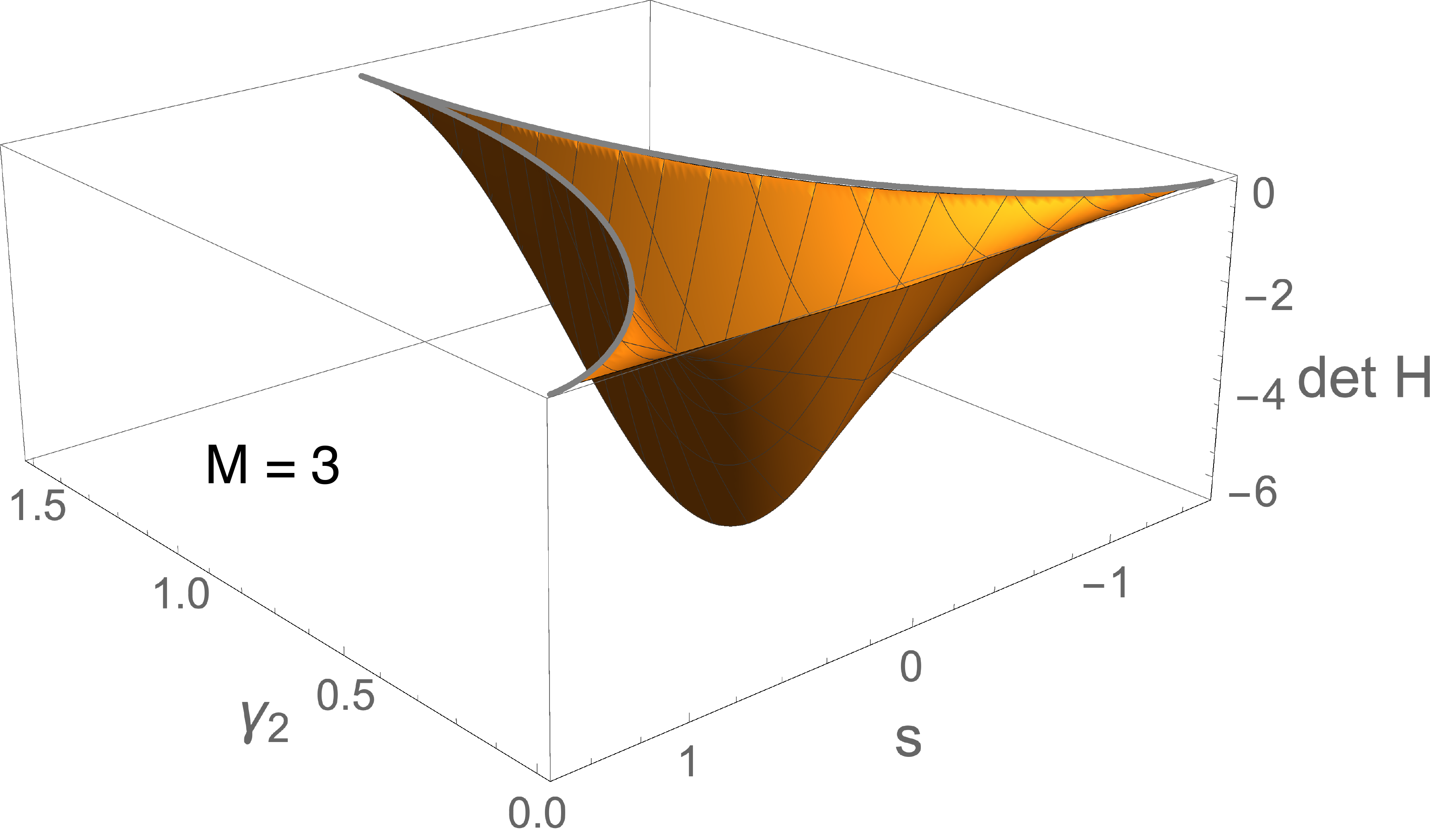}
	\caption{\label{fig:sum-sol-vac}The determinant of the
          Hessian, $\det H$, as a function of $s$ and $\gamma_2$ for $\gamma_1 =1$,
          $M=2$ (left) and  $M=3$ (right).
          The variable $s$ lies in the interval $[g(e_{\min})/2,g(-e_{\min})/2]$ when the
          saddle point equation has three solutions. The grey curves show the
          boundaries of this interval. For $M=2$ we give $\det H$ for the
          solution $(e^0, e^+)$ while for $M=3$ it is given for the
          solution  $(e^-,e^0, e^+)$. 
    }
\end{figure}

The question is which solutions of the saddle point equations are global
minima of $\wh V_{\rm eff}(e)$ so that we can conclude what kind of symmetry breaking patterns we can expect. For $M=2$, we can have either a solution with $ e_1 \ne  e_2$ leading to a symmetry
breaking pattern
$\U(2)\to\U(1)\times\U(1)$ or $ e_1 =  e_2$ with no symmetry breaking.
    For a second order phase transition to occur, the solution $( e^{0},  e^+)$
    should join smoothly with the solution $( e^{+},  e^+)$ as a function
    of $ \lambda$ (or $( e^{-}, e^-)$ joins with $( e^{-}, e^0)$).
    For this solution to be a global
    minimum, the Hessian at  $( e^{0},e^+)$ has to be positive definite.
In Appendix \ref{sec:locextimp} we have shown that for solutions with only
    one of the $g'(e_k) <0$, this is the case if
the determinant of the Hessian is positive.
In  figure \ref{fig:sum-sol-vac} we show the determinant of the
Hessian for $M=2$ and $M=3$       when $s$ varies from
$s_{\rm max}$ to $s_{\rm min}$. For $M=2$ the point where  $( e^{0},  e^+)$
coalesces with  $( e^{+}, e^+)$ is always at negative $s$. At this point
the determinant of the Hessian vanishes
(see grey curve in figure \ref{fig:sum-sol-vac}) and becomes negative
away from the minimum. This implies that a second order phase transition
does not occur for $M=2$ at $T=0$.

For $M>2$, $\gamma_1>0$ and $\gamma_2<3/2$, we can exploit insights
from the previous work \cite{Kanazawa:2019oxu} and the discussion 
in Appendix~\ref{sc:appA}. As is the case in the random matrix
theory  \cite{Kanazawa:2019oxu} on which the present model is based,
there are always the phases corresponding to the symmetry breaking pattern  $\U(M)\to\U(j)\times\U(M-j)$ with $j=0,\ldots,M$ (see Appendix \ref{sc:appA}).  
The integer $j$ is a
monotonously increasing function of ${\lambda}$ when the solution is given by $e=(e_a{\1}_{M-j},e_b{\1}_{j})$ with $e_a<e_b$.
The phase transition between the phase $j$ and $j+1$ with $j=1,\ldots,M-2$
evidently has to be of  first order because for a second order phase transition
to happen we need
$e_a=e_b$ at the transition point.

As we have seen in Appendix \ref{sec:nogo}
also the phase transitions between $j=0$ and $j=1$ as well as $j=M-1$ and $j=M$ are  of first order. 
This time the Hessian is positive semi-definite so that a second order phase transition might be possible,
but there is always a direction, namely the eigenvector of the Hessian with the eigenvalue $0$, whose leading term in the Taylor expansion becomes negative
(see Appendix~\ref{sec:nogo}). 

Although general arguments indicate \cite{Peskin:1980gc,Kogan:1984nb} that
the flavor breaking ground state should have maximum symmetry, more exotic
symmetry breaking patterns such as
$\U(M)\to\U(j)\times\U(M-j-1)\times\U(1)$ with $j=1,\ldots,M-1$, although
unlikely, can in principle appear. As can be shown from a plot of the Hessian
determinant as a function of $\gamma_2$ and $s$, the Hessian is always negative definite in such a phase, see figure~\ref{fig:sum-sol-vac} for $M=3$.
Therefore, there is no phase with the
symmetry breaking pattern  $\U(M)\to\U(j)\times\U(M-j-1)\times\U(1)$ at zero
temperature.

\section{\label{sc:temp}Nonzero temperature}
\subsection{Phase structure}
\begin{figure}[t!]
	\centering
	\includegraphics[width=0.8\textwidth]{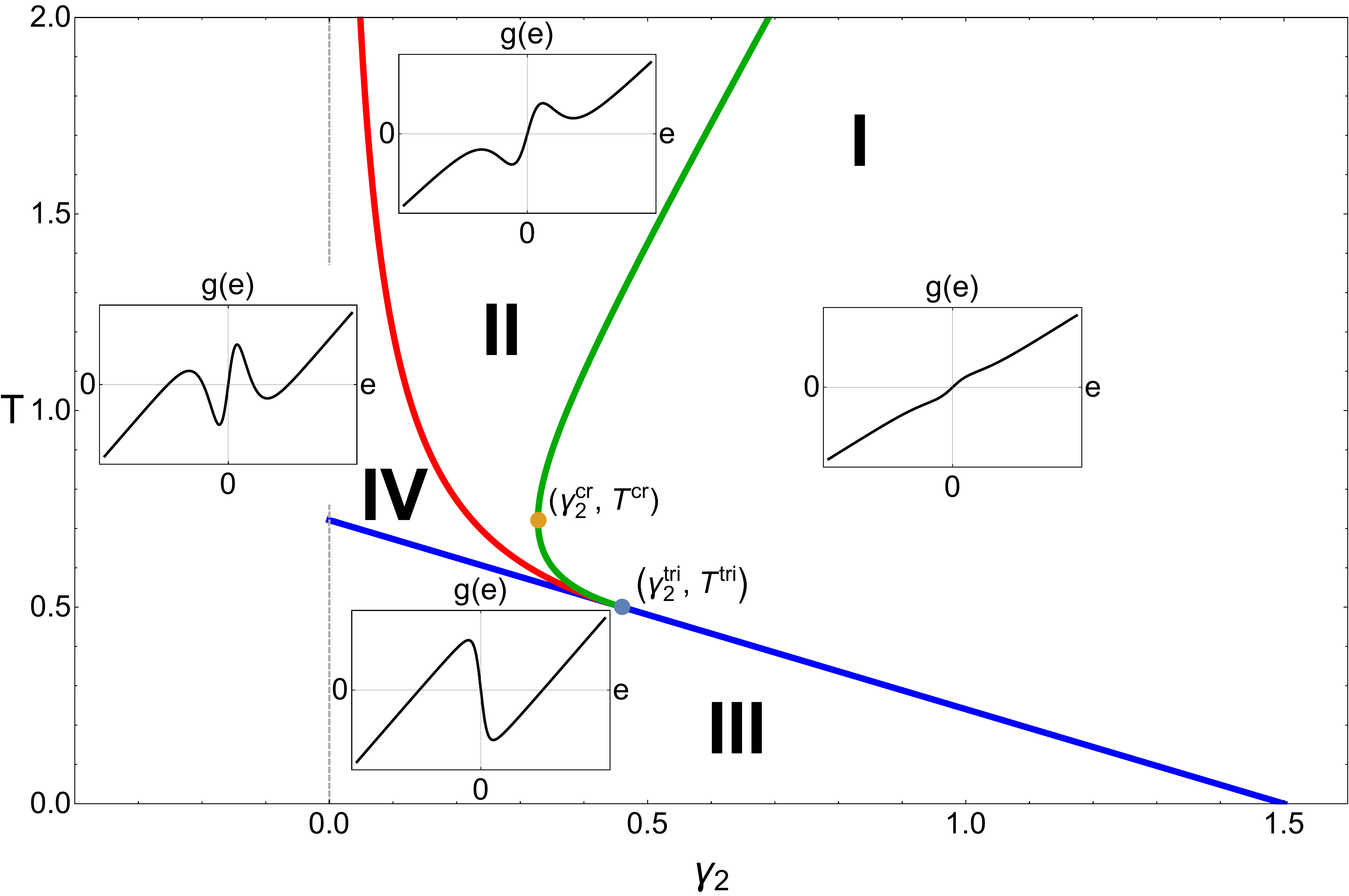}
	\caption{\label{fig:phase}Phase diagram of $g(e)$. For each region, labeled
          by Roman numerals, we included an inset with the  shape of $g(e) = v'(e)$.
          }
\end{figure}

The effective potential \eqref{eq:V00} at $T>0$ and $\mu=0$ can be evaluated analytically. Absorbing  $\Lambda$ in $T \to \Lambda^{3/2} T$, we have
\ba
	\frac{V_{\rm eff}(E)}{\Lambda^3} =\; & 
	\frac{\wt{g}_1^2}{\wt{g}_2^2-M\wt{g}_1^2}\mkakko{\sum_{k=1}^{M}e_k-\wt{\lambda}}^2
	+ \sum_{k=1}^{M}\Bigg\{e_k^2+\frac{4}{3\pi}|\wt{g}_2e_k|^3 - \frac{1}{6\pi}(1+4\wt{g}_2^2e_k^2)^{3/2} 
	\notag
	\\
	& 
	+ \frac{{T}^3}{\pi}\kkakko{
		\frac{2|\wt{g}_2e_k|}{{T}}\;\text{Li}_2\mkakko{-\rme^{-2|\wt{g}_2e_k|/{T}}}
		+ \text{Li}_3 \mkakko{-\rme^{-2|\wt{g}_2e_k|/{T}}}
	}\Bigg\},
	\label{eq:VT}
\ea
where $\displaystyle\text{Li}_s(z)=\sum_{k=1}^{\infty}\frac{z^k}{k^s}$ is the polylogarithm function.%
\footnote{This series is convergent for $|z|<1$.  The values for $|z|\geq 1$ are defined by analytic continuation.} In the notation of~\eqref{new-units}, we find
\begin{equation}\label{pot.mu.finite}
\begin{split}
\widehat{V}_{\rm eff}(e)=\ &\gamma_1\mkakko{\sum_{k=1}^{M}e_k-\lambda}^2 
	+ \sum_{k=1}^{M}\biggl\{\gamma_2e_k^2+|e_k|^3- (1+e_k^2)^{3/2}\\
	&+6{T}^3\biggl[\frac{|e_k|}{{T}}\text{Li}_2\mkakko{-\rme^{-|e_k|/{T}}}+\text{Li}_3\mkakko{-\rme^{-|e_k|/{T}}}\biggl]\biggl\}.
\end{split}
\end{equation}
Thus, the saddle point equation becomes
\begin{equation}\label{saddle-temp}
-2s=g(e)\quad{\rm with}\quad g(e)=2\gamma_2 e+3e\mkakko{|e|-\sqrt{1+e^2}}+6{T} e\,{\rm log}\left(1+e^{-|e|/{T}}\right)
\end{equation}
and $s=\gamma_1(\sum_{k=1}^{M}e_k-\lambda)$  the same as before. The possible shapes are depicted in the insets of  figure~\ref{fig:phase}.

Depending on the temperature and $\gamma_2$ we can distinguish 4 different domains
depending on the maximum number of different solutions of $g(e)= -2s$, see
insets in figure
\ref{fig:phase}.
The domains are separated by the following curves:

\begin{itemize}
\item[i)] The vanishing of the slope at $e=0$ (blue line in figure \ref{fig:phase}),
\be
g'(e=0) = -3 + 2 \gamma_2+6 T \log 2=0. 
\ee
Because the asymptotic behavior of $g(e) \approx 2\gamma_2 e$, if the
slope at 0 is negative
$g(e)$ cannot be a monotonic function, and the equation $g(e) = -2s$ can have three solutions
for $T<(3-2\gamma_2)/(6\log 2)$.
\item[ii)] The curve in the $(\gamma_2, T)$ plane (red curve in figure \ref{fig:phase}) with
  \be 
  g(e) = g'(e) = 0
  \ee
 separates region II and region IV.
  At those points a minimum of $g(e)$ touches the $e$-axes, and because $g(e)$ is an odd function,
  the equation $g(e) = -2s$ can have 5 possible real solutions in the region IV.

\item[iii)] The curve in the $(g_2,T)$ plane when a minimum and a maximum of $g(e)$ coincide
   (green curve in figure \ref{fig:phase}) is given by
  \be
  g'(e)=g''(e)= 0.
  \ee
  It indicates definitely a phase transition that splits the region above curve i) and ii) into  regions I and  II. In  region I
 the function $g(e) $ increases monotonously, while  in region II the equation $g(e) = -2 s$ can have
  at most three solutions despite $g(e)$ has two local minima and two local maxima.
\end{itemize}

At the tricritical point, the potential which had three minima in the region V,
joins the potential in regions
I and II, with one and two minima, respectively. Because the potential is even,
this has to happen at $e=0$. The condition for the tricritical point is thus
\be
g'(e=0)&=&-3 + 2 \gamma_2+6 T \log 2=0,\\
g''(e=0)&=&-\frac 32 +\frac 3{4T} =0,
\ee
which is solved by
\be
(\gamma_2^{\rm tri}, T^{\rm tri}) = 
\mkakko{\frac 32(1-\log 2), \frac 12}.
\ee
A second special point in the $(\gamma_2,T)$ plane is the point on the curve
$ g'(e)=g''(e)= 0$ where
\be
\left. \frac {\rmd g_2}{\rmd T}\right |_{\gamma_2^{\rm cr}, T^{\rm cr}} =0.
\ee
This point is at $\gamma_2^{\rm cr}=0.3278$ with $T^{\rm cr} = (3-2 \gamma_2)/(6 \log 2)$. For $ \gamma_2 < \gamma_2^{\rm cr} $ the system always experiences a cascade of phase transitions when varying $\lambda$.

\subsection{High temperature regime}\label{sec:hightemp}

At sufficiently high temperatures and fixed  $\gamma_2>0$, see region II
in figure~\ref{fig:phase}, the curve  $g(e)$ shows  a ``wiggle'' (a local maximum followed by a minimum) for large $|e|$.
Taking into account
the arguments of \cite{Kanazawa:2019oxu} and the discussion
in Appendix~\ref{sec:cascade}, we expect a cascade of phase transitions for sufficiently large $|\widehat{\lambda}|$.
This is indeed observed numerically, see the plots in figures~\ref{fg:pd_mu0_M2_broken_strip} as well as~\ref{fg:mu0_strongphase} for $M=2,3,4$. It shows as a strip which obeys
approximately a linear relation between ${T}$ and ${\lambda}$. The cascade of symmetry breaking patterns are those of $\U(M)\to\U(j)\times\U(M-j)$ where $j=0,1,\ldots,M-1$ changes by $1$.

In Appendices \ref{sec:cascade} and \ref{sec:nogo} we have argued that all phase transitions
for a locally double well shaped potential have to be of first order for $M\ge 3$. For $M=2$, a second
order phase transition is possible, but our numerics confirm that at high temperature
all transitions are first order.
 As we will see in the next section, a second order phase transition
does  occur for $M=2$  at lower temperatures.

\subsection{Low temperature regime}

At low temperature $T< (3-2\gamma_2)/(6\log2)$ and $\gamma_2 <\frac 32$
(region III in figure \ref{fig:phase})
we find a $g(e)$ in the shape of a wiggle,
this time about the origin.
When increasing the temperature we encounter three scenarios depending on the
value of $\gamma_2$ which will be discussed in the next three subsections.

\begin{figure}[tb]
	\centering
	\includegraphics[width=.5\textwidth]{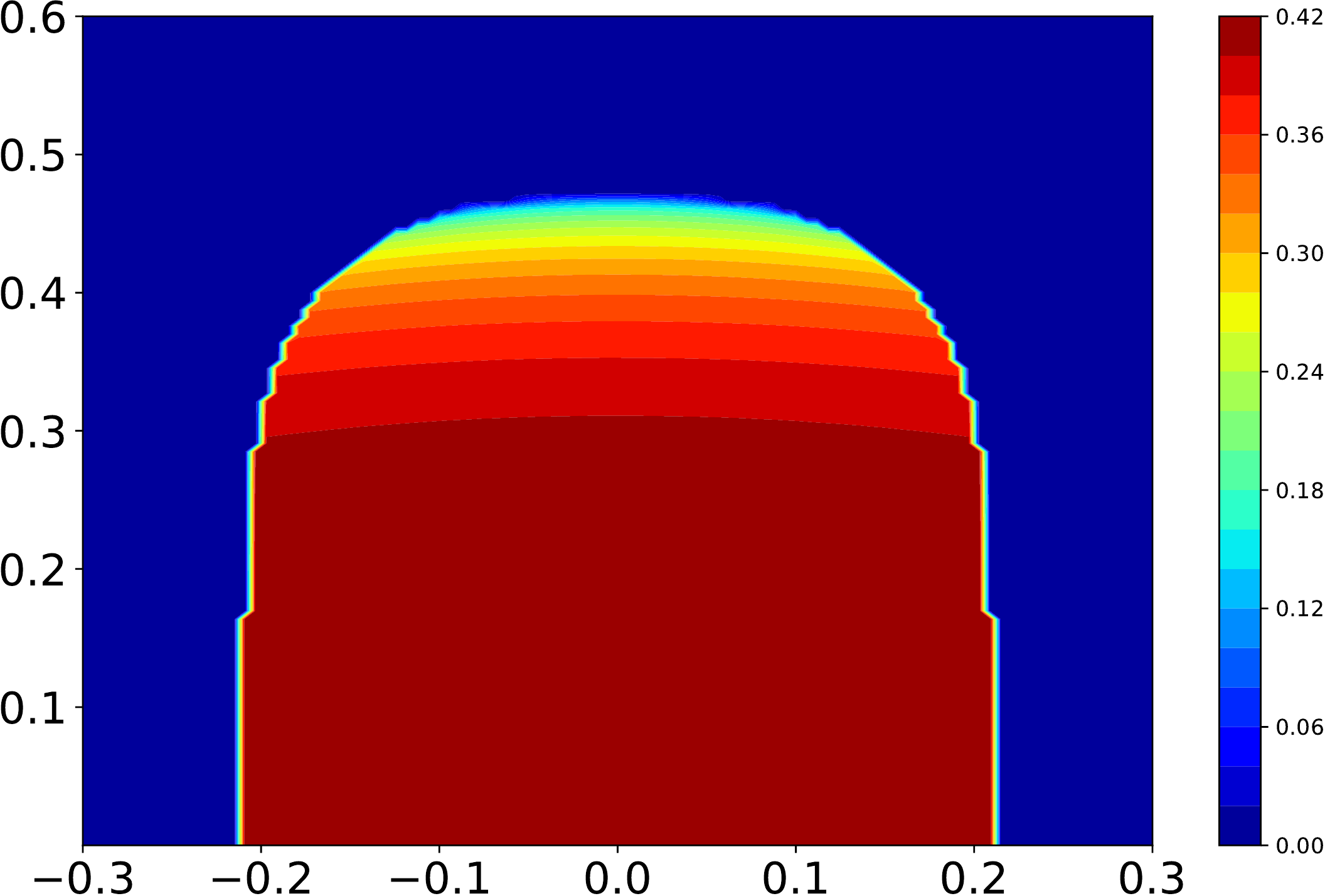}
	\quad 
	\raisebox{18pt}{\includegraphics[width=.39\textwidth]{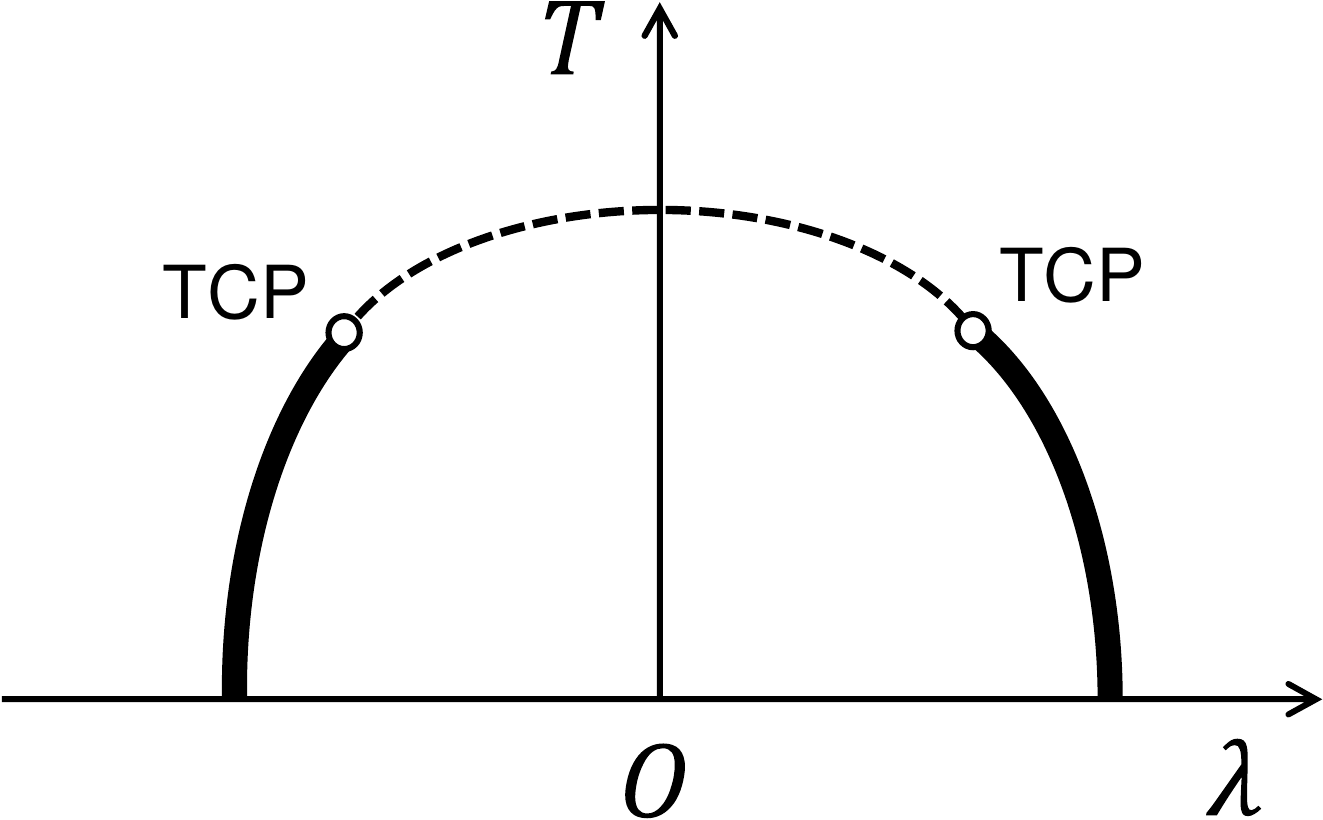}}
	\put(-300,-17){\large $\wt\lambda$}
	\put(-412,73){\large ${T}$}
	\put(-420,140){\Large (a)}
	\put(-170,140){\Large (b)}
	\caption{\label{fg:pd_mu0_M=2_weak_g2}(a)~The phase diagram for $M=2$ at $\mu=0$ with $\wt{g}_1=1$ and $\wt{g}_2=3$ in the large-$N$ limit. The magnitude of $|e_1-e_2|$ is plotted. (b)~A simplified sketch of (a). There are two tricritical points (TCP). The dashed line denotes a second-order transition and the thick solid line a first-order transition.  We have omitted the high temperature phase diagram where a strip
          of first order phase transitions starts 
at about ${T}=1.5$  and $\wt \lambda = 6.2$.}
\end{figure}

\subsubsection{Low temperature regime with
$\sqrt \pi <|\wt g_2| < \wt g_2^{\rm tri}$ 
($\gamma_2^{\rm tri} <\gamma_2<\frac 32$)}

For $\gamma_2^{\rm tri}<\gamma_2<\frac 32$, the function $g(e)$ becomes
a strict monotonously increasing function
for  $T> (3-2\gamma_2)/(6\log2)$, as we already
have seen at $T=0$ for $\gamma_2>\frac 32$
(or $|\widetilde g_2|<\sqrt \pi$). The second order phase
transition on the curve  $T=(3-2\gamma_2)/(6\log2)$ is at  $  \lambda=0$. For the
parameters of figure
\ref{fg:pd_mu0_M=2_weak_g2} this gives a critical temperature of
$T=0.469$. For $M=2$ the transition remains of second order until the tricritical
point in the $(\gamma_2, T)$ plane.
The location of this point can be extracted numerically. For  $\wt g_2 = 3$
($\gamma_2 = 0.5236$)
it is at $(\wt \lambda, T) = (\pm 0.1010, 0.457)$.
There is another second order phase transition point in the $(\gamma_2, T)$ plane at $ T = 1.499$
(not displayed in figure \ref{fg:pd_mu0_M=2_weak_g2}). 
This is the starting point of a strip of two close first order phase transitions
in the $(\wt \lambda,T)$ plane and begins  at $\wt \lambda = 6.2$. The first order transitions
are from a phase with  unbroken flavor symmetry 
to a phase with $\U(1) \times \U(1)$ breaking and back to the unbroken phase.
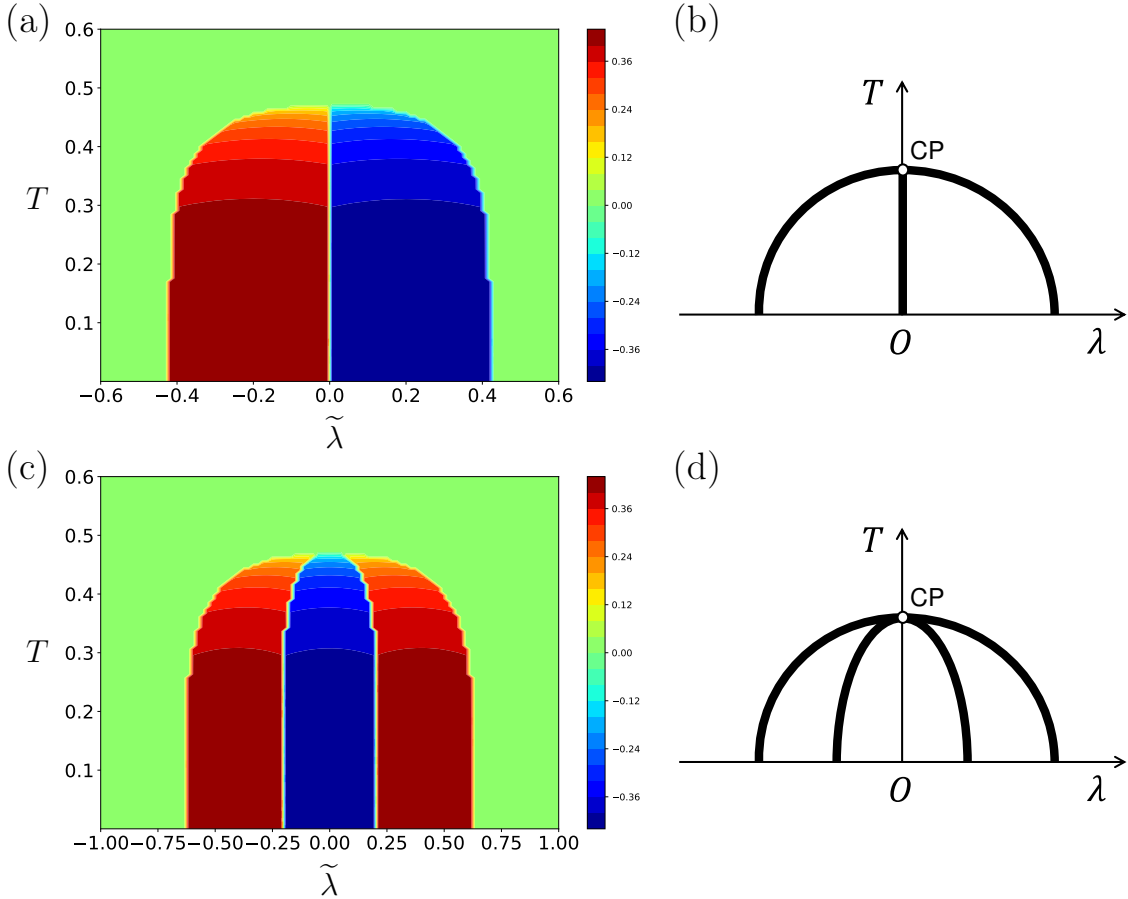
\begin{figure}[tb]
	\centering
	\includegraphics[width=.5\textwidth]{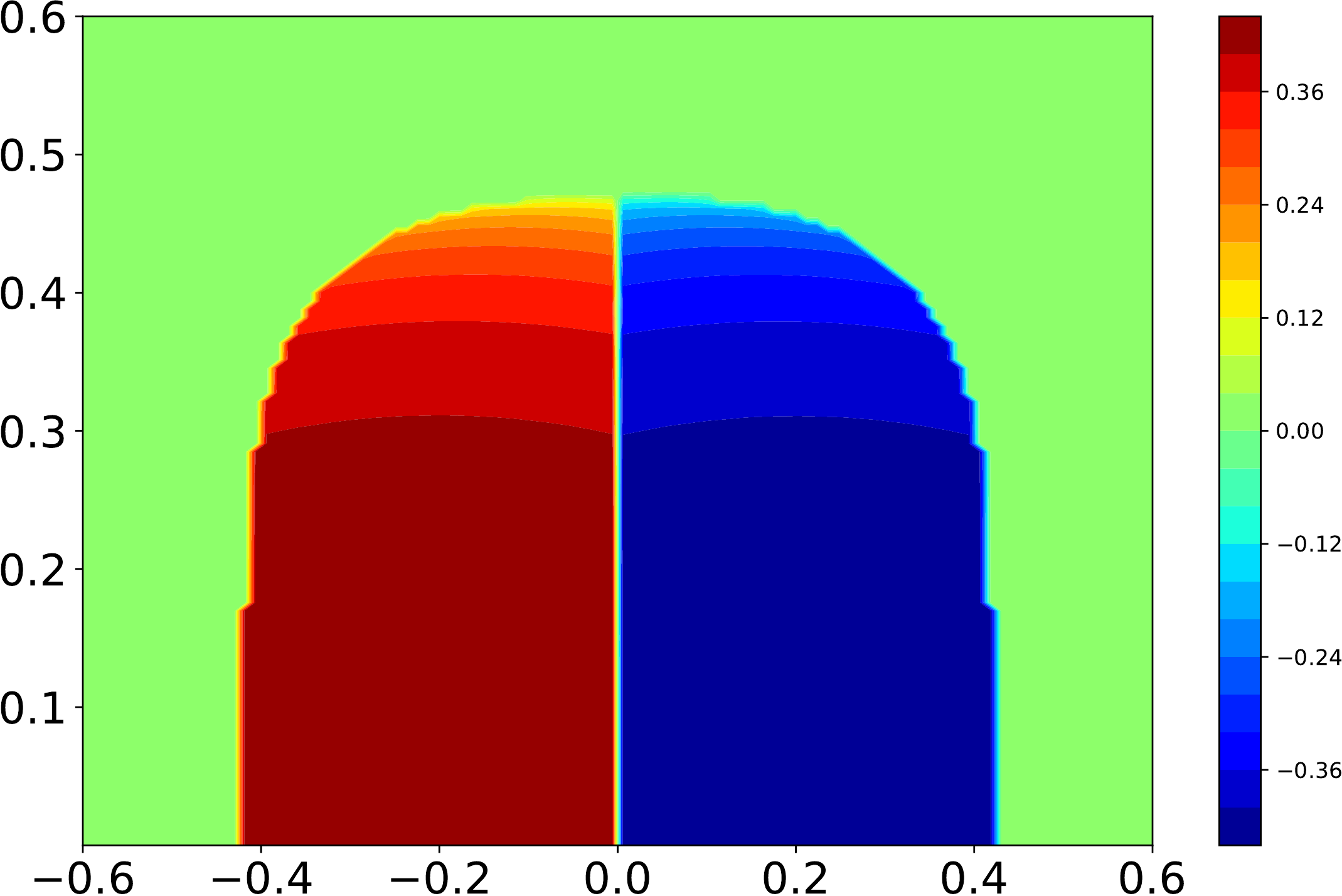}
	\quad 
	\raisebox{18pt}{\includegraphics[width=.39\textwidth]{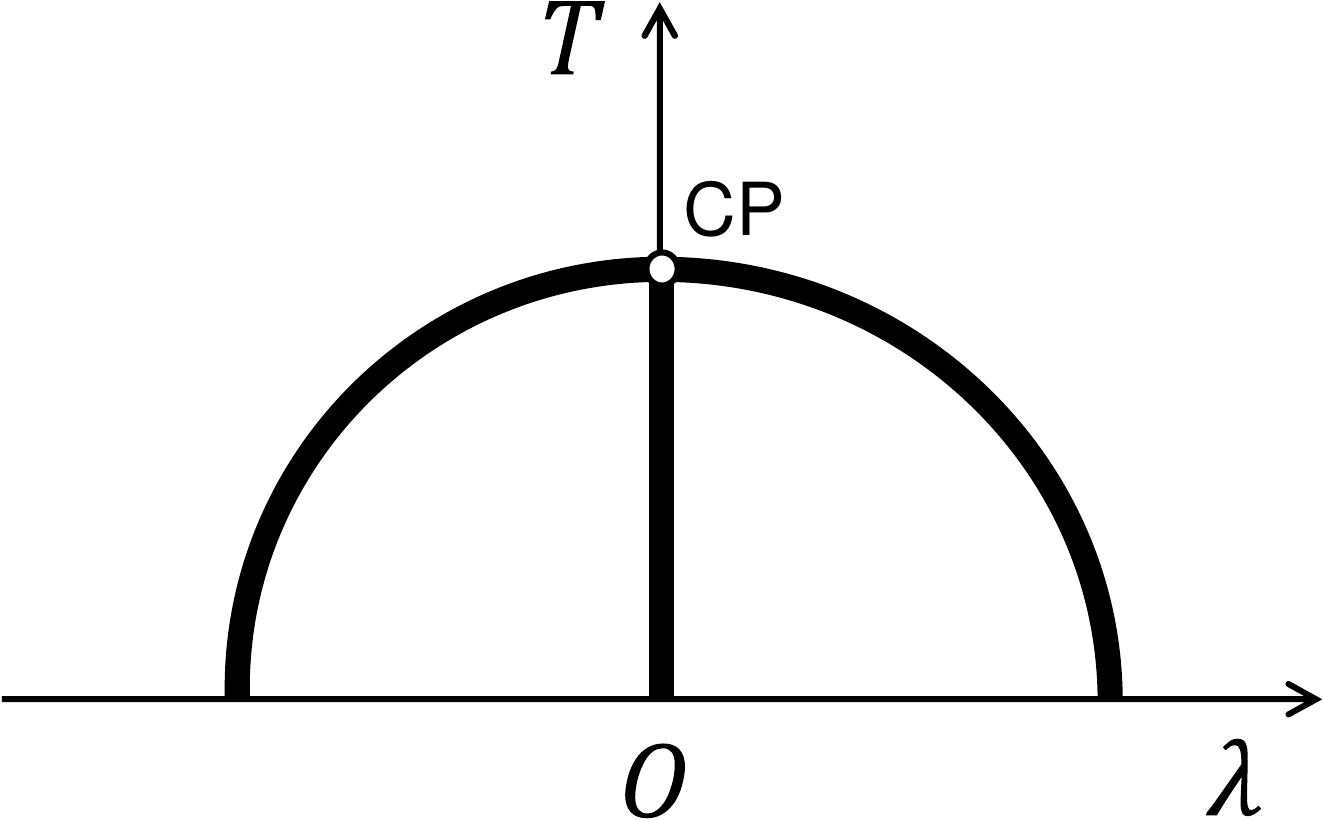}}
	\put(-300,-17){\large $\wt\lambda$}
	\put(-412,73){\large ${T}$}
	\put(-420,140){\Large (a)}
	\put(-170,140){\Large (b)}
	
	\centering
	\includegraphics[width=.5\textwidth]{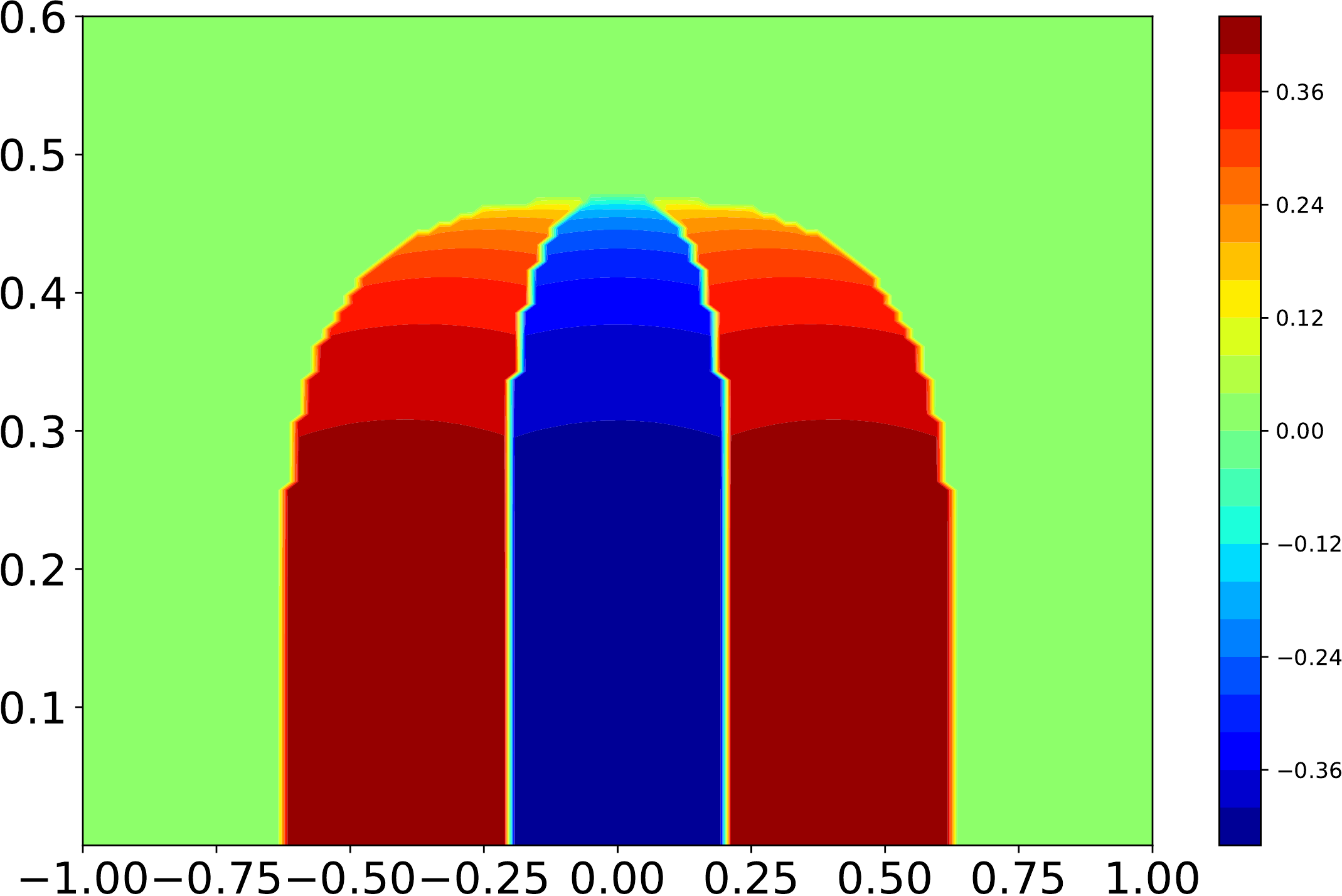}
	\quad 
	\raisebox{18pt}{\includegraphics[width=.39\textwidth]{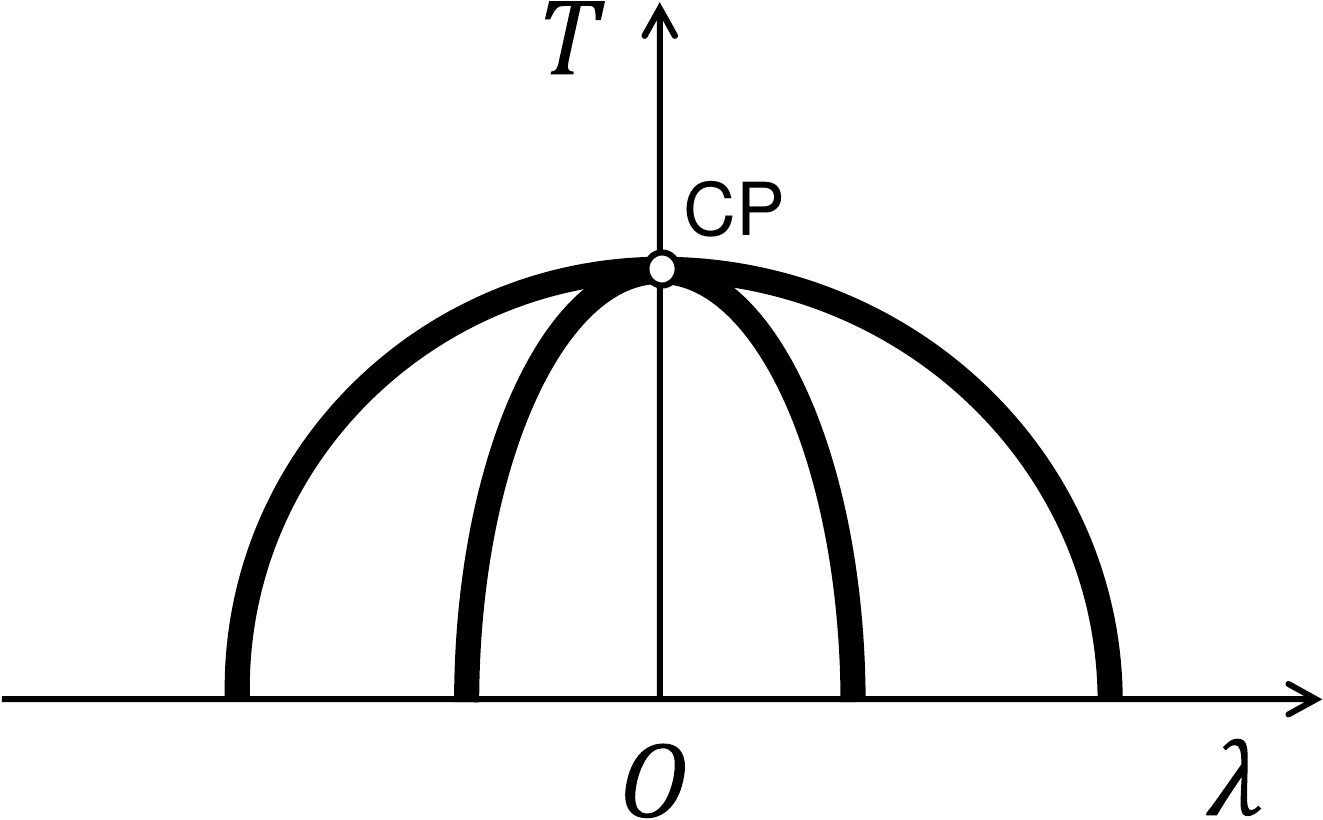}}
	\put(-302,-17){\large $\wt\lambda$}
	\put(-412,71){\large ${T}$}
	\put(-420,140){\Large (c)}
	\put(-170,140){\Large (d)}
	\caption{\label{fg:pd_mu0_M=3_weak_g2}(a)~The phase diagram for $M=3$ at $\mu=0$ with $\wt{g}_1=1$ and $\wt{g}_2=3$ in the large-$N$ limit. 
	The value plotted is $e_1-2e_2+e_3$, where the ordering $e_1\geq e_2\geq e_3$ is assumed. (c)~The phase diagram for $M=4$ at $\mu=0$ with $\wt{g}_1=1$ and $\wt{g}_2=3$ in the large-$N$ limit. 
	The value plotted is $e_1-2e_2+2e_3-e_4$, where the ordering $e_1\geq e_2\geq e_3 \geq e_4$ is assumed. The diagrams (b) and (d) are
        simplified versions of figures (a) and (c). There is a critical point (CP) at which all first-order transition lines meet. Again we have omitted the high temperature regime.}
\end{figure}

Adopting the order parameter
$\sum_{k=2}^{M}(-1)^k(e_{k-1}-e_k)$ (assuming $e_1\geq e_2\geq \cdots e_M$) the phase
diagram for $M=2$ in the $(\wt g_2,T)$ plane is mapped out in
figure~\ref{fg:pd_mu0_M=2_weak_g2}. The second order line extends from
$ \lambda =- \lambda^{\rm tri}$ to $ \lambda =  \lambda^{\rm tri}$.
We have omitted the high temperature regime in this figure which contains the
strip of first order phase transitions. It will be discussed in more detail
in the next subsection.

 As is discussed in Appendix~\ref{sec:nogo} for $M\ge 3$
there is no line of second order phase transitions in the $(\wt \lambda,T)$ plane
and the  only  second order point at $\wt \lambda =0$.

Moreover, we expect a cascade of phase transitions between phases with the symmetry breaking pattern $\U(M)\to\U(j)\times\U(M-j)$ with $j=0,1,\ldots,M$,
as in the high temperature phase, the system runs through all possible $j$ from $j=0$ to $j=M$ when increasing $\widehat{\lambda}$. We have corroborated this by numerical minimization of the potential~\eqref{eq:VT} for $M=3$ and $M=4$ (see figure~\ref{fg:pd_mu0_M=3_weak_g2}) where in both cases we have chosen $\wt{g}_2=3<|\wt{g}_2^{\rm tri}|$. Again we did not consider the high temperature phase.

\begin{figure}[tb]
	\centering
	\includegraphics[width=.45\textwidth]{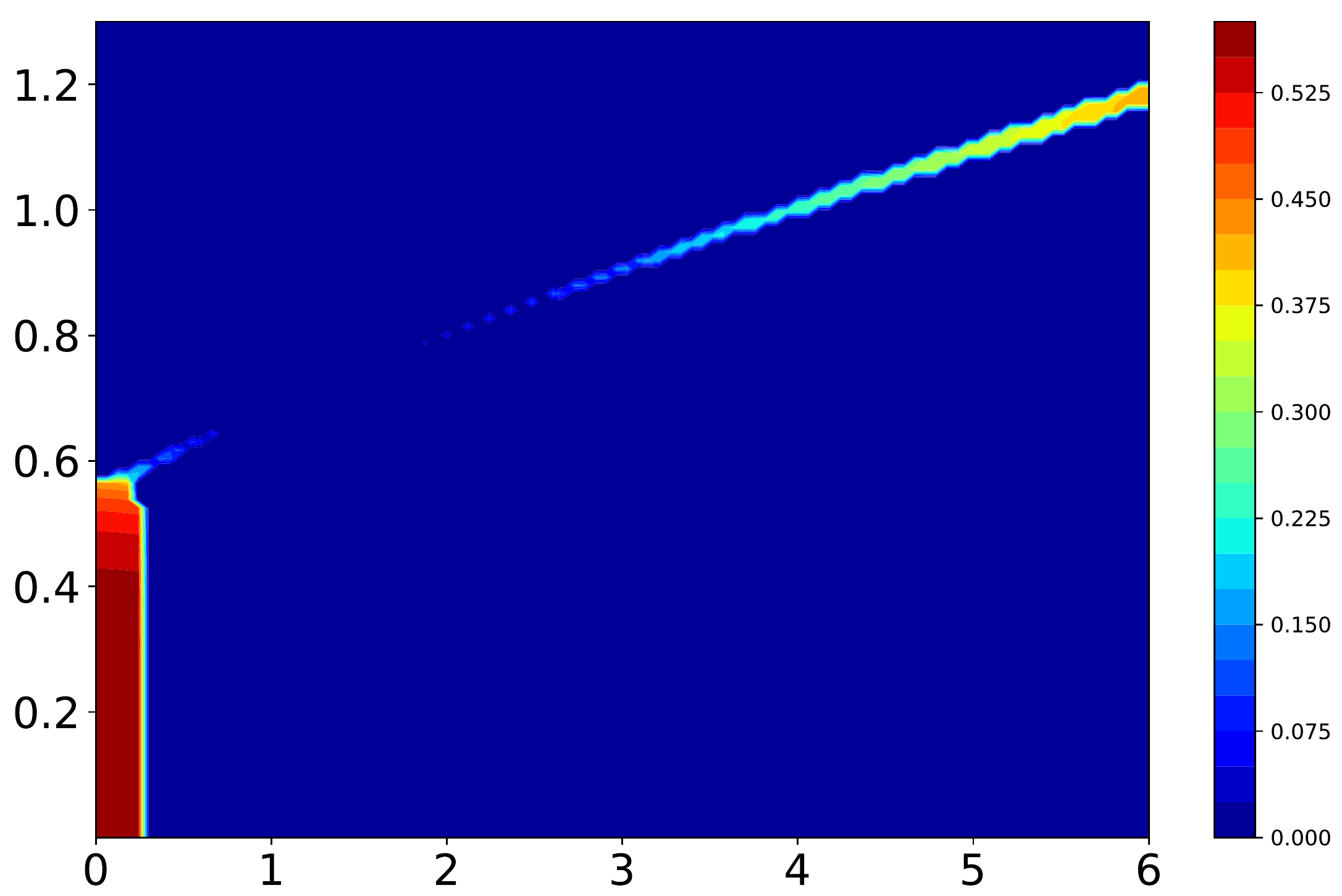}
	\quad 
	\includegraphics[width=.45\textwidth]{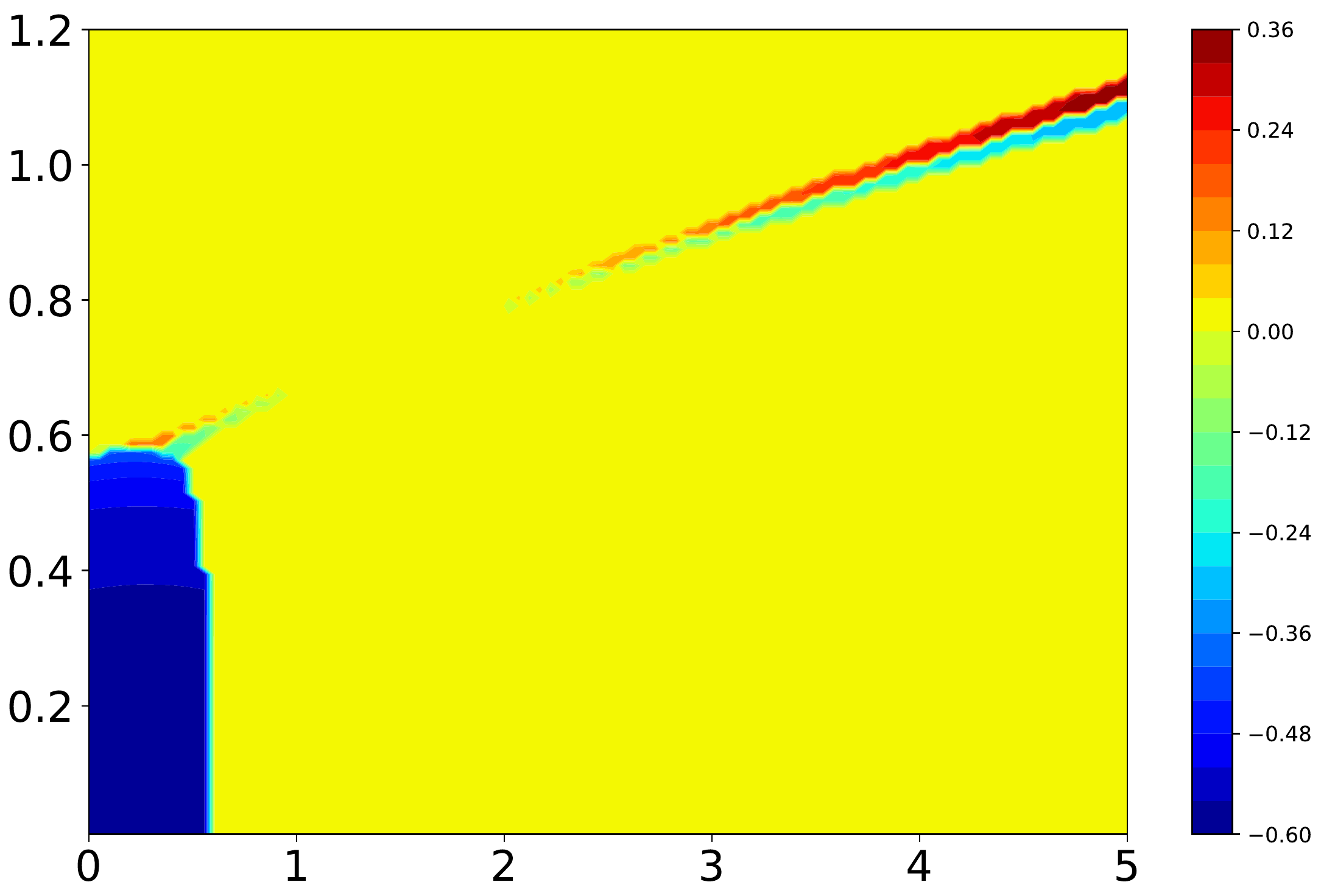}
	\put(-310,-17){\large $\wt\lambda$}
	\put(-412,73){\large ${T}$}
	\put(-103,-17){\large $\wt\lambda$}
	\put(-205,73){\large ${T}$}
	\put(-415,140){\Large (a)}
	\put(-195,140){\Large (b)}
	\put(-330,133){\Large $M=2$}
	\put(-125,133){\Large $M=3$}
	\vspace{2mm}\\
	\centering
	\includegraphics[width=.5\textwidth]{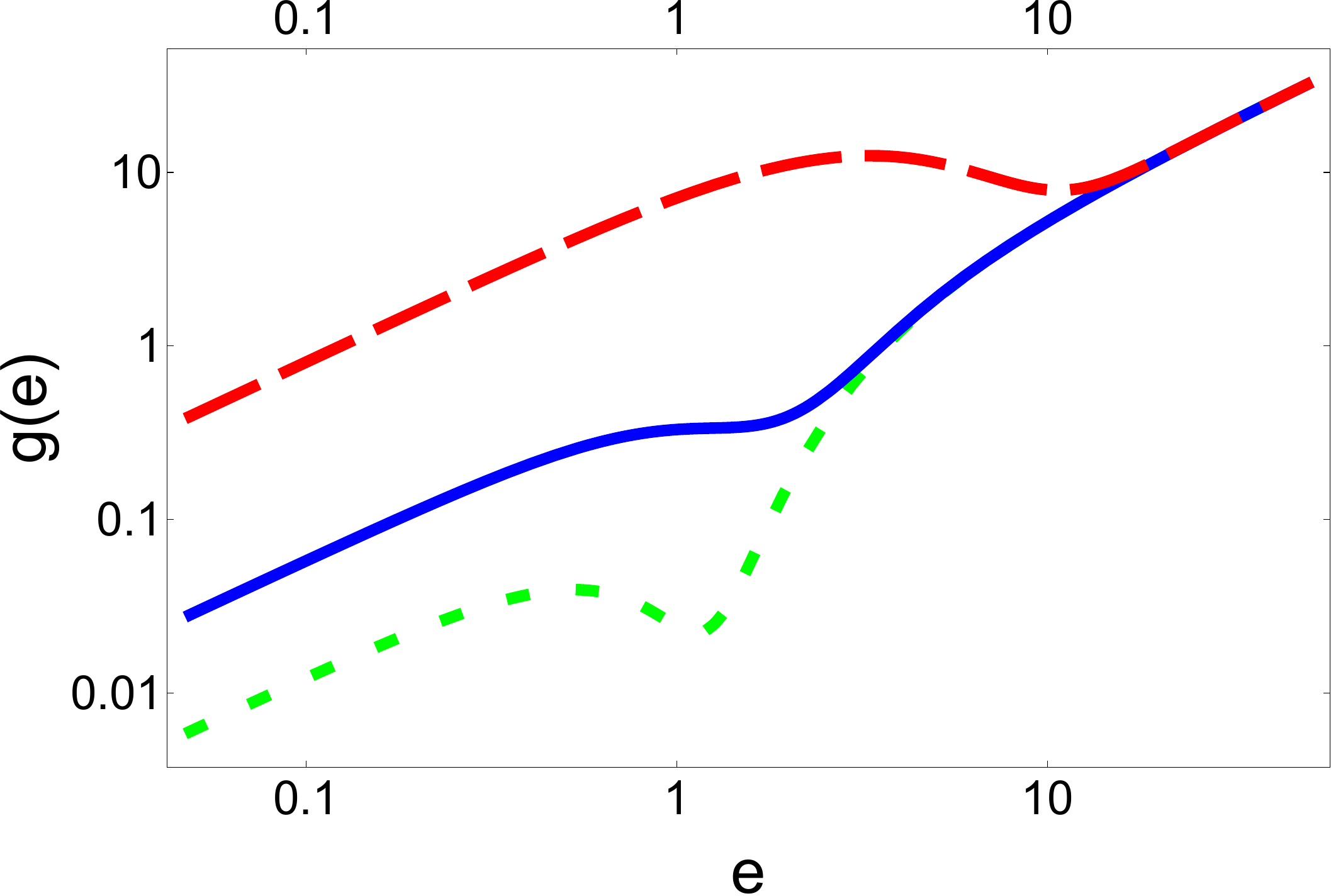}
	\put(-205,140){\Large (c)}
	\caption{(a)~The phase diagram for $M=2$ (a) and $M=3$ (b) at $\mu=0$ with $\wt{g}_1=1$ and $\wt{g}_2=3.75$ in the large-$N$ limit. The magnitude of $|e_1-e_2|$ is plotted. The strip of first order transitions 
for high  temperature is interrupted roughly between $\wt\lambda = 1$ and  $\wt\lambda = 2$, but is present close to the
broken phase around the origin and at high temperatures.
In figures  (c) we show a log-log-plot of the function $g(e)$ where $\gamma_2=3\pi/(2\wt{g}_2)\approx0.34$ for three different temperatures ${T}=0.59$ (green dashed curve), ${T}=0.7$ (blue solid curve), and ${T}=2.5$ (red dashed curve). The remnant of the strip close to the bulk of phase transitions at the origin can be explained by the existence of a ``wiggle'' of $g(e)$ which briefly dissolves for larger temperature and reappears anew. For smaller $\wt{g}_2$ (larger $\gamma_2$) the high temperature strip of phase transitions is completely separated from
the broken phase near the origin, cf. figures~\ref{fg:pd_mu0_M=2_weak_g2} and~\ref{fg:pd_mu0_M=3_weak_g2}.
}   \label{fg:pd_mu0_M2_broken_strip}
\end{figure}

\subsubsection{Low temperature regime with $\wt{g}_2^{\rm tri}<|\wt{g}_2|<\wt{g}_2^{\rm cr}$ ($\gamma_2^{\rm cr}<\gamma_2<\gamma_2^{\rm tri}$)}

In this regime we have a richer phase diagram which is mapped out in
figure \ref{fg:pd_mu0_M2_broken_strip} using $e_1-e_2$ as an order parameter.
The most notable feature is that the strip with the cascade of phase
transitions is split into two pieces. The strips end in second order points that are
visible in the $(\gamma_2, T)$ plane as the two transitions from region I to region II.
The two parts join each other at $\wt g_2 =\wt g_2^{\rm cr}$.
The function $g(e)$ is shown in figure~\ref{fg:pd_mu0_M2_broken_strip} for three
different temperatures, $ T =0.59$ corresponding to  the lower part of the
strip (green dotted curve), $ T = 0.7$
in between the two strips (blue solid curve) and $T=2.5$ corresponding to the upper
part of the strip (red dashed curve).

The transition between region III and region IV is first order. Since the curve
separating the regions III and IV  describes two minima of $v(e)$ coalescing with the
minimum at $e=0$,
one would expect a second order transition, and it may be accidental that the
position of the first order transition is located on this curve (it could also
be that our numerically accuracy is not sufficient). In the region IV we have
three first order transitions as a function of $ \lambda$ while there
are only two transitions in the region II which become second order at an
intermediate value of $T$.

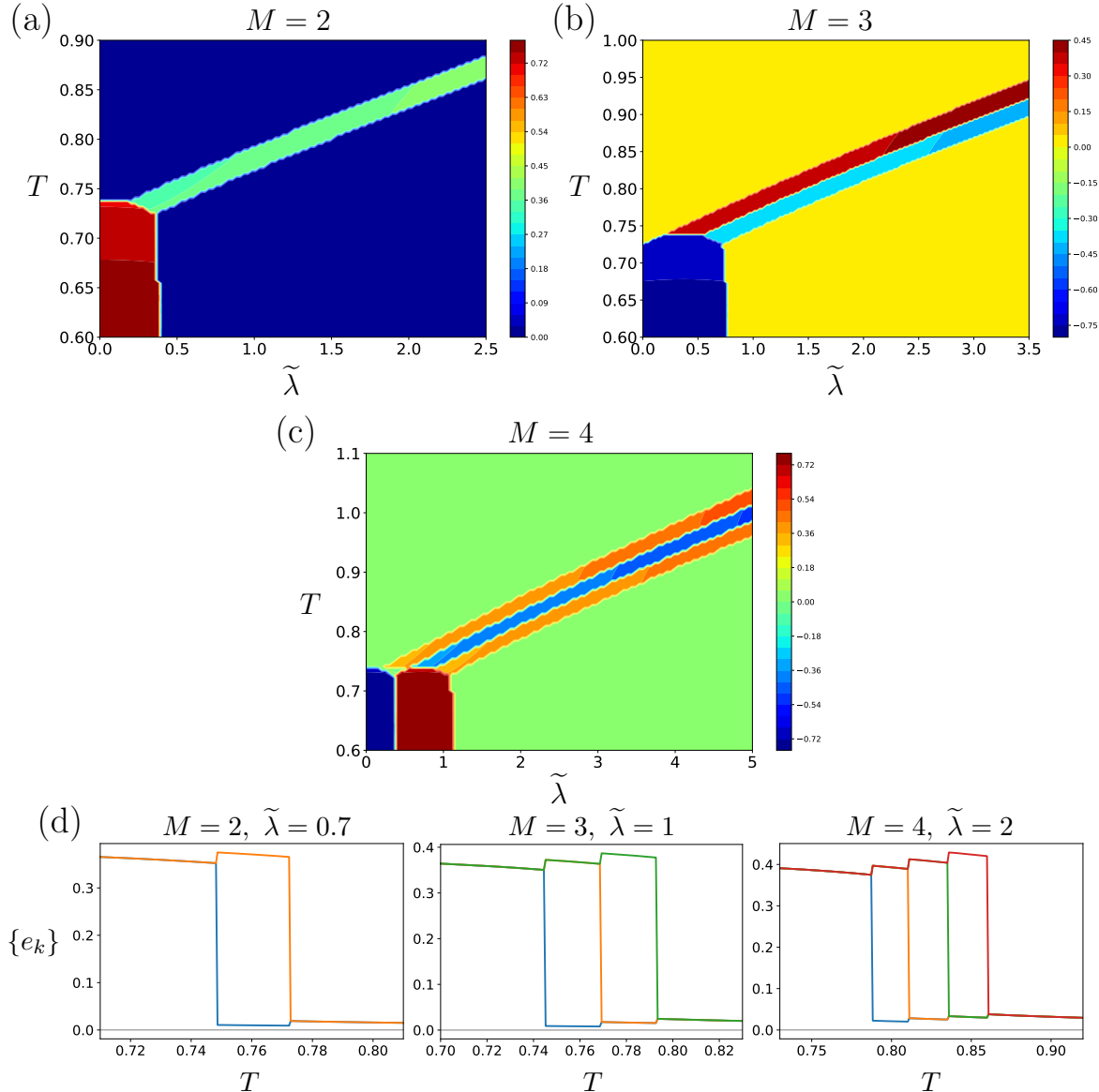
\begin{figure}[tb]
	\centering
	\includegraphics[height=.3\textwidth]{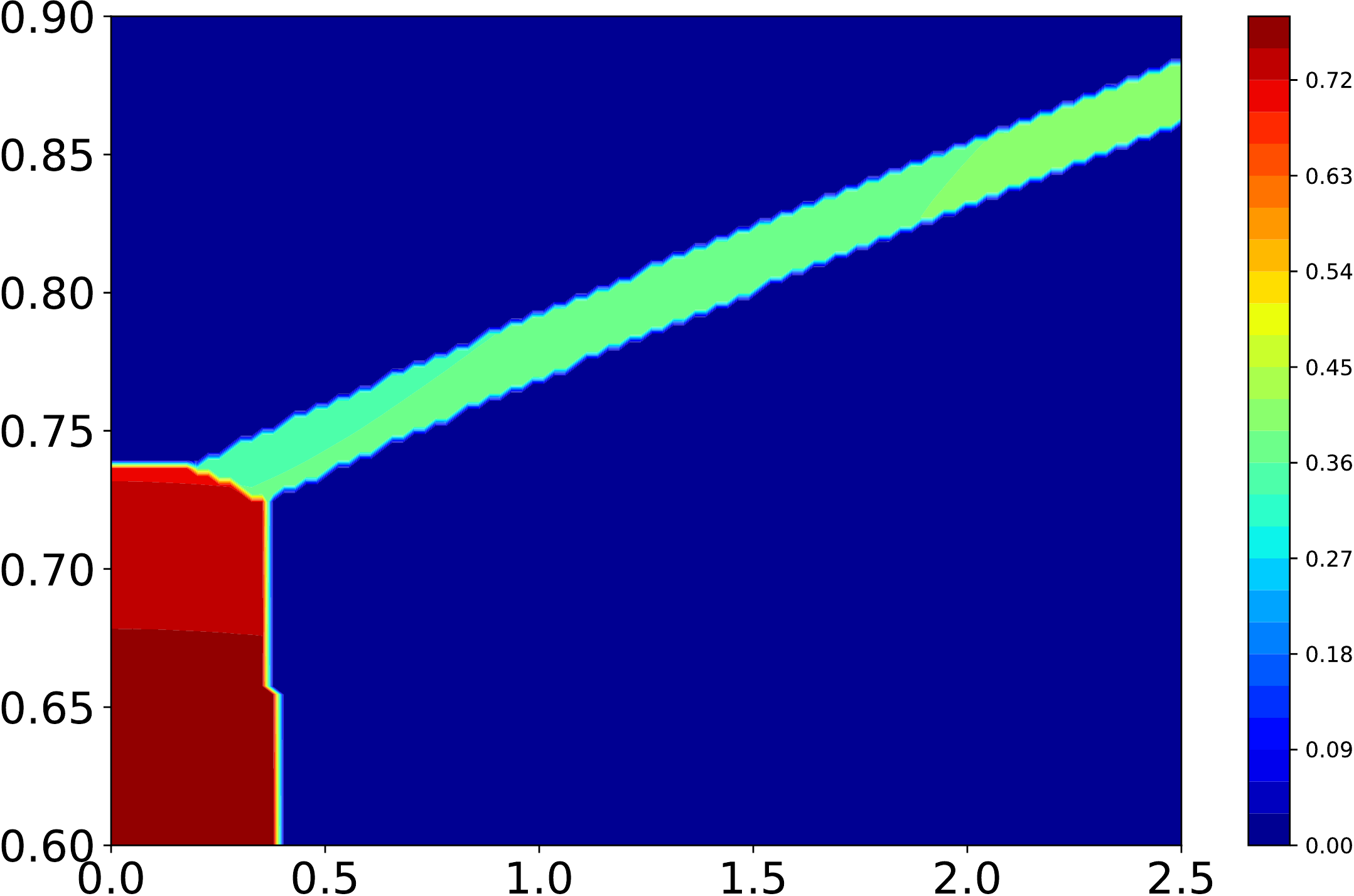}
	\put(-121,131){\large $M=2$}
	\put(-106,-16){\large $\wt{\lambda}$}
	\put(-208,64){\large ${T}$}
	\put(-215,131){\Large (a)}
	\qquad
	\includegraphics[height=.3\textwidth]{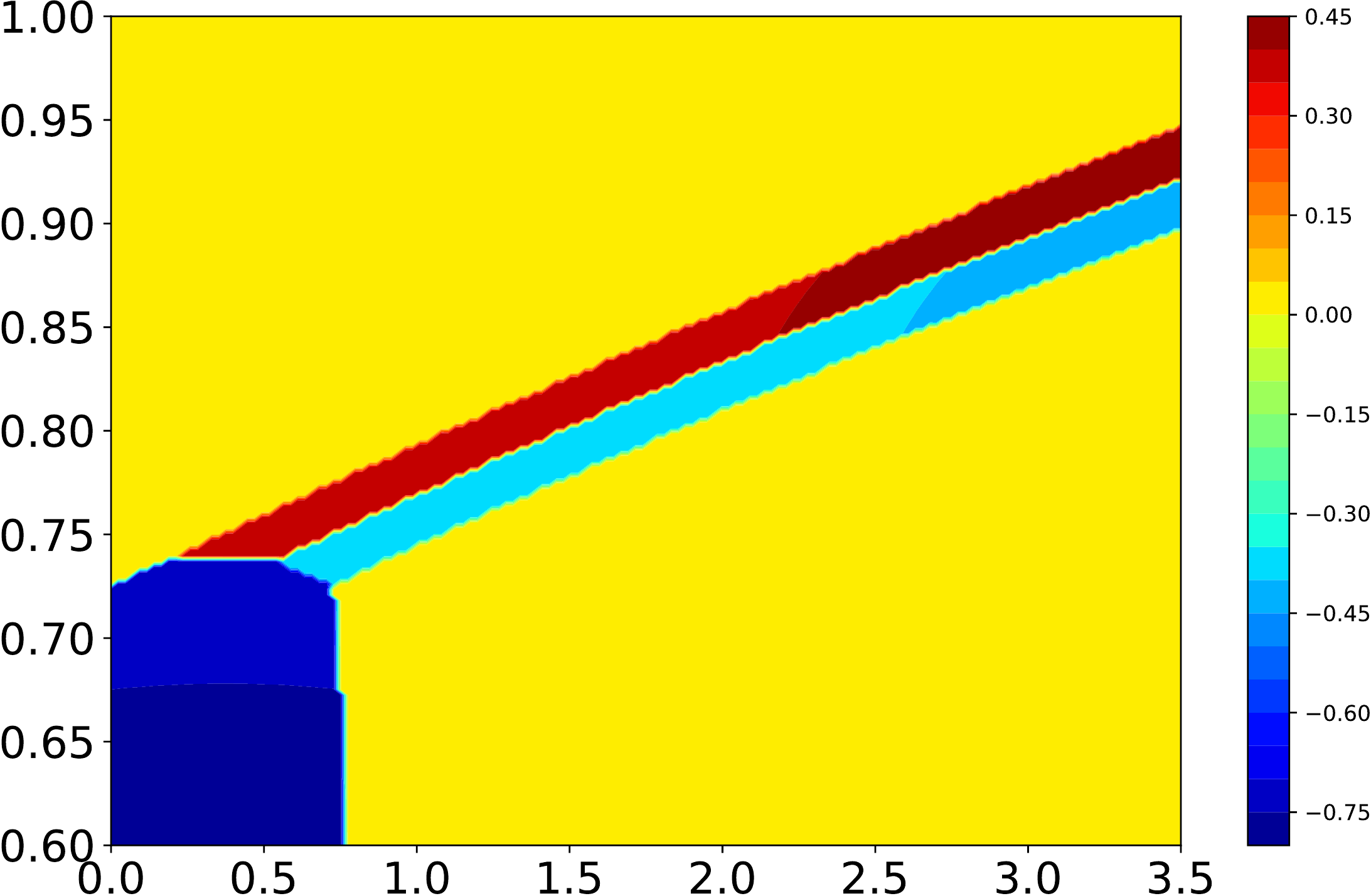}
	\put(-124,131){\large $M=3$}
	\put(-108,-16){\large $\wt{\lambda}$}
	\put(-211,64){\large ${T}$}
	\put(-218,131){\Large (b)}
	\vspace{.5\baselineskip}
	\\
	\centering
	\includegraphics[height=.3\textwidth]{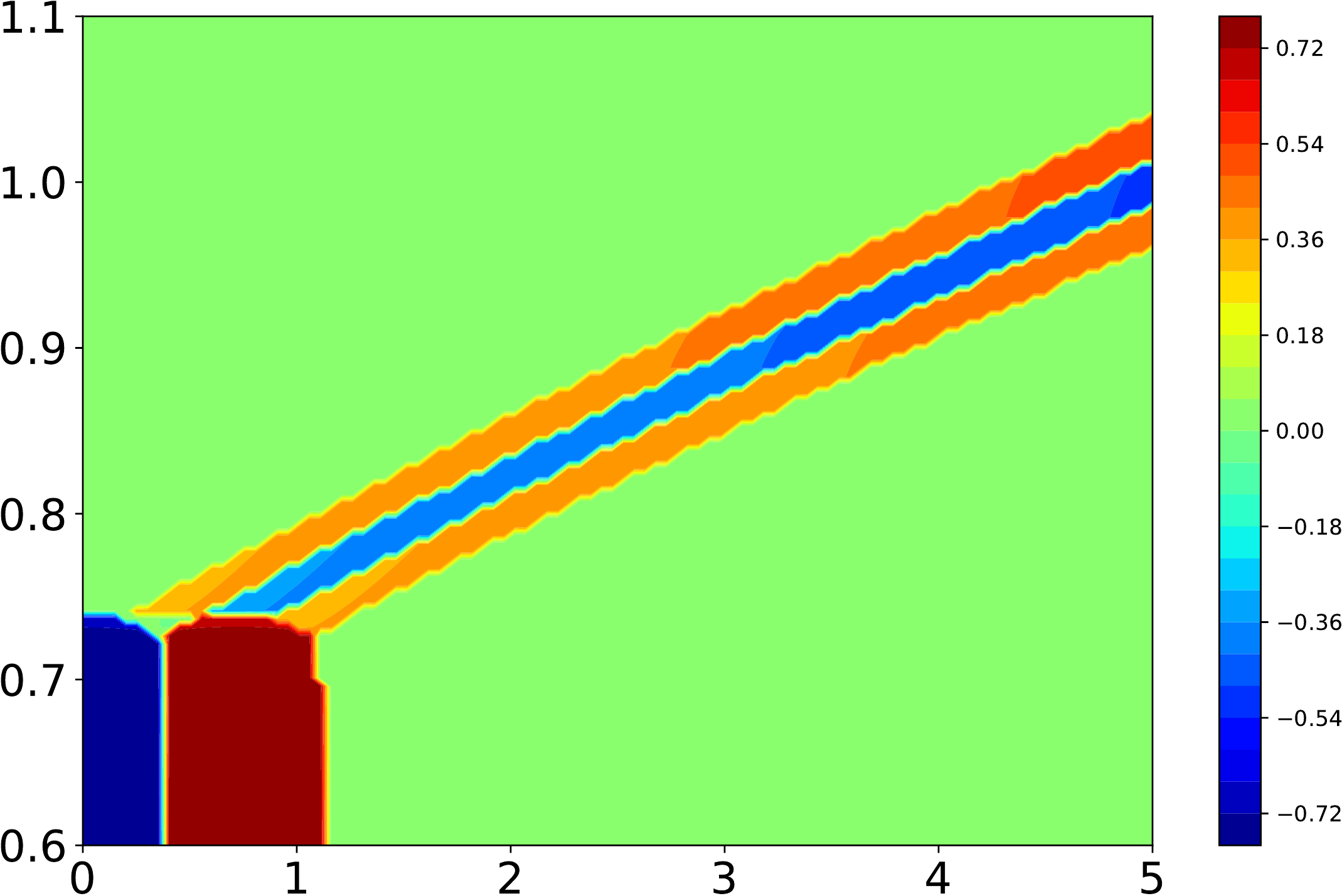}
	\put(-125,131){\large $M=4$}
	\put(-108,-14){\large $\wt{\lambda}$}
	\put(-208,62){\large ${T}$}
	\put(-218,131){\Large (c)}

	\centering
	\includegraphics[height=.2\textwidth]{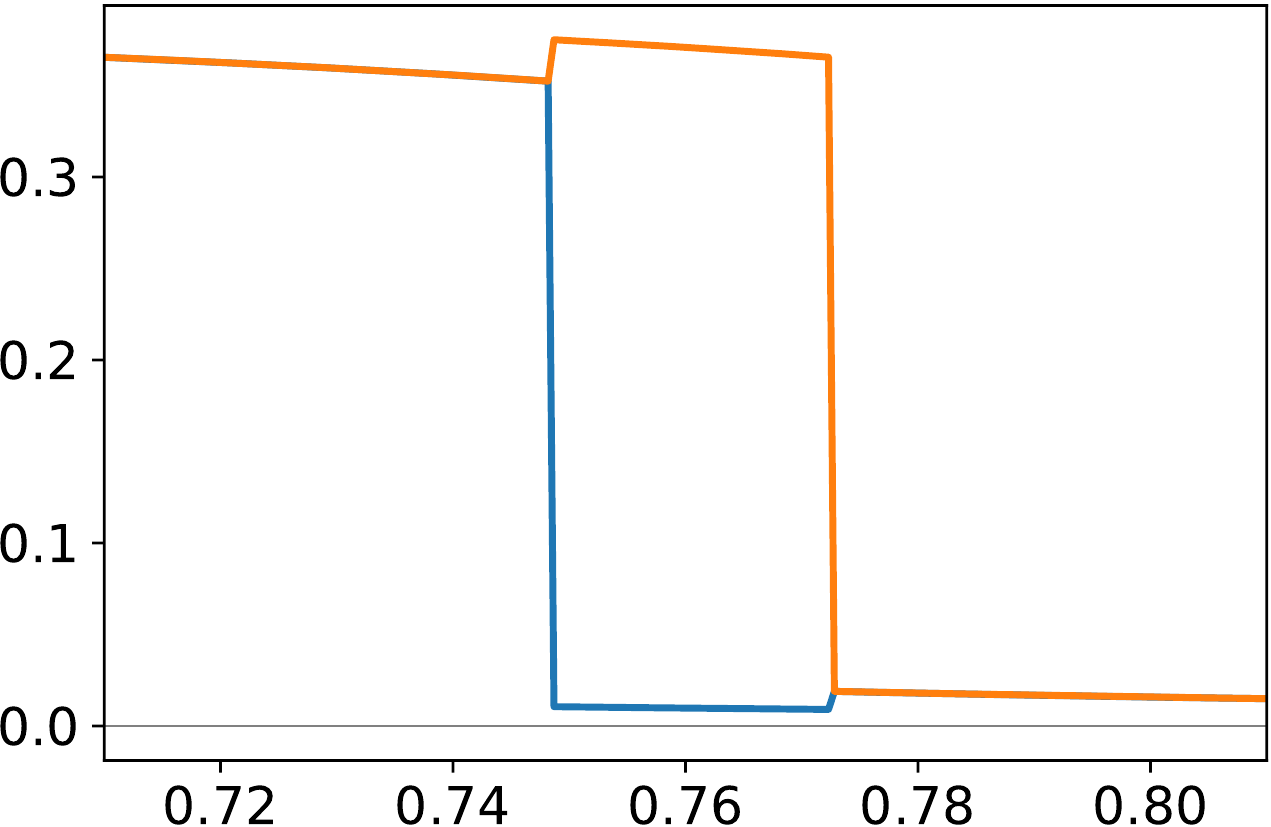}
	\includegraphics[height=.2\textwidth]{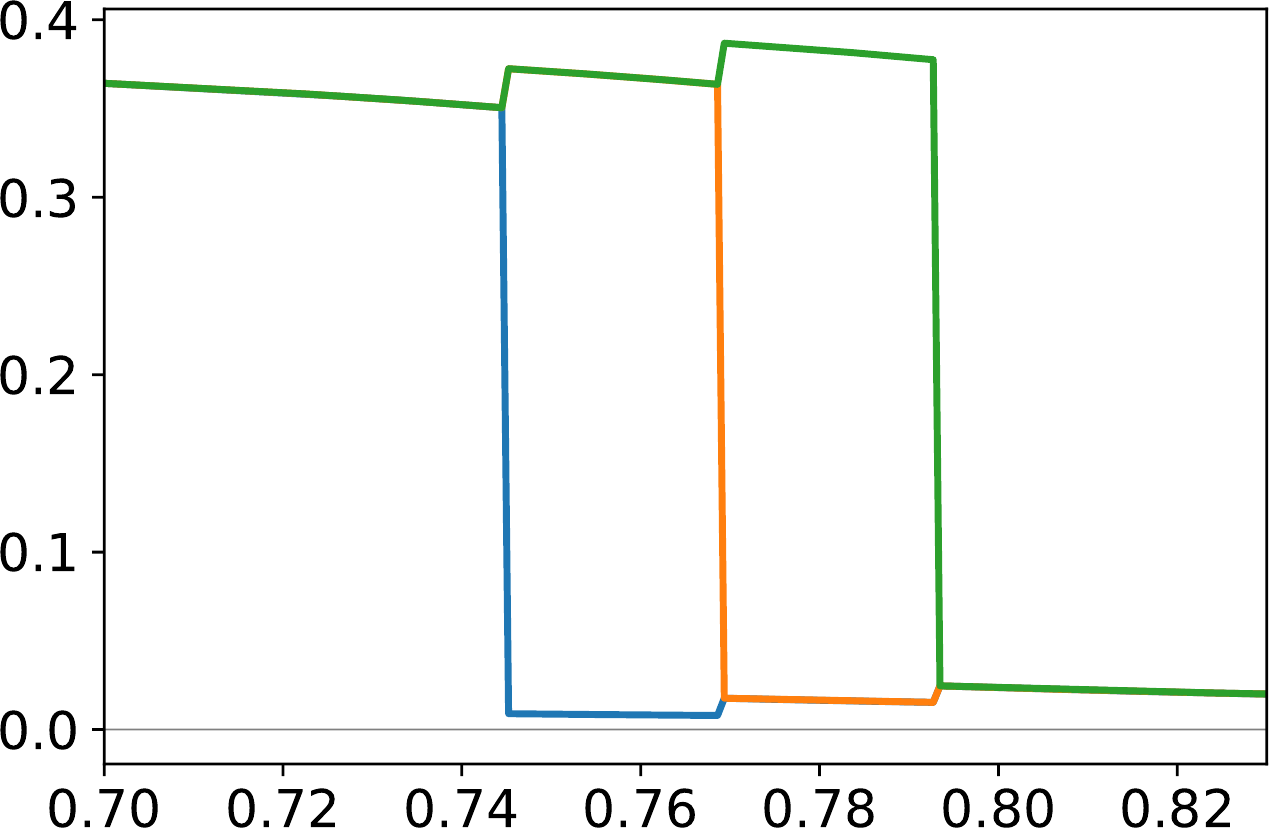}
	\includegraphics[height=.2\textwidth]{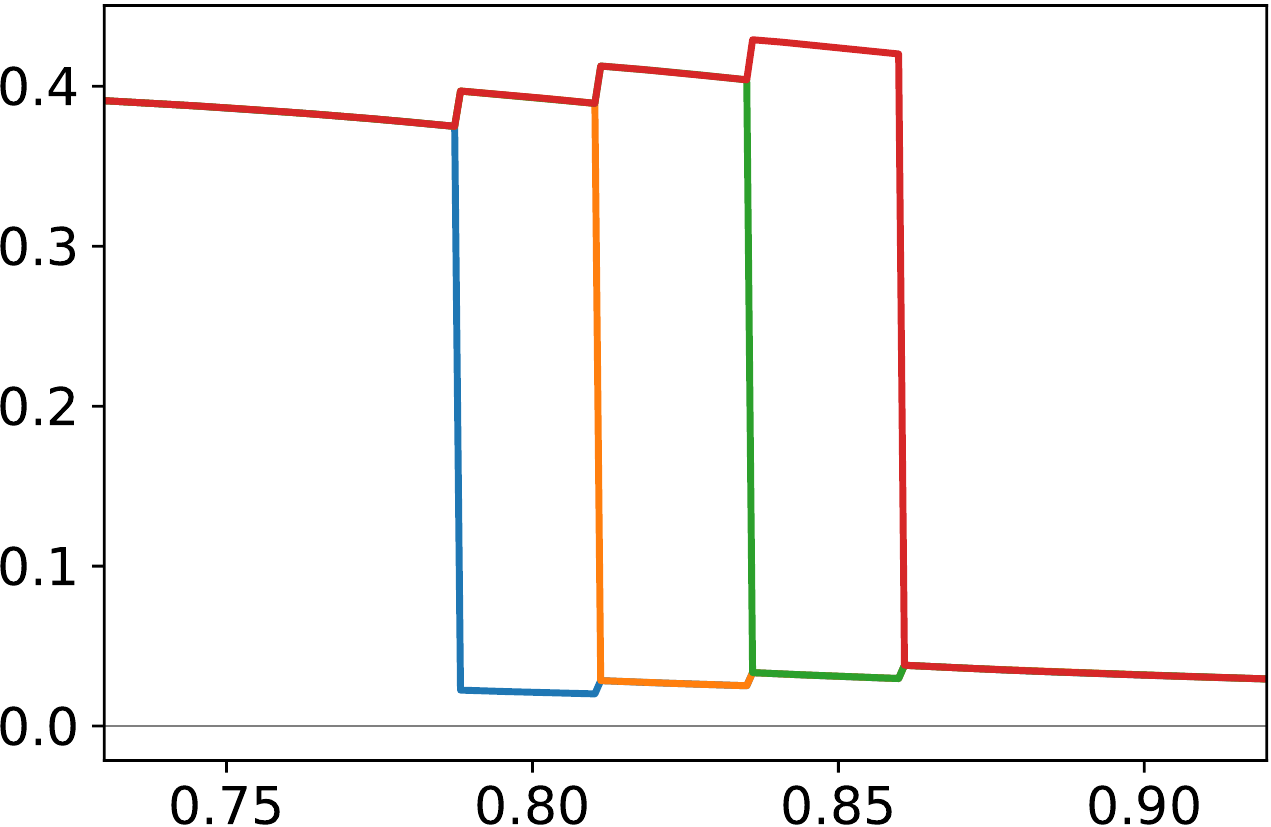}
	\put(-430,43){$\{e_k\}$}
	\put(-368,89){$M=2,~\wt\lambda=0.7$}
	\put(-230,89){$M=3,~\wt\lambda=1$}
	\put(-95,89){$M=4,~\wt\lambda=2$}
	\put(-337,-14){${T}$}
	\put(-200,-14){${T}$}
	\put(-65,-14){${T}$}
	\put(-418,90){\Large (d)}
	\caption{\label{fg:mu0_strongphase}
Figures  (a), (b) and (c) show the same plots as in
figures~\ref{fg:pd_mu0_M=2_weak_g2} and~~\ref{fg:pd_mu0_M=3_weak_g2}   but with $\wt{g}_2=5$. All phase transitions are first order. (d) The minimum of $V_{\rm eff}(E)$ with $\wt{g}_1=1$ and $\wt{g}_2=5$ at $\mu=0$. As the temperature rises, the eigenvalues drop sequentially through $M$ first order transitions.}
\end{figure}

\subsubsection{Low temperature regime with $|\wt{g}_2|>\wt{g}_2^{\rm cr}$ ($\gamma_2<\gamma_2^{\rm cr}$) }

For these values of $\wt g_2$ or $\gamma_2$, the system no longer enters region I with
increasing temperature so that the strip with the cascade of first order
phase transitions is no longer interrupted. 
We have numerically analyzed this regime for $M=2,3$ and $4$ at $\wt{g}_2=5$ in figure~\ref{fg:mu0_strongphase}. The cascade of phase transitions at high temperature has been visualized in figure~\ref{fg:mu0_strongphase}(d) where we have plotted the actual solutions $e_k$ at the global minimum of the potential~\eqref{eq:VT}.
To understand the nature of the two phases above and below this strip, we
interpret the $e_k$ as the effective masses of the fermions of the theory.
As shown in figure~\ref{fg:mu0_strongphase}, the low-${T}$ region is characterized by a large value of $|e_k|$ implying that the effective masses are heavy.
In contrast, in the high-${T}$ region above the strip all $|e_k|$ drop nearly to zero, making the fermions almost massless. The large bare mass $\kappa\bar\psi\psi$
of the constituent fermions
is dynamically canceled by interactions. This cancellation proceeds step by step across the strip. For a large fixed $\wt\lambda$, as the temperature goes up, there are $M$ first-order transitions; across each transition one of the $M$ species of fermions becomes light. After all transitions are traversed, all $M$ fermions become light. 

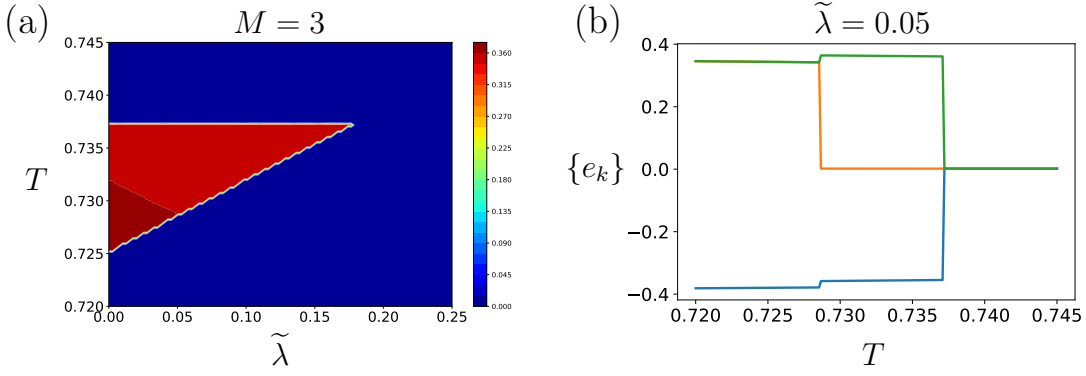
\begin{figure}[tb]
	\centering
	\includegraphics[height=.25\textwidth]{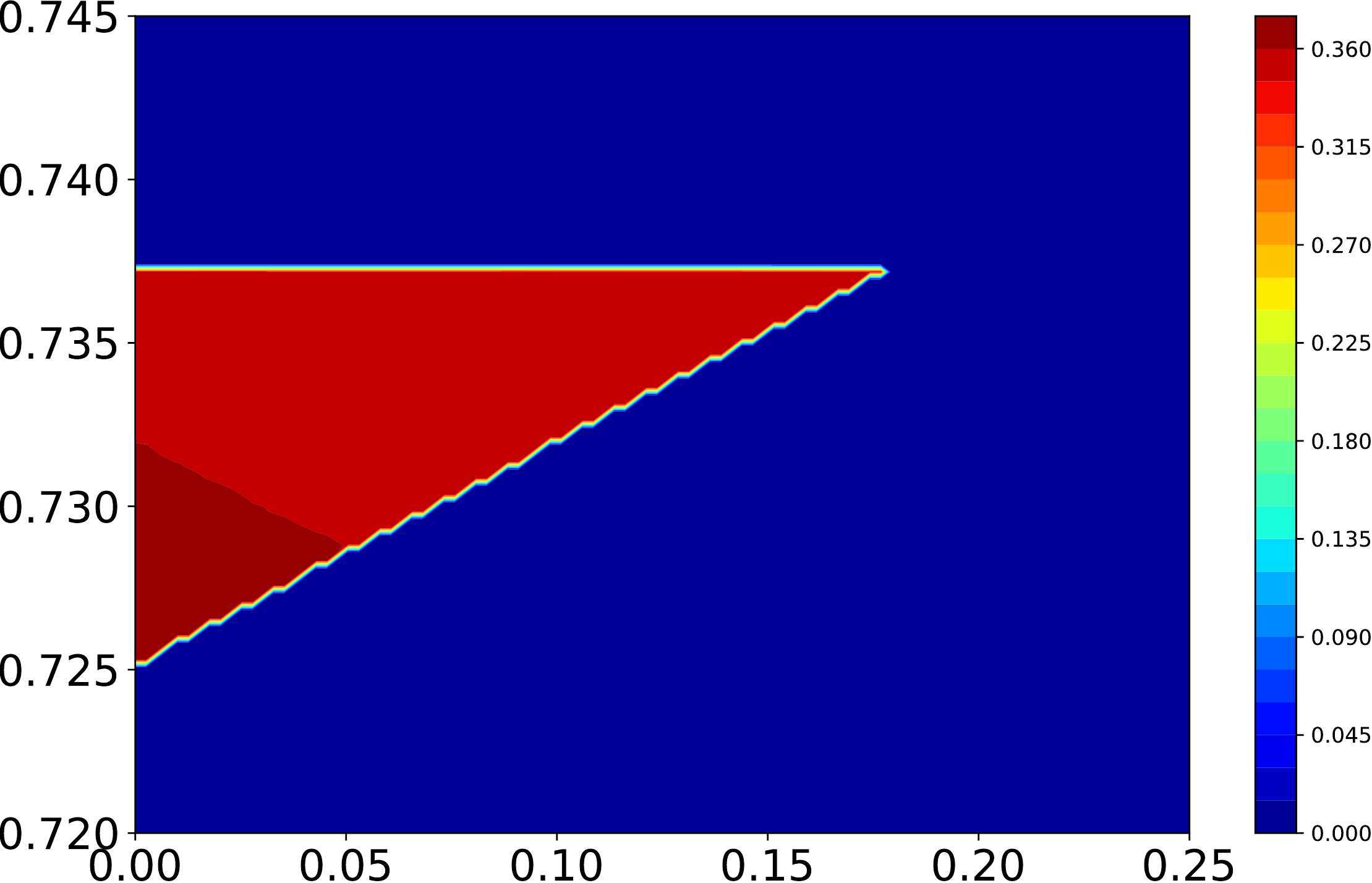}
	\put(-104,109){\large $M=3$}
	\put(-90,-16){\large $\wt{\lambda}$}
	\put(-182,50){\large ${T}$}
	\put(-190,109){\Large (a)}
	\qquad \qquad 
	\includegraphics[height=.25\textwidth]{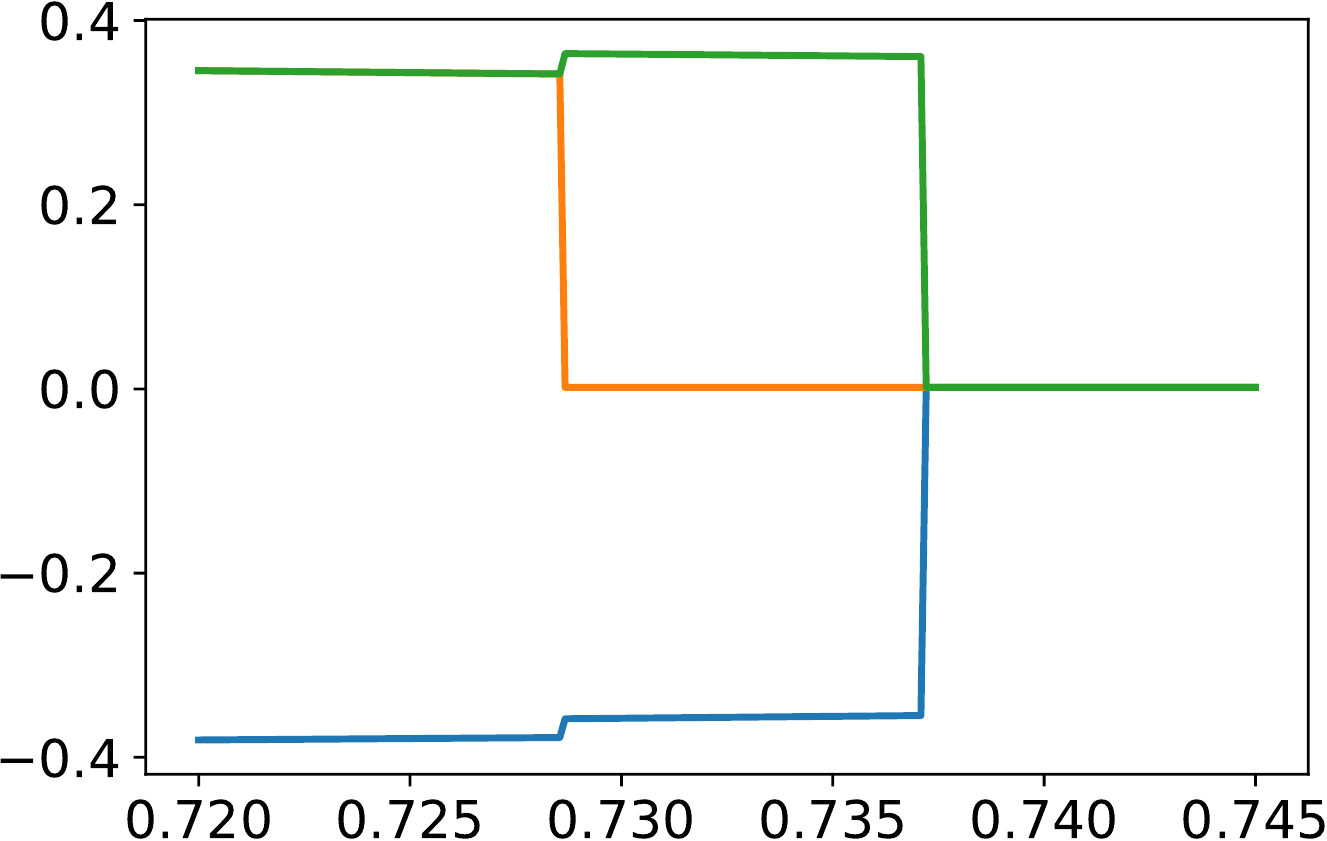}
	\put(-100,109){\large $\wt{\lambda}=0.05$}
	\put(-82,-16){\large ${T}$}
	\put(-193,55){\large $\{e_k\}$}
	\put(-190,109){\Large (b)}
	\caption{\label{fg:M3_U1U1U1}(a) The phase diagram for $M=3$ at $\mu=0$ with $\wt{g}_1=1$ and $\wt{g}_2=5$ in the large-$N$ limit. 
	The plotted observable is $\text{Min}(|e_1-e_2|,|e_2-e_3|,|e_3-e_1|)$. Within the red triangle the three $e_k$ differ from one another, indicating spontaneous symmetry breaking $\U(3)\to\U(1)^3$. (b) The $\wt{T}$-dependence of $\{e_k\}$ at $\wt\lambda=0.05$. There is a range of ${T}$ in which the three $e_k$ are all different.}
\end{figure}

In the same way as in the high temperature regimes, one can depict the phase transitions in the low $ T$ and low $\wt \lambda$ phase where we also find a cascade
of phase transitions.
A new phenomenon shows up for $|\wt{g}_2|>|\wt{g}_2^{\rm cr}|$ for parameter values
in region IV.
When zooming into figure~\ref{fg:mu0_strongphase}(b) there is a large region where  the symmetry breaking pattern is $\U(3)\to \U(2)\times\U(1)$ (namely, two of the three $e_k$ coincide). Yet, in a  tiny region, shown in figure~\ref{fg:M3_U1U1U1}, all $e_k$ are mutually distinct and break the symmetry  as $\U(3)\to\U(1)\times\U(1)\times\U(1)$. In this phase, one of the bosons is very light but  the other two are heavy.
Indeed when $|\wt{g}_2|>|\wt{g}_2^{\rm cr}|$ (or $|\gamma_2|<|\gamma_2^{\rm tri}|$), we find a different kind of transition in the shape of the function $g(e)$  in the region IV (see insets in figure~\ref{fig:phase}). One of the consequences is the occurrence of 
 exotic phases corresponding to the symmetry breaking patterns $\U(M)\to\U(j)\times\U(k)\times\U(M-j-k)$ because   $g(e)$ has three positive slopes  so that the potential
 $v(e)$ has three minima, see Appendix~\ref{sec:locextimp}.

\section{\label{sc:mu}Nonzero chemical potential}

The zero-temperature potential at $\mu>0$ can be readily found from \eqref{eq:mainV} as
\ba
	V_{\rm eff}(E) & = V_{\rm eff}(E)\Big|_{\mu=0} - \sum_{k=1}^{M}\int\frac{\rmd^2p}{(2\pi)^2}
	\mkakko{\mu-\sqrt{\mathbf{p}^2+4g_2^2E_k^2}}\Theta\mkakko{\mu-\sqrt{\mathbf{p}^2+4g_2^2E_k^2}}\nn	\\
	& = V_{\rm eff}(E)\Big|_{\mu=0} - \frac{1}{12\pi}\sum_{k=1}^{M}
	(\mu-2|g_2E_k|)^2(\mu+4|g_2E_k|)\Theta(\mu-2|g_2E_k|)
\ea
where $V_{\rm eff}(E)\Big|_{\mu=0}$ is as given in \eqref{eq:Vvac} and $\Theta(x)$ is the Heaviside step function. In dimensionless units where $\Lambda$ is absorbed in $\mu\to\Lambda^{3/2}\mu$
we have
\ba
	\frac{V_{\rm eff}(E)}{\Lambda^3} =\; & 
	\frac{\wt{g}_1^2}{\wt{g}_2^2-M\wt{g}_1^2}\mkakko{\sum_{k=1}^{M}e_k-\wt{\lambda}}^2
	+ \sum_{k=1}^{M}\bigg\{e_k^2+\frac{4}{3\pi}|\wt{g}_2e_k|^3 - \frac{1}{6\pi}(1+4\wt{g}_2^2e_k^2)^{3/2} 
	\notag
	\\
	& 
	- \frac{1}{12\pi}(\mu-2|\wt{g}_2e_k|)^2(\mu+4|\wt{g}_2e_k|)\Theta(\mu-2|\wt{g}_2e_k|)
	\bigg\} .
\ea 
For analytical considerations, we adopt again the notation of~\eqref{new-units}, where the potential becomes
\begin{equation}
\begin{split}
\widehat{V}_{\rm eff}(e)=\ &\gamma_1\left(\sum_{k=1}^{M}e_k-\lambda\right)^2 
	+ \sum_{k=1}^{M}\biggl\{\gamma_2e_k^2+|e_k|^3- (1+e_k^2)^{3/2}\\
	&-\frac{1}{2}({\mu}-|e_k|)^2({\mu}+2|\f_k|)\Theta({\mu}-|e_k|)\biggl\}.
\end{split}
\end{equation}
with the saddle point equation
\begin{equation}\label{saddle-chem}
-2s=g(e)\quad {\rm with}
\quad g(e)=2\gamma_2 e+3e\mkakko{|e|-\sqrt{1+e^2}}+3e({\mu}-|e|)\Theta({\mu}-|e|)
\end{equation}
and $s=\gamma_1(\sum_{k=1}^{M}e_k-\lambda)$.

\begin{figure}[t!]
	\centering
	\includegraphics[width=\textwidth]{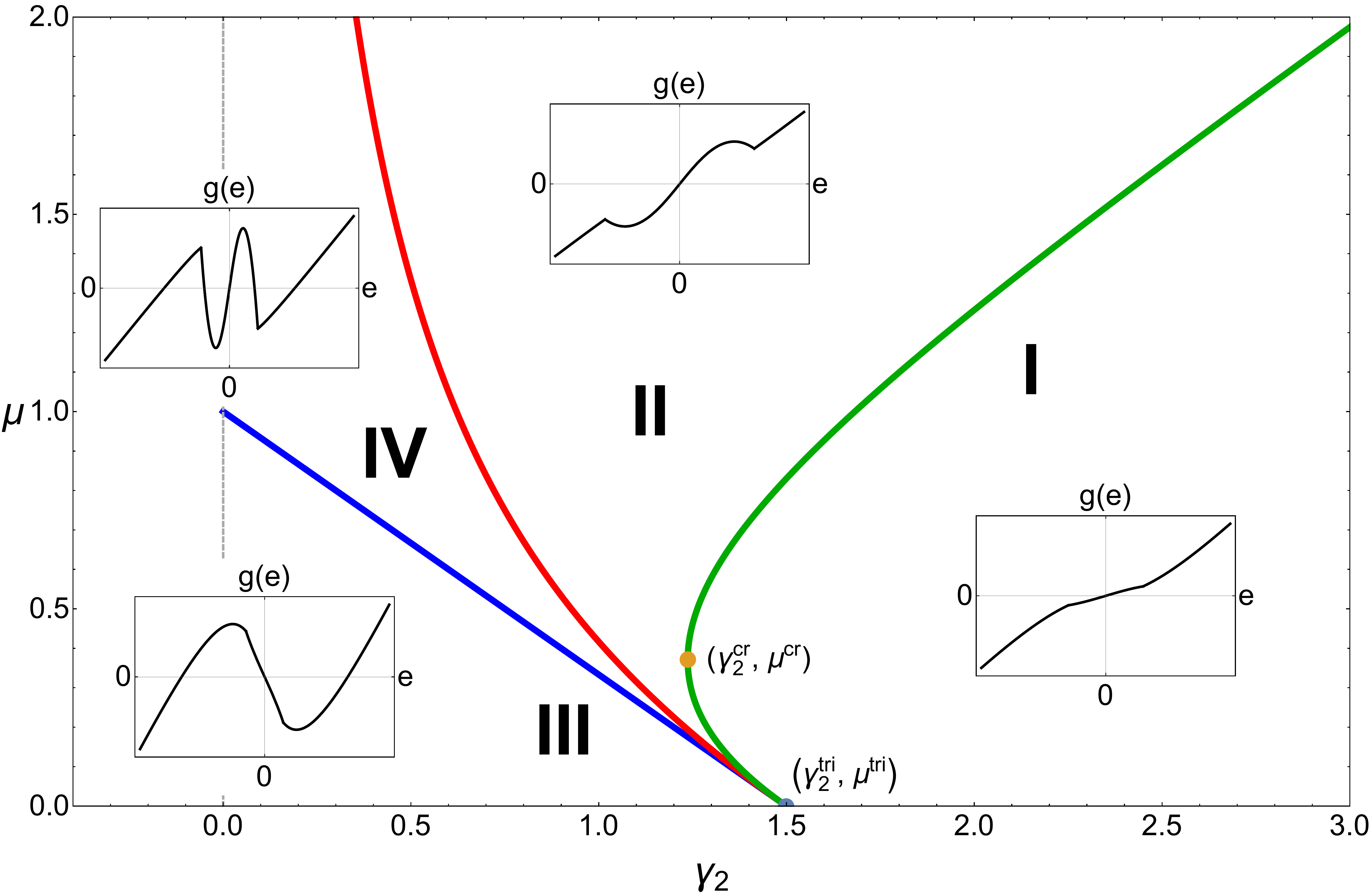}
	\caption{\label{fig:curve-chem}Phase diagram of the derivative of the potential, $v'(e)=g(e)$, in the $(\gamma_2,\mu)$ plane. The insets show the different shapes of the function $g(e)$.}
\end{figure}

The insets in figure \ref{fig:curve-chem} show the different shapes of the
function $g(e)$ in the $(\gamma_2, \mu)$ plane.
Taking into account the Heaviside $\Theta$ function, in a similar way as at $\mu =0$ and nonzero temperature, the regions are separated
by the following three curves:
\begin{itemize}
\item[i)] The vanishing of the slope at $e=0$,
\be
  g'(e=0)=2\gamma_2+3{\mu}-3=0,
  \ee
see blue line in figure \ref{fig:curve-chem}.

\item[ii)] The curve (red curve
  in figure \ref{fig:curve-chem}) defined by
  \be
  g(\mu)=0,
  \ee
  has the explicit solution
  is given by
  \be
  \gamma_2=\frac 32 \left (\sqrt{1+\mu^2} - \mu\right ).
  \ee
  
\item[iii]) The third curve is given by
    \be
  \lim_{\epsilon\to0^+}g'(\mu-\epsilon)=0.
  \ee
  The limit is introduced so that the Heaviside $\Theta$ function is equal to one (green curve in figure
  \ref{fig:curve-chem}). This equation can be solved explicitly with the solution given by
  \be
  \gamma_2=\frac{3(1+2\mu^2)}{2\sqrt{1+\mu^2}}-\frac{3}{2}\mu.
  \ee
 From that expression we obtain the critical point
  \be
  \frac {\rmd \gamma_2}{\rmd \mu} =\frac{3\mu(3+2\mu^2)}{2(1+\mu^2)^{3/2}}-\frac{3}{2}=0,
  \ee
  which is solved by $\mu^{\rm cr}=0.3703$ with a corresponding value of $\gamma_2$ given
  by
  \be
  \gamma_2^{\rm cr} = 1.2370.
  \ee
  This point is indicated by the yellow point in figure \ref{fig:curve-chem}.
  
  \end{itemize}
These three curves partition the $(\gamma_2, \mu)$ plane into four regions which are
anew referred to by Roman numerals, see figure \ref{fig:curve-chem}. The curves
meet at the tricritical point where the minima of the potential coalesce.
Combining $2\gamma_2+3\mu =3$ with $2\gamma_2=3(\sqrt{1+\mu^2}-\mu)$ we find
  that the tricritical point is at $\mu^{\rm tri} =0$ and $ \gamma_2^{\rm tri} = 3/2$ (blue point in figure \ref{fig:curve-chem}). 
  
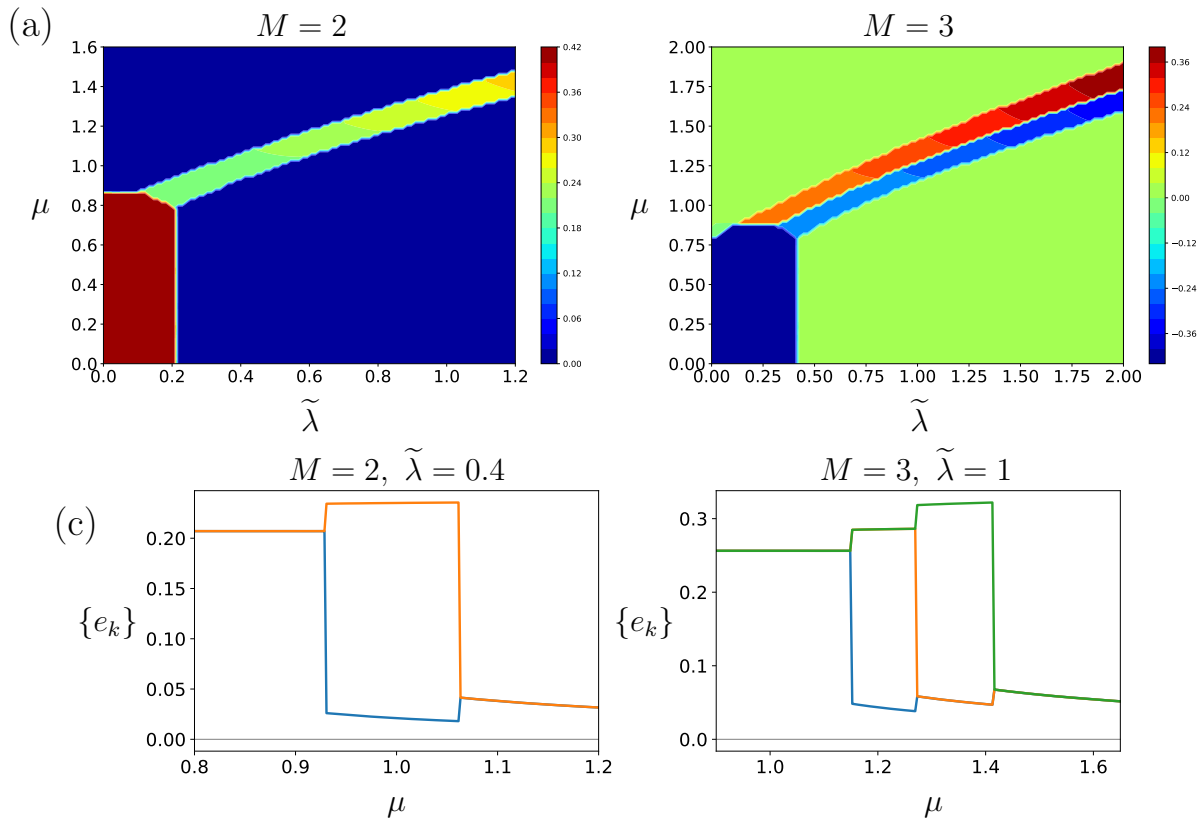
\begin{figure}[tb]
	\centering
	\includegraphics[height=.3\textwidth]{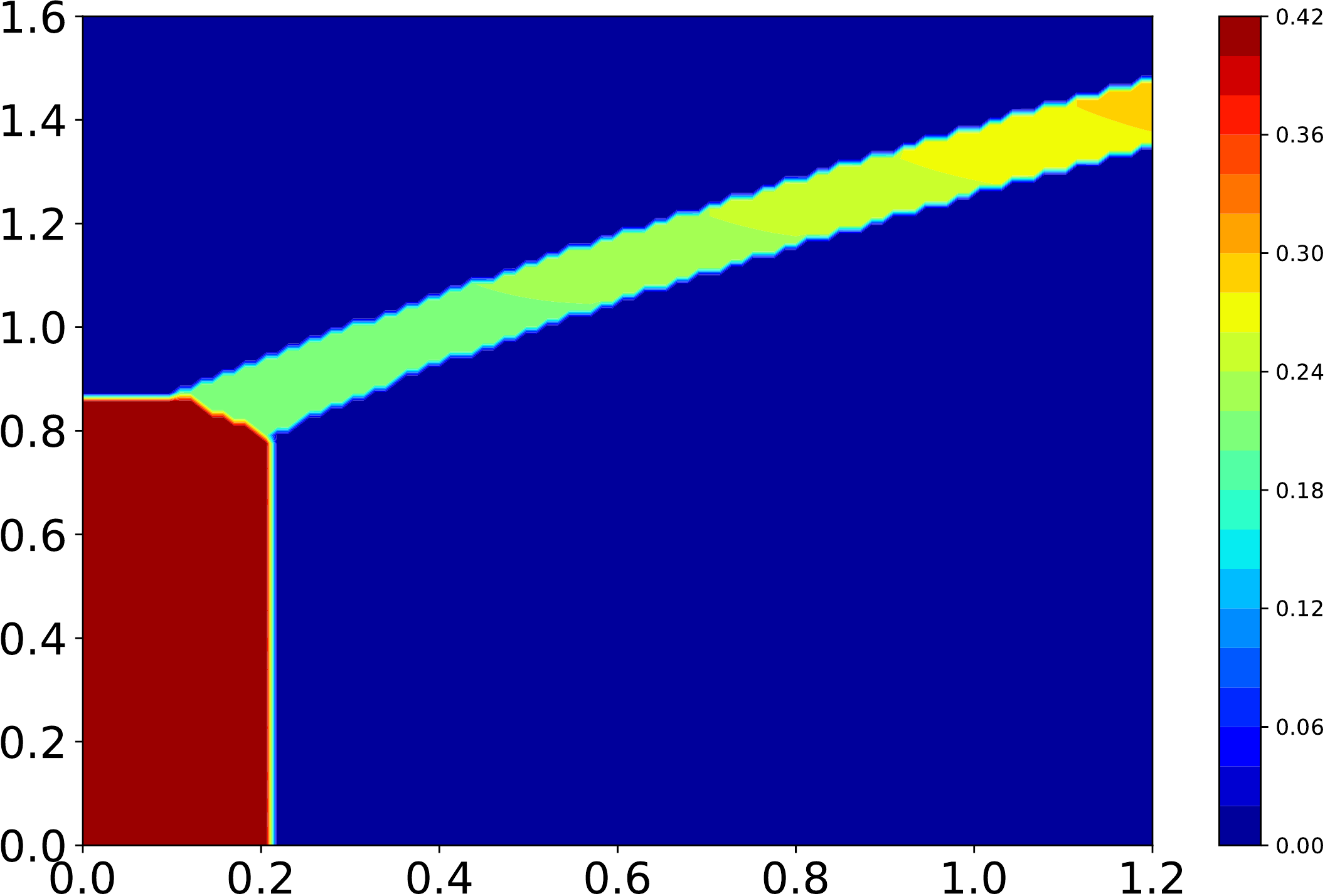}
	\put(-122,131){\large $M=2$}
	\put(-105,-18){\large $\wt{\lambda}$}
	\put(-206,64){\large ${\mu}$}
	\put(-215,131){\Large (a)}
	\qquad \quad
	\includegraphics[height=.3\textwidth]{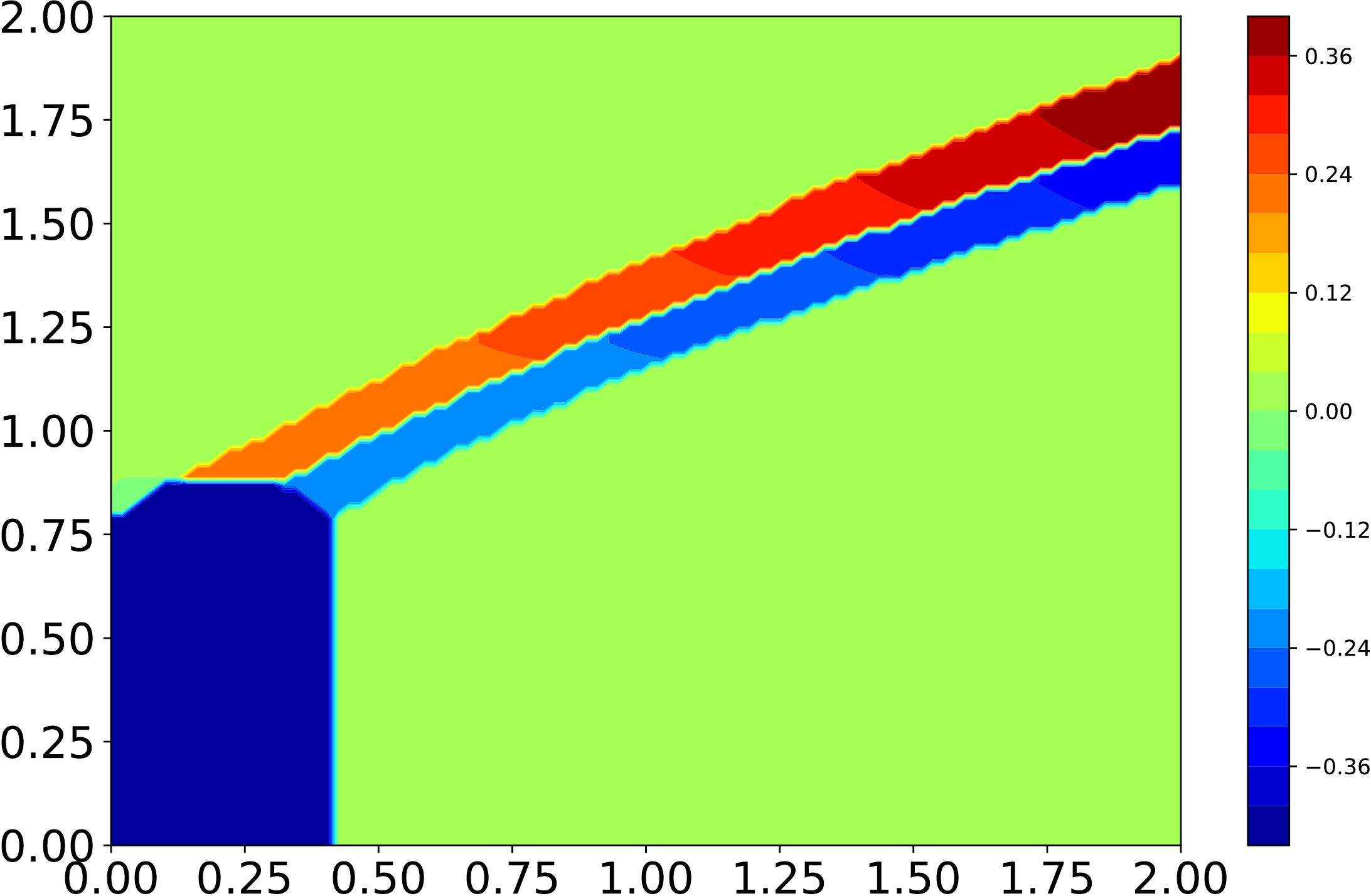}
	\put(-125,131){\large $M=3$}
	\put(-107,-18){\large $\wt{\lambda}$}
	\put(-212,64){\large ${\mu}$}
	\vspace{2mm}\\
	\centering
	\includegraphics[height=.25\textwidth]{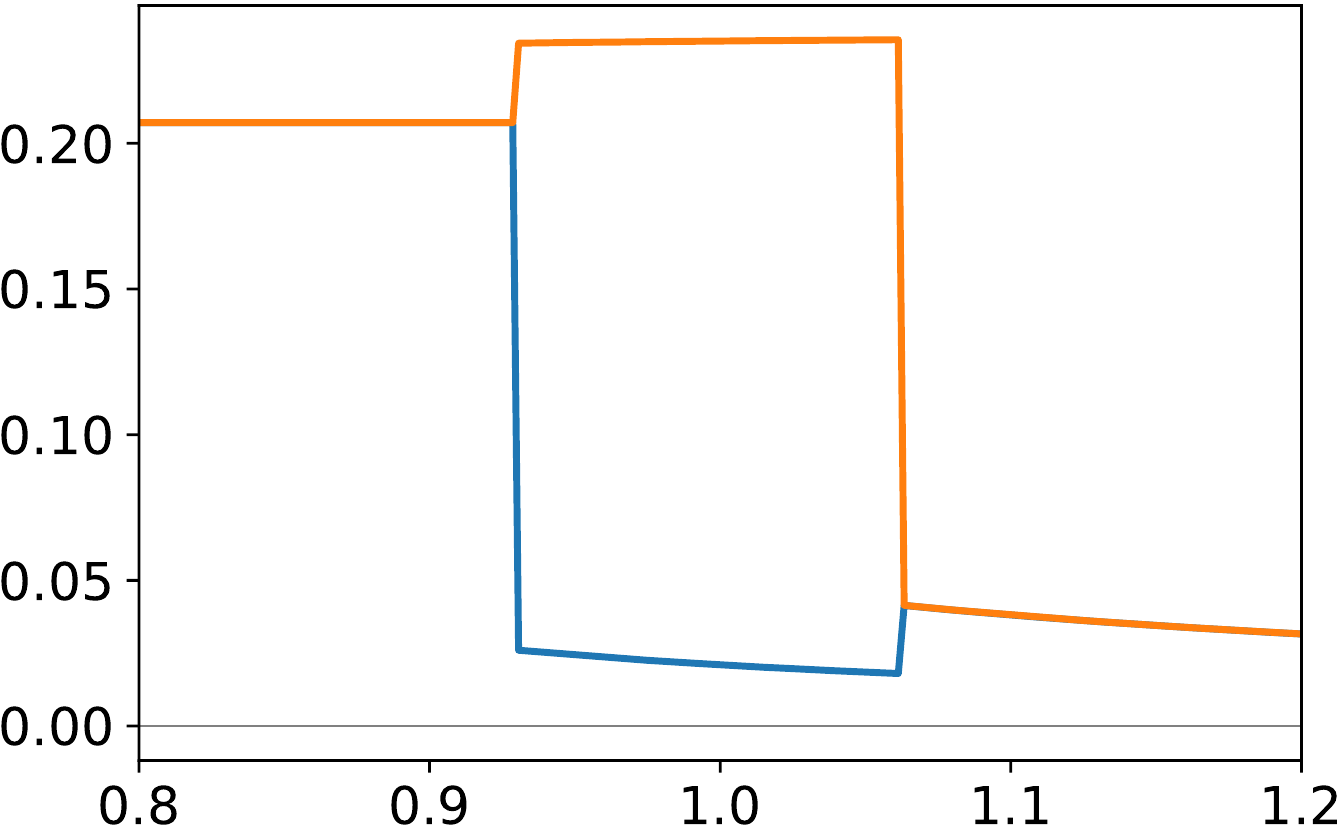}
	\qquad
	\includegraphics[height=.25\textwidth]{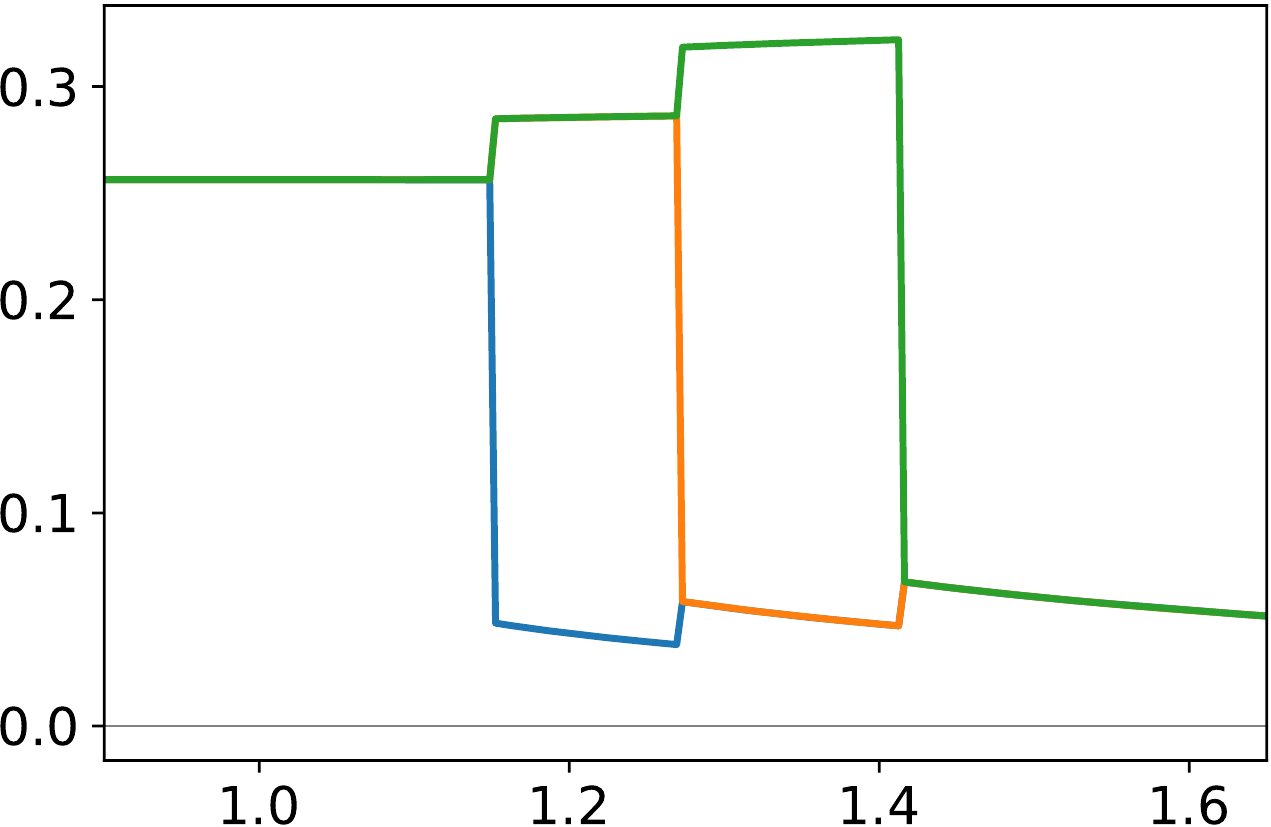}
	\put(-391,54){\large$\{e_k\}$}
	\put(-190,54){\large$\{e_k\}$}
	\put(-312,111){\large$M=2,~\wt\lambda=0.4$}
	\put(-113,111){\large$M=3,~\wt\lambda=1$}
	\put(-275,-14){\large${\mu}$}
	\put(-74,-14){\large${\mu}$}
	\put(-400,90){\Large (c)}
	\caption{\label{fg:pd_mu}The $T=0$ phase diagram
at nonzero chemical potential
          for (a) $M=2$ and (b) $M=3$ with $\wt{g}_1=1$ and $\wt{g}_2=3$. The value plotted is $|e_1-e_2|$ in (a) and $e_1-2e_2+e_3$ in (b) with $e_1\geq e_2\geq e_3$ assumed. All phase transitions are first order. (c) The minimum of $V_{\rm eff}(E)$ with $\wt{g}_1=1$ and $\wt{g}_2=3$ at $T=0$. As the chemical potential increases, the eigenvalues drop sequentially through $M$ first-order transitions.}
\end{figure}

  Since the shapes of $g(e) $ at nonzero $\mu$  and $T=0$ are similar to the shapes
  of $g(e)$ for $\mu=0$ and nonzero $T$, we expect a similar phase diagram where we
  again can distinguish three regions depending on the value of $\gamma_2$
  relative to $\gamma_2^{\rm cr}$ and  $\gamma_2^{\rm tri}$.
  Especially, we see a cascade of phase transitions between phases of the form $\U(M)\to\U(j)\times\U(M-j)$. We have checked this numerically for $M=2$ and $M=3$
  with $\wt g_2 =3$ ($\gamma_2 =0.5236$), see figure~\ref{fg:pd_mu} (a) and (b). As is the case at
  nonzero $T$ and $\mu =0$ for $\gamma_2<\gamma_2^{\rm cr} $ the strip of first
  order phase transitions is connected. Both inside this strip and in the region around
  $(\mu,\wt\lambda) =(0,0)$ we observe a cascade of first order phase transitions.

\begin{figure}[tb]
	\centering
	\includegraphics[height=.25\textwidth]{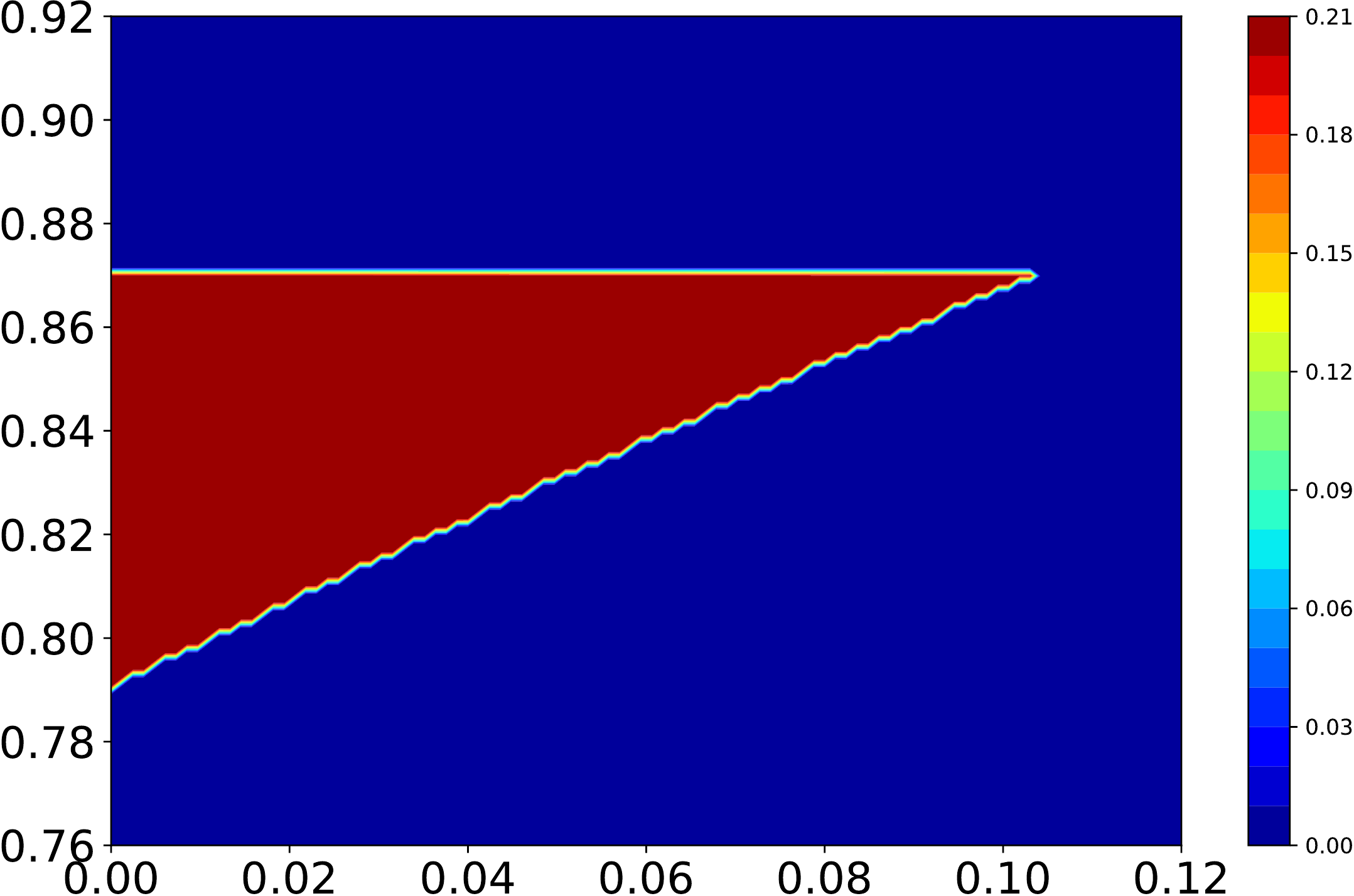}
	\put(-104,109){\large $M=3$}
	\put(-90,-16){\large $\wt{\lambda}$}
	\put(-176,53){\large ${\mu}$}
	\put(-185,109){\Large (a)}
	\qquad \qquad 
	\includegraphics[height=.25\textwidth]{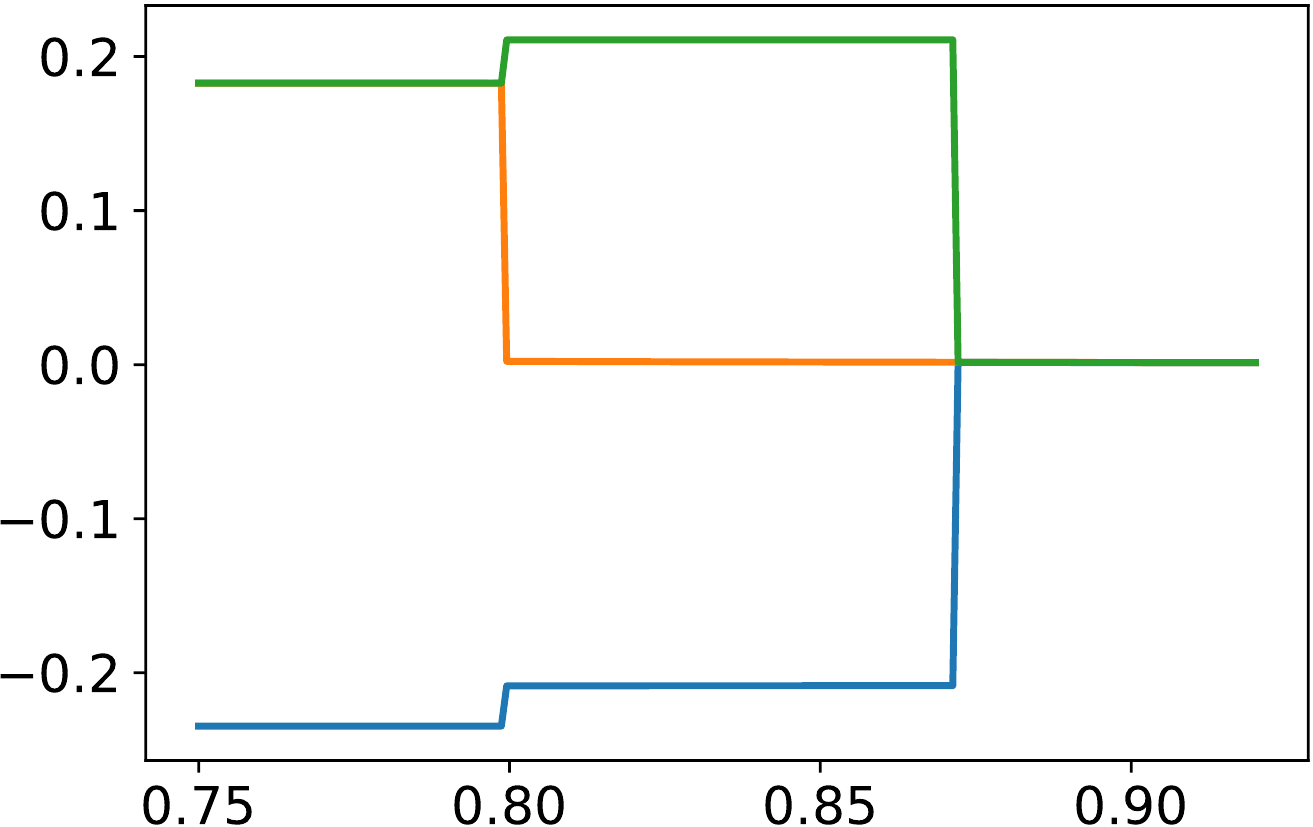}
	\put(-100,110){\large $\wt{\lambda}=0.01$}
	\put(-82,-12){\large ${\mu}$}
	\put(-193,55){\large $\{e_k\}$}
	\put(-190,109){\Large (b)}
	\caption{\label{fg:M3_mu_U1U1U1}(a) The $T=0$ phase diagram for $M=3$ with $\wt{g}_1=1$ and $\wt{g}_2=3$. 
	The plotted observable is $\text{Min}(|e_1-e_2|,|e_2-e_3|,|e_3-e_1|)$. Within the red triangle the three $e_k$ differ from one another, indicating spontaneous symmetry breaking $\U(3)\to\U(1)^3$. (b) The ${\mu}$-dependence of $\{e_k\}$ at $\wt\lambda=0.01$. There is a range of $\mu$ in which the three $e_k$ are all different.}
\end{figure}
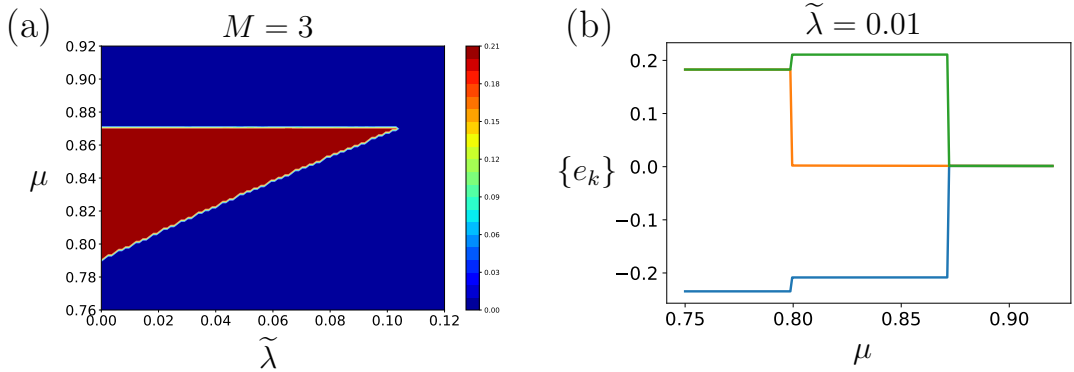

When increasing the chemical potential at fixed $|\wt{g}_2|>\wt g_2^{\rm cr}$ (or $\gamma_2<\gamma_2^{\rm cr}$)  we enter region IV, where $v(e)$ has three minima.
This  opens the possibility of phases with the symmetry breaking pattern $\U(M)\to\U(j)\times\U(k)\times\U(M-j-k)$. This has been indeed observed by us for $M=3$, see figure~\ref{fg:M3_mu_U1U1U1}. The region where this kind of symmetry breaking pattern happens is very narrow as it has been the case for finite temperature.  Note the similarity of figure~\ref{fg:M3_mu_U1U1U1} to figure~\ref{fg:M3_U1U1U1}. We repeated the same analysis for $M=4$ and $M=5$. For $M=4$ we found an exotic phase in which $\U(4)$ is broken down to $\U(2)\times\U(1)\times\U(1)$. For $M=5$ we found a phase with the breaking pattern $\U(5)\to\U(2)\times\U(2)\times\U(1)$.

When $\gamma_2 <\gamma_2^{\rm cr} $ the strip of first order phase transitions
is connected to the broken region around the origin of the $(\lambda, \mu)$ plane.
For $\gamma_2^{\rm cr}<\gamma_2<\gamma_2^{\rm tri} $ the strip is interrupted exactly
as in the finite $T$ and $\mu = 0$ case. The part that is connected to the broken
region about the origin disappears for $\gamma_2<\gamma_2^{\rm tri} $.

The strip divides the $\U(M)$-symmetric part of the phase diagram into two regions. 
The qualitative distinction between these two regions is clear from the behavior of the  $\{e_k\}$ (figure~\ref{fg:pd_mu}). They start with a large value at low $\wt\mu$ and then successively drop to a small value as $\wt\mu$ increases. 

At nonzero chemical potential the region between the tricritical point and
$\gamma_2 = 3/2$, which gives a second order phase transition for $M=2$
at low temperature, is absent. Therefore, we do not expect second order transitions
at $\mu\ne 0$ and $T=0$, even for $M=2$. Only at the end points of the cascades of phase transitions where all first order transitions run together, we expect second order phase transitions.

Finally, we consider the fermion number density $n$. Since it is proportional to $N$ it is useful to divide it by $N$. In dimensionless units, we have
\ba
	\frac{n}{N\Lambda^2} & = \frac{1}{4\pi}\sum_{k=1}^M 
	(\mu^2-4\wt{g}_2^2e_k^2)\Theta(\mu-2|\wt{g}_2e_k|)\,.
\ea
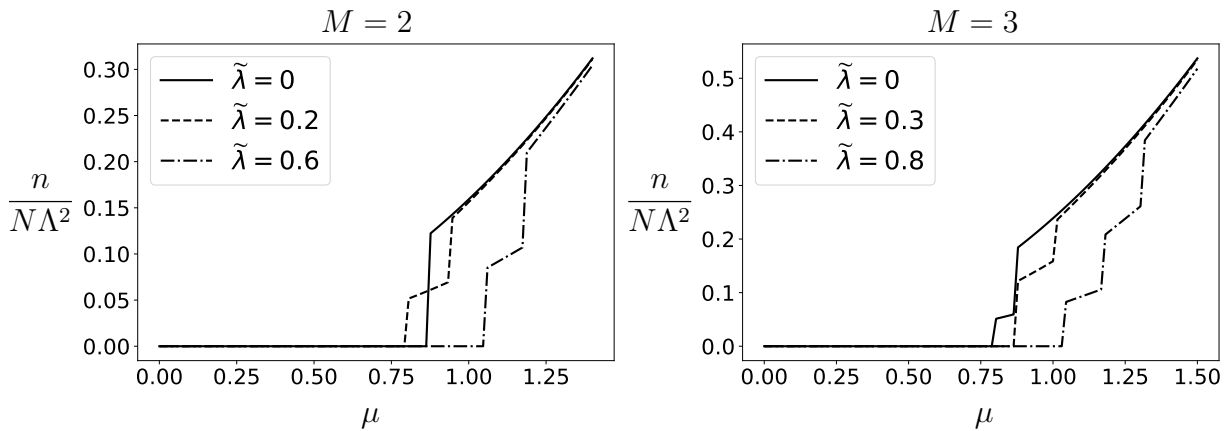
\begin{figure}[tb]
	\centering
	\includegraphics[height=.3\textwidth]{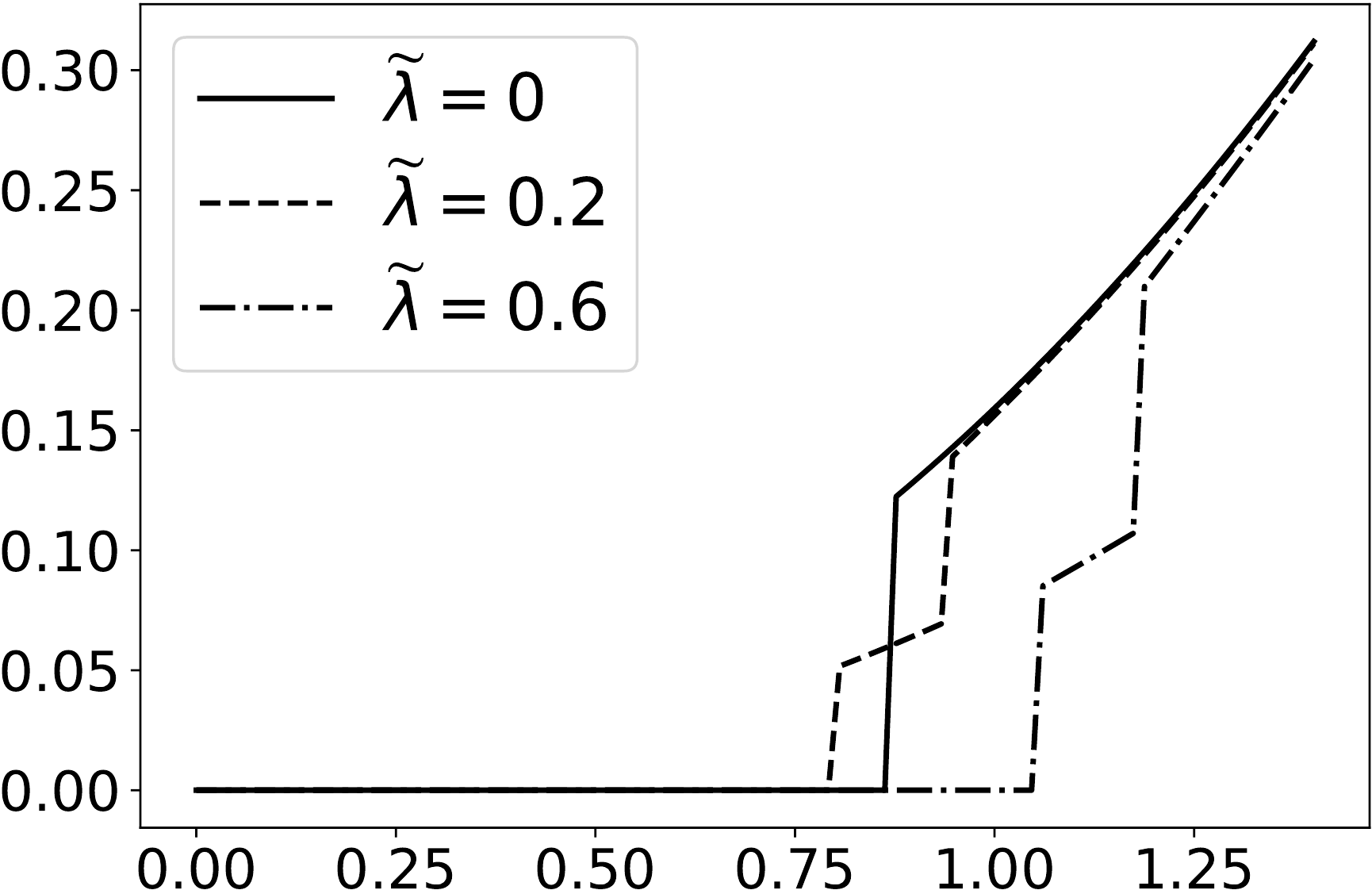}
	\put(-110,133){\large $M=2$}
	\put(-95,-14){\large ${\mu}$}
	\put(-228,66){\large $\displaystyle \frac{n}{N\Lambda^2}$}
	\qquad \quad
	\includegraphics[height=.3\textwidth]{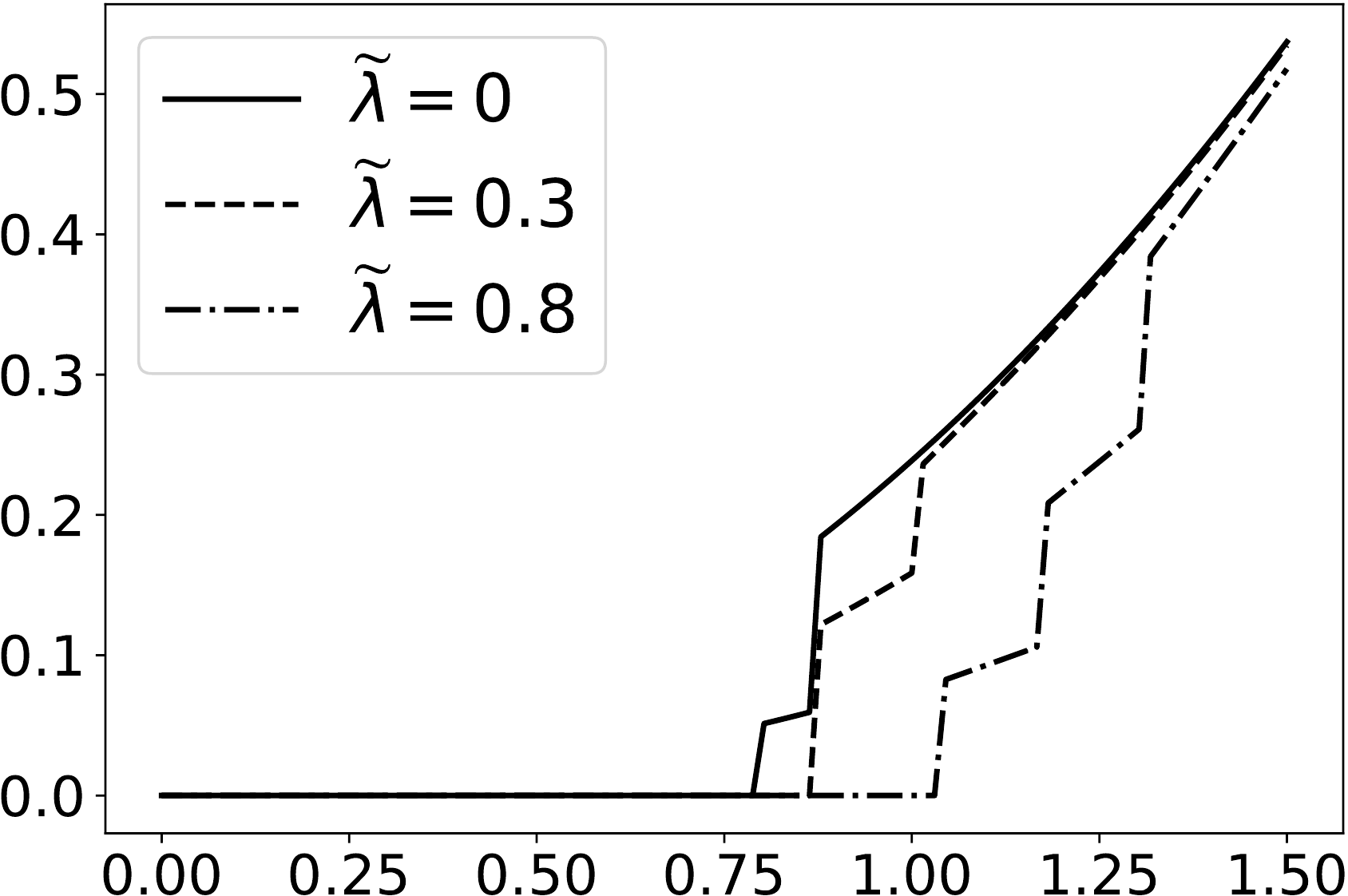}
	\put(-110,133){\large $M=3$}
	\put(-92,-14){\large ${\mu}$}
	\put(-223,66){\large $\displaystyle \frac{n}{N\Lambda^2}$}
	\caption{\label{fg:num_density}Fermion number density at $T=0$ with $\wt{g}_1=1$ and $\wt{g}_2=3$.}
\end{figure}%

In figure~\ref{fg:num_density} we show the $\mu$-dependence of the fermion number density. It jumps at first-order transitions. The plots for $\wt{\lambda}=0.6~(M=2)$ and $\wt{\lambda}=0.8~(M=3)$ reveal that the two $\U(M)$-symmetric phases at small $\mu$ and large $\mu$ are physically quite distinct, as was suggested in figure~\ref{fg:pd_mu} as well. At small $\mu$, fermions are rather heavy and the number density is quenched to zero. In contrast, at large $\mu$, fermions are nearly massless and their density is enhanced: the large bare mass
originating from $\kappa\bar\psi\psi$ in the Lagrangian is dynamically screened by interactions.

\section{\label{sc:fullpd}\boldmath Phases at nonzero $T$ and $\mu$}

Finally, in this section we investigate the effect of simultaneously nonzero $T$ and $\mu$. 
From \eqref{eq:mainV} the potential in this case is found to be
\ba
	\frac{V_{\rm eff}(E)}{\Lambda^3} =\; & 
	\frac{\wt{g}_1^2}{\wt{g}_2^2-M\wt{g}_1^2}\mkakko{\sum_{k=1}^{M}e_k-\wt{\lambda}}^2
	+ \sum_{k=1}^{M}\Bigg\{e_k^2+\frac{4}{3\pi}|\wt{g}_2e_k|^3 - \frac{1}{6\pi}(1+4\wt{g}_2^2e_k^2)^{3/2} 
	\notag
	\\
	& 
	+ \frac{{T}^3}{2\pi}\Bigg[  
	\frac{2|\wt{g}_2e_k|}{\wt{T}}\text{Li}_2\mkakko{-\rme^{-\frac{2|\wt{g}_2e_k|+{\mu}}{{T}}}} + \text{Li}_3\mkakko{-\rme^{-\frac{2|\wt{g}_2e_k|+{\mu}}{{T}}}}  
	\notag
	\\
	& 
	+ \frac{2|\wt{g}_2e_k|}{{T}}\text{Li}_2\mkakko{-\rme^{-\frac{2|\wt{g}_2e_k|-{\mu}}{{T}}}} + \text{Li}_3\mkakko{-\rme^{-\frac{2|\wt{g}_2e_k|-{\mu}}{{T}}}}
	\Bigg] \Bigg\} .
\ea
To speed up numerical minimization, we used the Taylor expansion of $\text{Li}_2(z)$ and $\text{Li}_3(z)$ around $z=-1$ up to 11th order to compute values for $-1\leq z<0$, and then used functional identities relating $\text{Li}_s(z)$ to $\text{Li}_s(1/z)$ \cite{wolfram_Li2, wolfram_Li3} to compute values for $z<-1$.

\begin{figure}[t!]
	\centering
	\includegraphics[width=\textwidth]{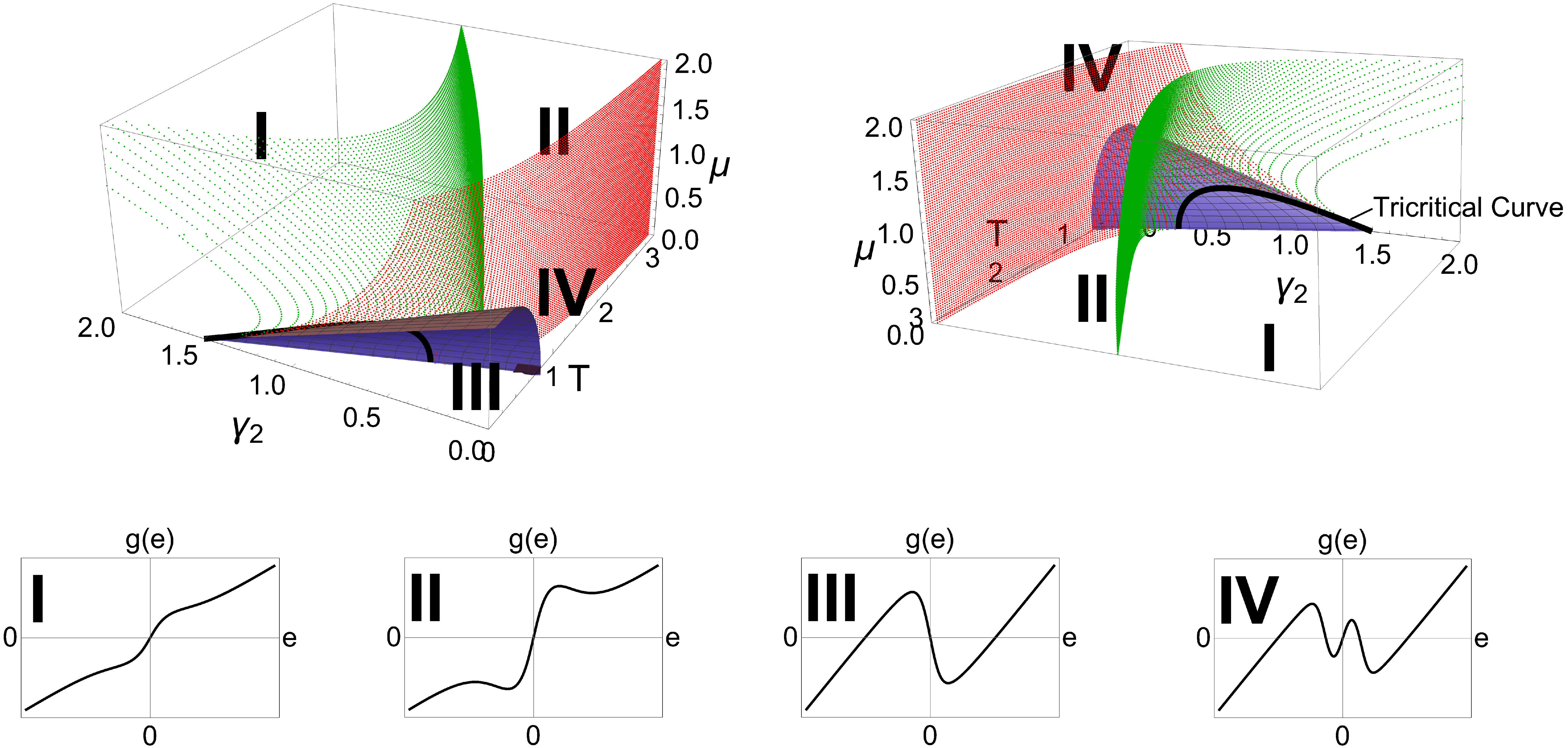}
	\caption{\label{fig:curve-chem-temp} Three-dimensional
          phase diagram of $g(e)$ in the $(\gamma_2, \mu, T)$ space.
          The insets show the possible shapes of the derivative of the potential
          $g(e)$, see~\eqref{saddle-chem-temp}, at finite chemical potential and temperature.} 
\end{figure}

De novo we use the notation of~\eqref{new-units} in which the potential reads
\begin{equation}
\begin{split}
\widehat{V}_{\rm eff}(\f)=\ &\gamma_1\left(\sum_{k=1}^{M}\f_k-\lambda\right)^2 
	+ \sum_{k=1}^{M}\biggl\{\gamma_2\f_k^2+|\f_k|^3- (1+\f_k^2)^{3/2}\\
	&+3{T}^3\biggl[\frac{|\f_k|}{{T}}{\rm Li}_2\left(-e^{-(|\f_k|+{\mu})/{T}}\right)+{\rm Li}_3\left(-e^{-(|\f_k|+{\mu})/{T}}\right)\\
	&+\frac{|\f_k|}{{T}}{\rm Li}_2\left(-e^{-(|\f_k|-{\mu})/{T}}\right)+{\rm Li}_3\left(-e^{-(|\f_k|-{\mu})/{T}}\right)\biggl]\biggl\}.
\end{split}
\end{equation}
with the saddle point equation
\begin{equation}\label{saddle-chem-temp}
\begin{split}
-2s=&g(\f_k)\ {\rm with}\ g(e)=2\gamma_2 e-3e\sqrt{1+e^2}+3{T}e\,\log\left(2\cosh\left[\frac{e}{{T}}\right]+2\cosh\left[\frac{{\mu}}{{T}}\right]\right)
\end{split}
\end{equation}
and $s=\gamma_1(\sum_{k=1}^{M}\f_k-\lambda)$. The possible four shapes of the function $g(e)$, see figure~\ref{fig:curve-chem-temp}, are basically smoothened versions of the zero temperature but finite chemical potential setting, compare with figure~\ref{fig:curve-chem}. This also implies that the discussion of the phase diagram in the $(\wt{\lambda},{\mu})$-plane looks essentially the same with one exception,
namely the onset of a second order phase transition for $M=2$ which results from the finite temperature picture in section~\ref{sc:temp}.

The phase transitions can  again be  understood via a Taylor expansion of the function $g(e)$ about the origin, i.e.,
\begin{equation}
\begin{split}
g(e)\approx&\left(2\gamma_2+3{T}\,\log\left[2\left(1+{\rm cosh}\left[\frac{{\mu}}{{T}}\right]\right)\right]-3\right)e+\frac{3}{2}\left(\frac{1}{T\mkakko{1+{\rm cosh}\left[{\mu}/{T}\right]}}-1\right)e^3\\
&+\frac{1}{8}\left(\frac{{\rm cosh}\left[{\mu}/{T}\right]-2}{{T}^3\mkakko{1+{\rm cosh}\left[{\mu}/{T}\right]}^2}+3\right)e^5+o(e^5).
\end{split}
\end{equation}
The coefficient of the linear term  determines the plane (blue dotted plane
in figure \ref{fig:curve-chem-temp})
that separates region
III from regions I and IV, which is explicitly given by
 \begin{equation}
\bar \gamma_2(T,\mu)=\frac{3}{2}\left(1-{T}\,\log\left[2\left(1+{\rm cosh}\left[\frac{{\mu}}{{T}}\right]\right)\right]\right).
\label{6.5}
 \end{equation}
The intersections of the blue surface with the $\mu=0$ and $T=0$ planes
are given by the blue curve in 
figures \ref{fig:phase} and \ref{fig:curve-chem}, respectively.
For $M = 2$ and $\lambda=0$, the part of this surface between regions I and III allows second order phase transitions. It continues to exist for not too large values of the chemical potential.
 When $\gamma_2< \bar \gamma_2(T,\mu)$
 the phase will have the symmetry breaking pattern $\U(2)\to\U(1)\times\U(1)$, and for $\gamma_2>3/2$ the flavor symmetry remains unbroken  at low temperature and chemical potential.

 There are two additional planes that divide the $(\mu, T, \gamma_2)$ space. As is the case
 at zero chemical potential, the first one is given by
 (green dotted surface in figure \ref{fig:curve-chem-temp})
 \be
 g'(e)=g''(e) =0.
 \ee
 On this surface, that separates regions I and II, the extrema of $g(e)$ in region
 II join so that $g(e)$ in region I becomes monotonous.

 The second plane is given by the equation (red dotted plane in figure \ref{fig:curve-chem-temp})
 \be
 g(e)=g'(e) =0.
 \ee
 On this plane the minimum of $g(e>0)$ touches the $e$-axis so that the corresponding potential will have three minima in region IV.
 
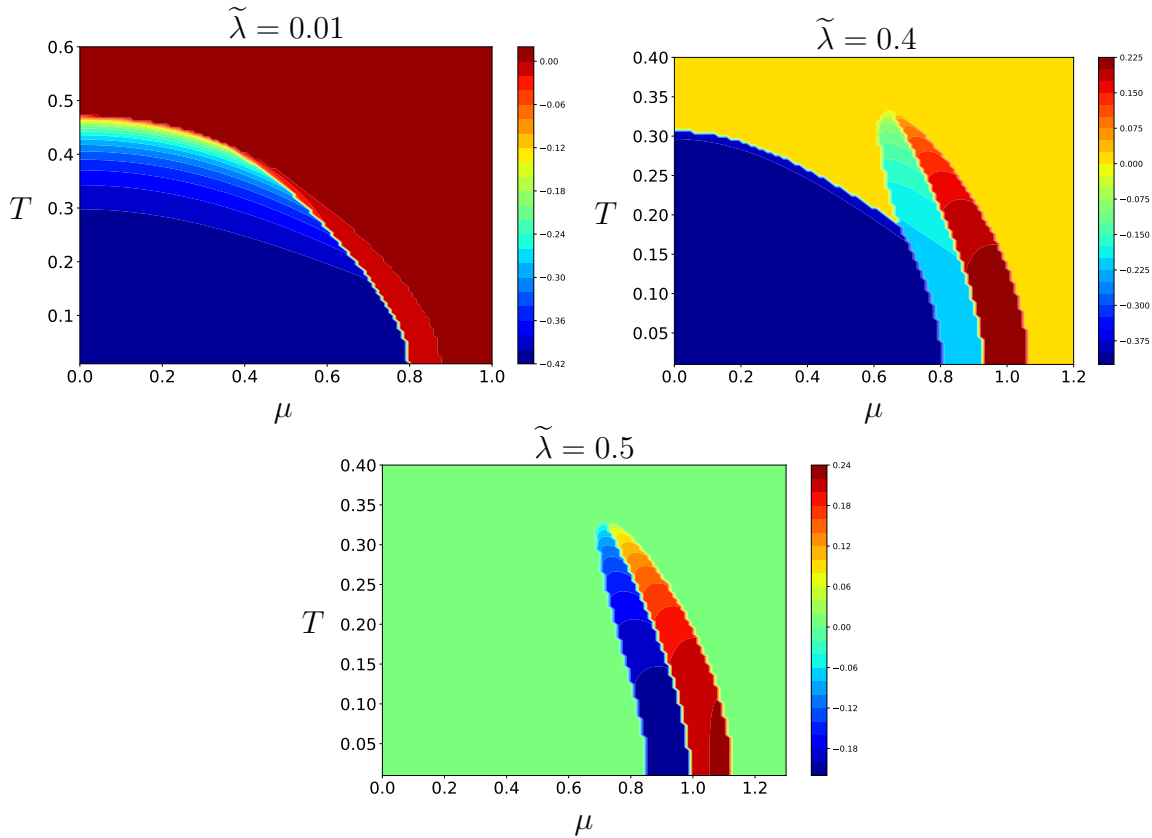
\begin{figure}[tb]
	\centering
	\includegraphics[width=.45\textwidth]{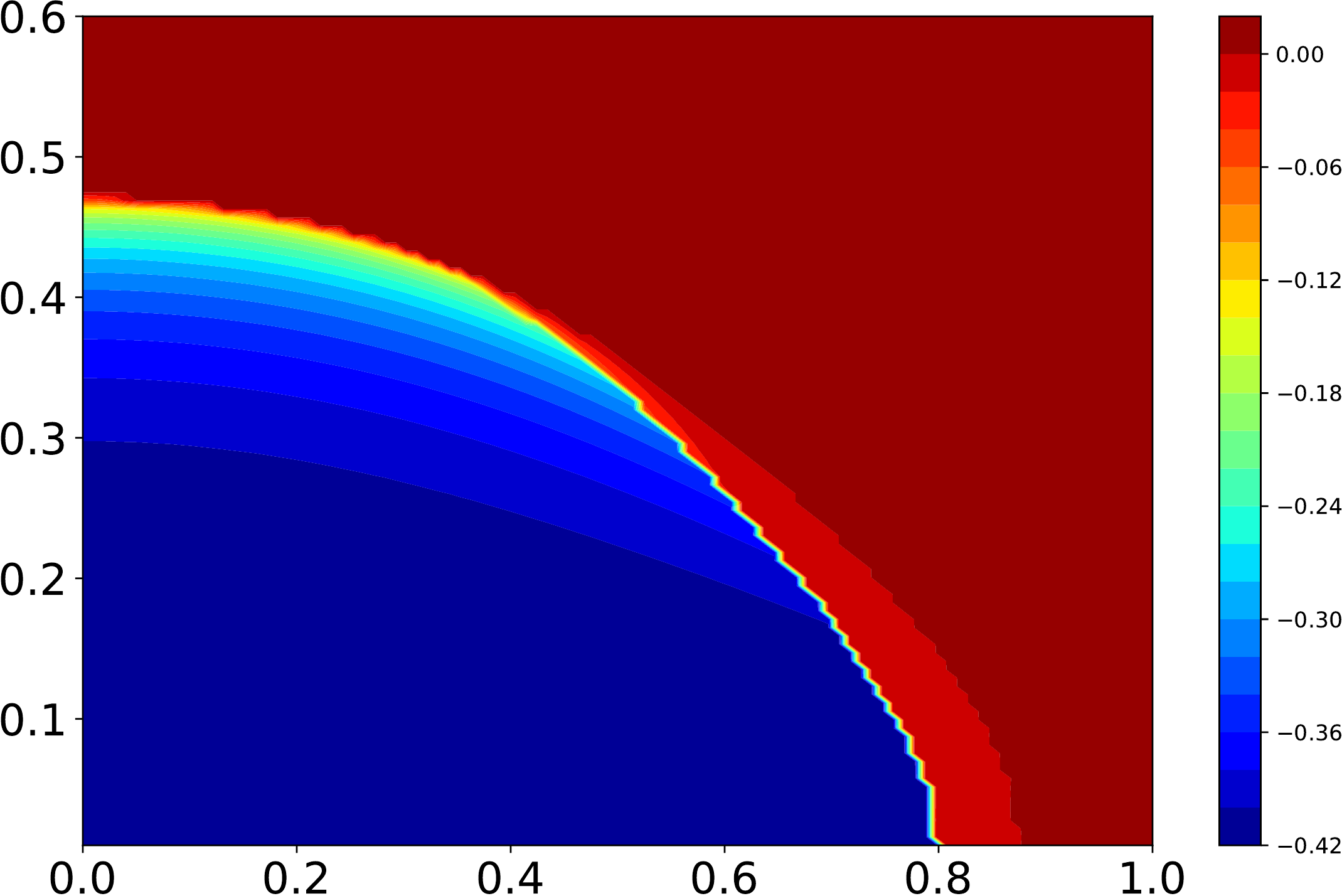}
	\qquad
	\includegraphics[width=.45\textwidth]{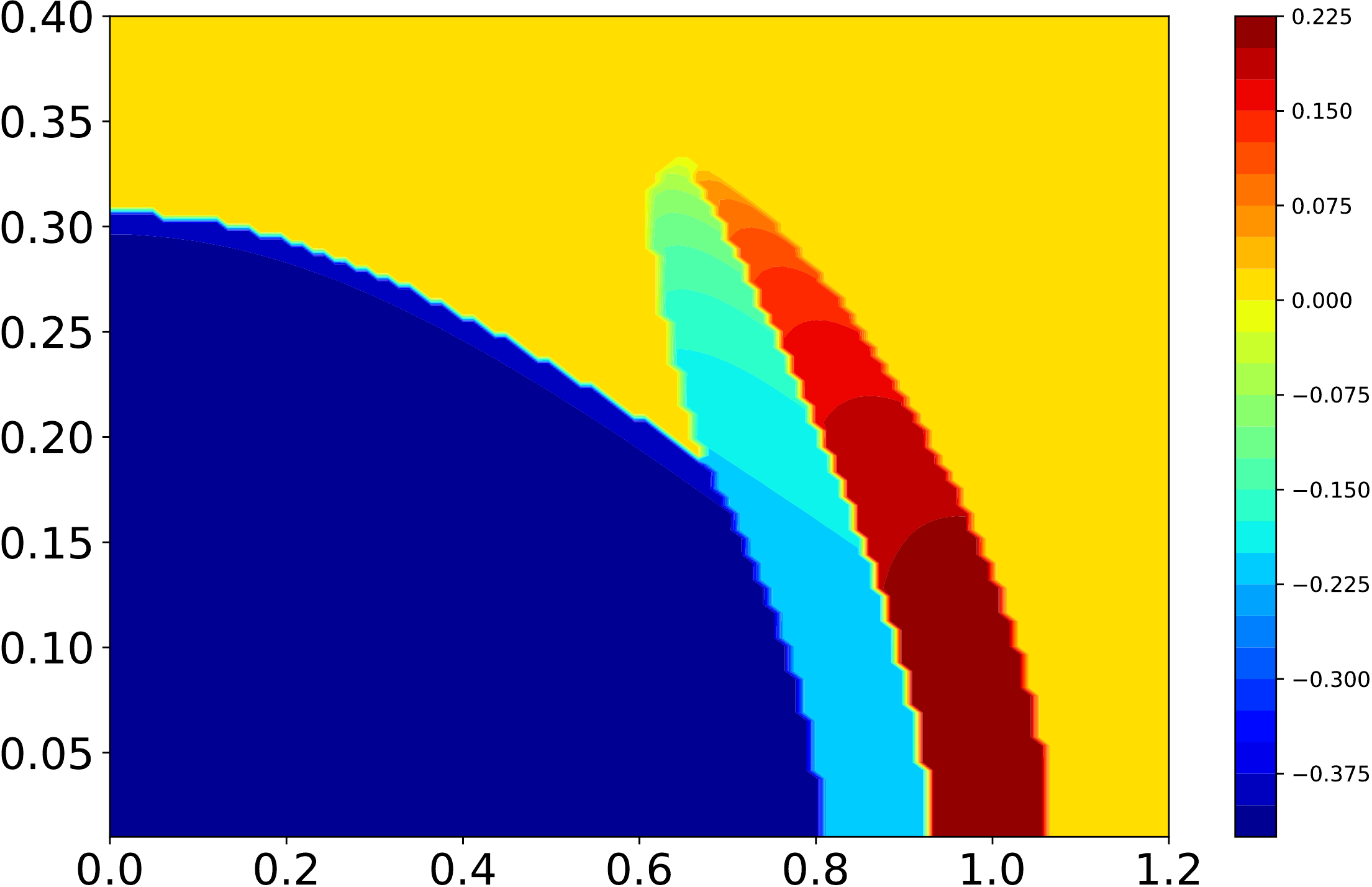}
	\put(-427,62){\large ${T}$}
	\put(-208,60){\large ${T}$}
	\put(-328,-13){\large ${\mu}$}
	\put(-107,-13){\large ${\mu}$}
	\put(-345,130){\large $\wt{\lambda}=0.01$}
	\put(-125,126){\large $\wt{\lambda}=0.4$}
	\\
	\centering
	\includegraphics[width=.45\textwidth]{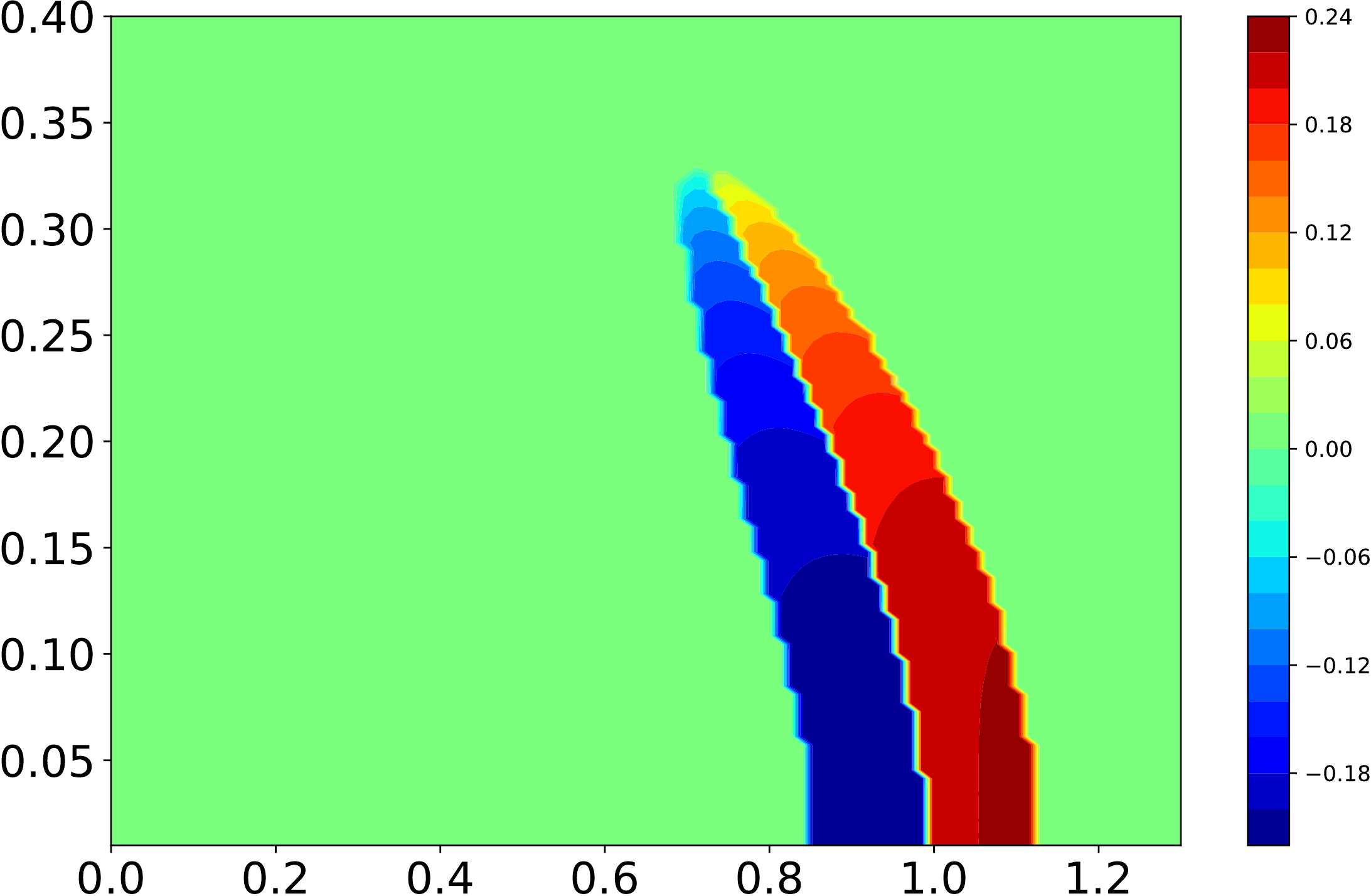}
	\put(-121,127){\large $\wt{\lambda}=0.5$}
	\put(-106,-13){\large $\mu$}
	\put(-208,61){\large ${T}$}
	\vspace{-.5\baselineskip}
	\caption{\label{fg:pd_final_g2=3}The $M=3$ phase diagram in the large-$N$ limit with $\wt{g}_1=1$ and $\wt{g}_2=3$ for various $\wt{\lambda}$. The plotted observable is $e_1-2e_2+e_3$ with $e_1\geq e_2 \geq e_3$ assumed.}
\end{figure}

The tricritical points at ${T}^{\rm tri}=1/2$ for ${\mu}=0$ becomes
now a tricritical curve
(see black curve in figure \ref{fig:curve-chem-temp}
for finite ${\mu}$, namely when the cubic term in $g(e)$ is also vanishing, which is at
\begin{equation}\label{tri-crit-line}
{\mu}^{\rm tri}={T}^{\rm tri}{\rm arccosh}\left[\frac{1-{T}^{\rm tri}}{{T}^{\rm tri}}\right]\quad\Rightarrow\quad \gamma_2^{\rm tri}=\frac{3}{2}\left(1-{T}^{\rm tri}\,\log\left[\frac{2}{{T}^{\rm tri}}\right]\right).
\end{equation}
There are bounds for the location of this curve, particularly ${T}^{\rm tri}\in[0,1/2]$, $\gamma_2^{\rm tri}\in[3/2(1-\log2),1.5]$ and ${\mu}^{\rm tri}\in[0,0.45]$ (the latter number is an approximation for the maximum of the right hand side of~\eqref{tri-crit-line}).
Whenever ${\mu}<{\mu}^{\rm tri}$ for $M=2$ the system experiences a second order phase transition at $\gamma_2=\bar{\gamma_2}$
(see eq. \eref{6.5}).
Larger values of $M$ remain untouched and all phase transitions are of first order apart from the critical points where first order transition lines  end.
Also for suitably large $\wt{\lambda}$ all second order phase transitions will vanish and what remains are first order transitions.

In region IV at fixed $\gamma_2$ the potential
will have three minima and we can again expect exotic phases
 of the form $\U(M)\to \U(j)\times\U(k)\times\U(M-j-k)$. They will certainly appear only for small $\mu$ since this has been already the case for either ${T}=0$ or ${\mu}=0$.

For suitably large $\gamma_2$, ${\mu}$ and ${T}$, we find either a strictly monotonous $g(e)$ (region I) or one which has local maxima and minima symmetrically about the origin in two separate quadrants (region II). The latter signals again the existence of a strip of cascades of phase transitions in the high ${T}$ and ${\mu}$ region. Yet, the shape of $g(e)$ in region II can be also found for a small region for $\gamma_2<\gamma_2^{\rm tri}$ which will show itself
as a remaining appendix of this strip of phase transitions at the phase region about the origin, cf. figure~\ref{fg:pd_mu0_M2_broken_strip}.

\begin{figure}[tb]
	\centering
	\includegraphics[width=.45\textwidth]{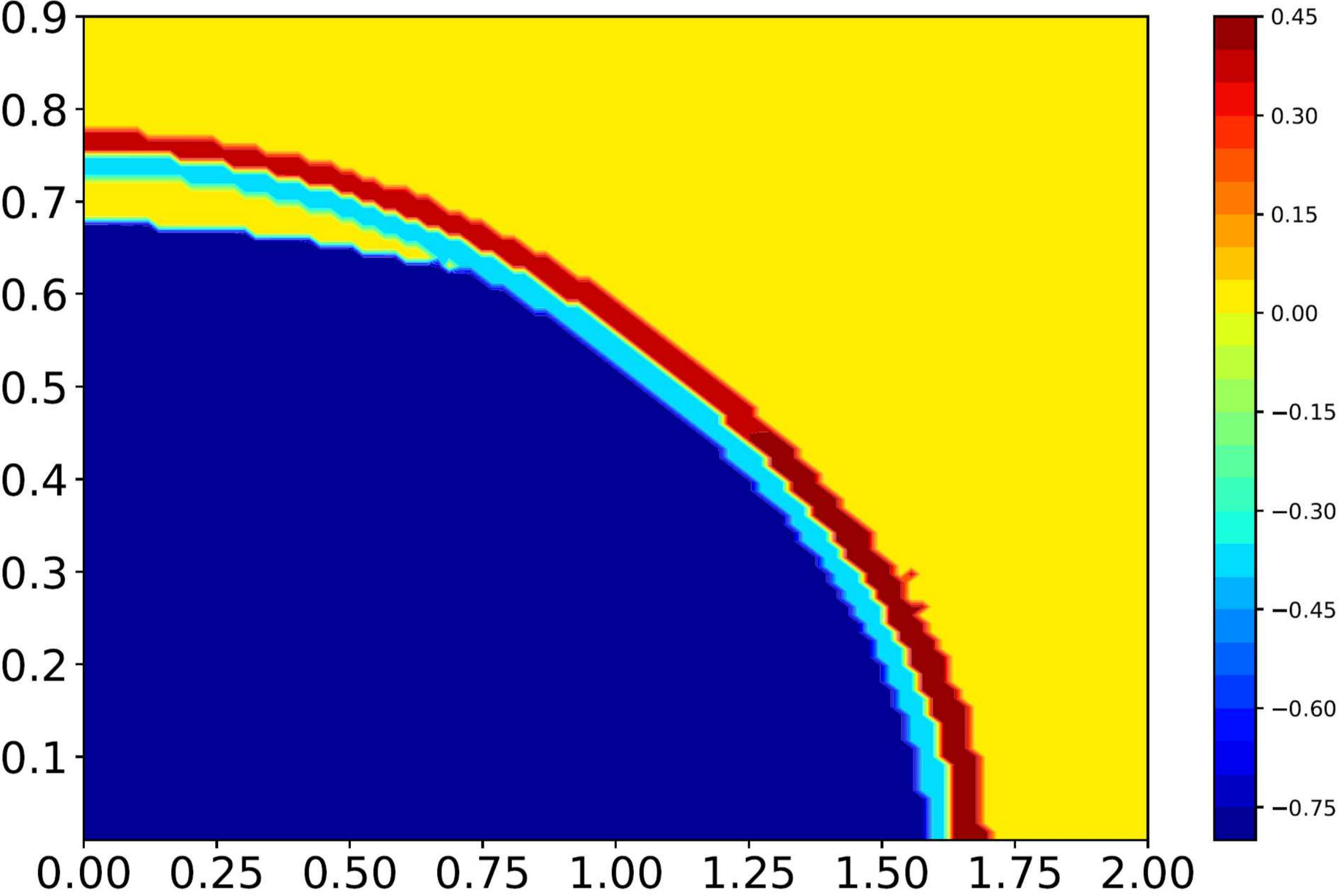}
	\qquad
	\includegraphics[width=.45\textwidth]{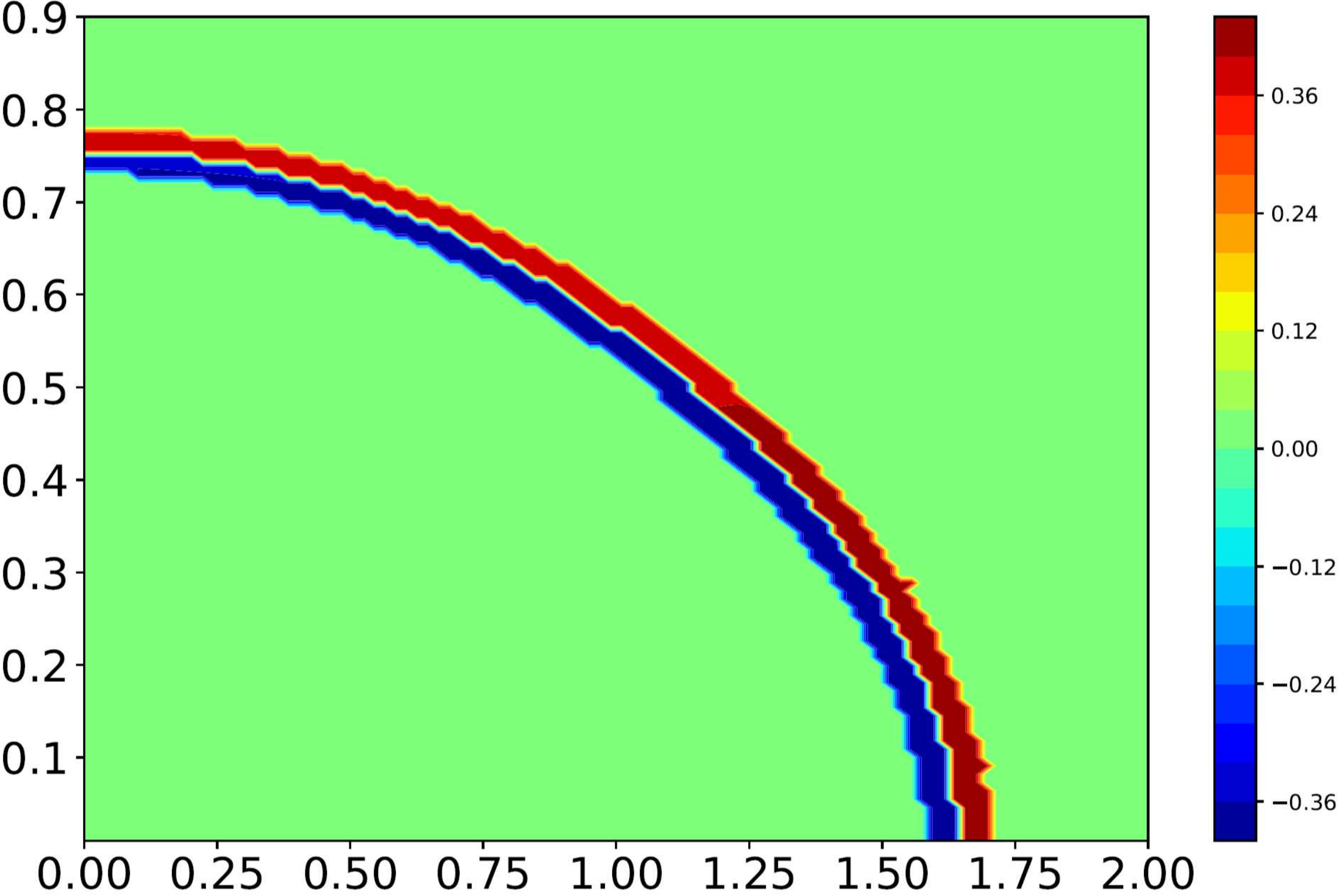}
	\put(-427,62){\large ${T}$}
	\put(-208,62){\large ${T}$}
	\put(-328,-13){\large ${\mu}$}
	\put(-107,-13){\large ${\mu}$}
	\put(-345,130){\large $\wt{\lambda}=0.75$}
	\put(-125,130){\large $\wt{\lambda}=0.8$}
	\vspace{-.5\baselineskip}
	\caption{\label{fg:pd_final_g2=5}Same as figure~\ref{fg:pd_final_g2=3} but with $\wt{g}_2=5$.}
\end{figure}
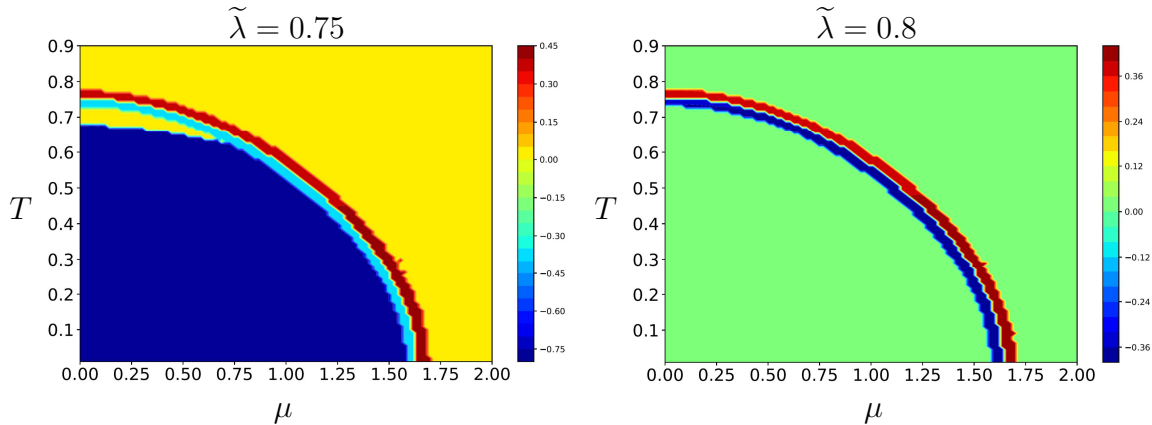

For simplicity of exposition we limit our numerical analysis to $M=3$.
Our main results are summarized in figure~\ref{fg:pd_final_g2=3} (for $\wt{g}_2=3$) and figure~\ref{fg:pd_final_g2=5} (for $\wt{g}_2=5$).
Figure~\ref{fg:pd_final_g2=3} shows that the phase structure depends on $\wt\lambda$ in a nontrivial way. At small $\wt\lambda$, there is a large region at low ${T}$ and low ${\mu}$ in which $\U(3)$ is spontaneously broken to $\U(2)\times\U(1)$. Along the boundary of this phase, there is a narrow strip in which $\U(3)$ is broken to $\U(1)\times\U(1)\times\U(1)$ (cf.~figure \ref{fg:M3_mu_U1U1U1}). As $\wt\lambda$ increases, this strip gradually disappears. At $\wt\lambda=0.4$ there are three symmetry-broken phases: they have the same symmetry ($\U(2)\times\U(1)$) but are separated by first-order phase transitions. As $\lambda$ increases further, the symmetry gets restored in the low-${T}$ low-$\mu$ region but remains broken in the cold dense region. 

At stronger coupling a qualitatively new feature emerges. In figure~\ref{fg:pd_final_g2=5} we observe that the symmetry-broken phase forms a thin annulus, separating the low-${T}$ low-${\mu}$ region from the high-${T}$ high-$\mu$ region. This annulus never disappears even at very large $\lambda$, although it is shifted to higher ${T}$ and $\mu$ gradually. By monitoring the behavior of $|e_k|$ we found that the $\U(3)$-symmetric phase below the annulus is characterized by very heavy fermions, while the other $\U(3)$-symmetric phase above the annulus is characterized by massless fermions. Although these two phases cannot be distinguished by symmetries, they host quite different physics.

\section{\label{sc:conc}Conclusions and outlook}

In the present article, we investigated various aspects of Dirac fermions with nonstandard quartic interactions in two spatial dimensions. We showed within the mean-field approximation that the model experiences a cascade of phase transitions when the flavor-symmetric parity-breaking mass is varied, in a way quite analogous to the behavior of QCD$_3$ \cite{Komargodski:2017keh,Armoni:2019lgb}. At
nonzero temperature and chemical potential we provided
analytical  and numerical arguments that show how a complicated phase diagram embellished by exotic symmetry breaking patterns emerges. In particular, we showed (in figures~\ref{fg:mu0_strongphase}, \ref{fg:pd_mu} and \ref{fg:pd_final_g2=5}) that, at strong coupling, the low-$(\mu, T)$ phase with heavy bosons is separated from the high-$(\mu, T)$ phase with almost massless bosons by a series of $M$ phase transitions, through which $M$ species of fermions become light one after another. At finite temperature there is a subtlety about symmetry breaking due to enhanced infrared singularities and we gave a speculative comment on this. Summarizing above, our results shed light on previously unnoticed novel dynamics of Dirac fermions in $2+1$ dimensions and have potential implications for planar gauge theories as well as planar condensed matter systems. 

There are several directions in which this work can be extended. First, the present analysis in the large-$N$ limit could be generalized to incorporate finite-$N$ corrections. Fluctuations of bosonic fields can be conveniently included by employing methods such as the functional renormalization group \cite{Wetterich:1992yh}. At finite $N$, the Jacobian (the squared Vandermonde determinant) associated with the diagonalization of the matrix field can no longer be neglected and will affect ground state properties. Secondly, it would be interesting to see what happens if our assumption $g_2^2>Mg_1^2$ is relaxed. Thirdly, various topological excitations arise in our model. For instance, in the phases depicted in figures~\ref{fg:M3_U1U1U1} and \ref{fg:M3_mu_U1U1U1}, $\pi_2(\U(3)/\U(1)^3)=\mathbb{Z}\times\mathbb{Z}$, implying there are two kinds of Skyrmions.  Fourthly, while we have only considered fermion-anti-fermion condensates, a di-fermion condensate may form at high density \cite{Buballa:2003qv,Alford:2007xm}.  The competition of two kinds of condensates may be an interesting subject of research. Finally, it would be challenging but quite important to take into account the possibility of an inhomogeneous condensate that spontaneously breaks translation symmetry. While the existence of such a condensate has been firmly established in some $(1+1)$-dimensional models at finite density \cite{Thies:2006ti,Basar:2008ki,Basar:2009fg}, the situation is elusive in higher dimensions \cite{Buballa:2014tba,Hidaka:2015xza}.

\acknowledgments
This work was in part supported (JV)  by  U.S. DOE Grant
No. DE-FAG-88FR40388. 
MK acknowledges support from the ARC grant DP210102887.

\appendix

\section{\label{sc:appA}Phase diagram of the effective potential and a toy model}

In this Appendix, we  outline the general strategy for analyzing
the phases for an effective potential of the general structure considered
in the main body of the text given by the sum of a confining ($\gamma_1>0$) harmonic collective potential
and confining ``single-particle'' terms with potential $v(x)$,
\begin{equation}
{V}(e_1,\cdots,e_M, {\lambda}) =\gamma_1\left(\sum_{j=1}^M {e}_j- {\lambda}\right)^2+\sum_{j=1}^M {v}({e}_j).
\label{vgen}
\end{equation}
Generically, we assume that $v(x)$ has the shape of a double well potential
that increases faster than linear for large $|e|$ (i.e. $\lim_{|e|\to\infty} v(e)/|e| =\infty)$).
Such potentials show a similar behavior such as the cascade of phase transitions and the kind of symmetry breaking patterns we have found
in the physical system. Moreover, the mechanism when and how the system experiences a second order phase transition is very similar for different ${v}(e)$.

In the last part of this Appendix, we illustrate the general arguments with the properties of a much simpler toy model (indicated by the subscript {\rm tm})
 with the confining potential
\ba
	v_{\rm tm}({e}_j;T)=({e}_j^2+T-1)^2.
	\label{eq:def_toy}
\ea
This model is motivated by the analysis in section~\ref{sc:temp} and its numerical observations. It is essentially a truncation of the expansion of the single particle part of the potential \eqref{pot.mu.finite} to fourth order, which is expected to capture some essential part of the physics in the vicinity of a second-order phase transition. Regardless of its simplified form, this toy model already exhibits generic features for general ${v}(e)$. Hence, we would like to underline that most conclusions apply for a more general confining potential  ${v}(e)$.

\subsection{Saddle point equation and its asymptotic solutions}\label{sec:saddeqasymp}

What has to be studied are the $M$ saddle point equations
 \begin{equation}\label{sadd-toy}
 -2s=-2\gamma_1\left(\sum_{j=1}^M{e}_j-{\lambda}\right)=g({e}_k)={v}'({e}_k).
 \end{equation}
 The extrema of the potential ${V}({e}, {\lambda})$ are determined by the
 intersections of $g(e)$ with  $-2s$, which also select
 the  possible phases, especially, which symmetry breaking patterns, the system can exhibit as a function of $\lambda$ and the parameters of the potential. Note
 that $s$ can have different values for the same values of the parameters. The reason is that the solution of the saddle point equations
 is not unique.

One particular ingredient is the asymptotic behavior of the solutions of~\eqref{sadd-toy} for large $|{\lambda}|$. The asymptotic super-linear growth of ${v}(e)$ implies also an asymptotic growth of $|g(e)|$. Particularly we have three cases to consider where the asymptotic value
\begin{equation}
\lim_{e\to\pm\infty}\frac{g(e)}{e}=c_{\pm }
\end{equation}
can be either vanishing ($c_{\pm}=0$), be finite ($0<c_{\pm}<\infty$), or diverge ($c_{\pm}(e)=\infty$). Depending on which case the asymptotic solution for ${e}_j$ becomes unique and takes the form
\begin{equation}
{e}_k\approx\left\{\begin{array}{ccl} \displaystyle \frac{{\lambda}}{M} -\frac{1}{2\gamma_1 M} g\left(\frac{{\lambda}}{M}\right), &\quad & c_{{\rm sign}({\lambda})}=0, \\  \displaystyle \frac{2\gamma_1{\lambda}}{2\gamma_1M+c_{{\rm sign}({\lambda})}}, & \quad & 0<c_{{\rm sign}({\lambda})}<\infty, \\  \displaystyle g^{-1}(2\gamma_1{\lambda}), & \quad & c_{{\rm sign}({\lambda})}=\infty, \end{array}\right.
\end{equation}
for all $k=1,\ldots,M$. The function $g^{-1}$ is the inverse of $g$ in the asymptotic regime; for instance when $g(e)$ grows like $e^L$, the inverse is essentially $e^{1/L}$.  In the physical system in the main text we have the situation of  a finite $c_{+}=c_{-}$ while the toy model~\eqref{eq:def_toy} leads to $c_{{\rm sign}({\lambda})}=\infty$. An asymptotic behavior with $c_+$ and $c_-$ in
a different class  is possible, but we do not consider that in the present work.

Regardless which case the asymptotic satisfies, we obtain the same conclusions
for the global minimum of the potential.
First, all ${e}_k$ are degenerate for suitably large $|{\lambda}|$. Second, the modulus of the auxiliary parameter $s$ in~\eqref{sadd-toy} also grows asymptotically, and its sign is the opposite of ${\lambda}$ and ${e}_k$. 
As a physical conclusion we find that for suitably large  $|{\lambda}|$
we always have a solution with all $e_k$ equal which has unbroken
flavor symmetry.

\subsection{Local extrema of $g(e)$ and implications on the possible phases}\label{sec:locextimp}

The solutions of the saddle point equation can be either minima or maxima for $V(e,\lambda)$.
For a saddle point with  $g'(e_k)>0$  for all ${e}_k$  the potential has certainly a  minimum, not necessarily the global one we are looking for. This follows from the fact that the Hessian at the saddle point is given by
\begin{equation}
H=\{\partial_{{e}_k}\partial_{{e}_j}{V}({e}, {\lambda})=2\gamma_1+g'({e}_k)\delta_{jk}\}_{j,k=1,\ldots,M}
\end{equation}
The Hessian is positive definite if the determinants
\be\label{positivity-condition}
\det(H_{jk})_{j,k=1,\ldots,n}=\det \{2\gamma_1+g'({e}_k)\delta_{jk}\}_{j,k=1,\ldots,n}>0,
\  {\rm for\ all}\ n\le M.
\ee
The term proportional to $2\gamma_1$ is of rank $1$ which simplifies the evaluation of the determinant drastically and it is equal to
\be
\det(H_{jk})_{j,k=1,\ldots,n}=\left (1+ 2\gamma_1 \sum_{l=1}^n \frac 1{g'(e_l)}\right )\prod_{k=1}^n g'(e_k),\  {\rm for\ all}\ n\le M.
\ee

In general, we may have a solution with $L$ different $e_k$. 
 At most one of the $e_k$ may have 
$g'(e_k) <0$. The reason is
that the term of the Hessian that is proportional to $\gamma_1$ is of rank one.
A rank one addition can maximally switch one eigenvalue of a matrix from positive to negative and vice versa, regardless how large its prefactor is.
Let us label the intersection with  $g'( e_k) <0$ as $e_M$. Then all subdeterminants up to $n=M-1$ are positive, and the condition for the positive
definiteness of the Hessian matrix is given by the positivity of its determinant,
\be
\det \{2\gamma_1+g'({e}_k)\delta_{jk}\}_{j,k=1,\ldots,M}>0\ \Leftrightarrow\ 1+ 2\gamma_1 \sum_{l=1}^M \frac 1{g'(e_l)}<0.
\ee
If the solution with $L$ different $e_k$ is a global minimum,
this would result
in the symmetry breaking pattern $\U(M)\to \U(j_1)\times\cdots\times\U(j_l)$ with $\sum_{l=1}^Lj_l=M$. The unbroken symmetry associated with $e_k$ with
$g'(e_k)<0$ can only be a $\U(1)$ factor.
However, not all of
these saddle points are global minima of $ {V}({e}, {\lambda})$ which is the hard part of the  analysis.

  The simplest case is $M=2$. Then the only possible flavor symmetry breaking
  pattern is $\U(2) \to \U(1) \times \U(1)$, that means a 
  transition from a solution with $e_1 =e_2$ to a solution with $e_1\ne e_2$.
  This transition
  can only be of second order if the solutions join continuously to the point
  where  the determinant of the Hessian vanishes, i.e.
  for parameter values with  $g'(e_1)=g'(e_2)=0$ (or at
  $g'(e_1)=g'(e_2)=-4\gamma_1$, but that is not allowed for $\gamma_1>0$). 
  There is no second order phase transition  when $g(e)$ increases monotonically. More generally, the phase transition is first order when the global minimum
  is not a continuous function of $\lambda$.

  As is shown for the potential in the main text and for the toy
  model~\eqref{eq:def_toy} a second order phase transition is realized
  for  $M=2$. However, as we will show below, for
  $M>2$ the phase transition for a potential of the form \eref{vgen}
  is always first order.

\subsection{Cascade of phase transitions}
\label{sec:cascade}

In this subsection, we discuss the phases of the general potential~\eref{vgen}
with $v(e)$ having locally the shape of a confining (not necessarily symmetric) double well. The notion ``locally'' means that there is region for $s$ bounded by a local maximum and local minimum of $g(e)$ where the saddle point equation~\eqref{sadd-toy} has only three solution when fixing $s$. The integral of $g(e)$ in this region looks like a double well potential.

For the solutions of the saddle point equations
we use the Ansatz ${\diag(e)}=\diag(X_1\1_{M-j},X_2\1_{j})$, with $X_1<X_2$ without restriction of generality. The two variables $X_1$ and $X_2$ must satisfy the equations
\begin{equation}\label{sadd-toy.M3}
  -2\gamma_1((M-j)X_1+j X_2-{\lambda})=g(X_1)\quad{\rm and}\quad -2\gamma_1((M-j)X_1+jX_2-{\lambda})=g(X_2).
\end{equation}
These equations can be solved for a real $\widehat{j}\in\mathbb{R}$. We would like to highlight the difference of $\widehat{j}\in\mathbb{R}$ and $j=0,\ldots,M$; while the former is real and can take an optimal position minimizing
\begin{equation}
  {V}_{\widehat{j}}(X_1,X_2,{\lambda})=\gamma_1((M-\widehat{j})X_1+\widehat{j}X_2-{\lambda})^2+(M-\widehat{j}){v}(X_1)+\widehat{j}{v}(X_2)
  \label{v-ann}
\end{equation}
the latter can be only an integer and only approximately minimizes the potential.
The saddle point equation for $\wh j$ is given by
\begin{equation}
  \label{toy-j-der}
0=\frac{\partial{V}_{\widehat{j}}}{\partial\widehat{j}}=2\gamma_1(X_1-X_2)((M-\widehat j)X_1+\widehat{j}X_2-\widehat{\lambda})+{v}(X_1)-{v}(X_2),
\end{equation}
which yields a unique minimum for $\widehat{j}$ in terms of $X_1$ and $X_2$.
This follows from   the second derivative in $\widehat{j}$ which is always positive when $X_1\neq X_2$, 
\begin{equation}
\frac{\partial^2{V}_{\widehat{j}}}{\partial\widehat{j}^2}=2\gamma_1(X_1-X_2)^2>0.
\end{equation}
Note that for $X_1=X_2$ the saddle point equation for $\widehat{j}$ is satisfied trivially.

The minimizer $\widehat{j}$ will generally not lie on one of the integers $j=0,\ldots,M$. Yet, the convexity of ${V}_{\widehat{j}}(X_1,X_2,{\lambda})$ in $\widehat{j}$ shows that only those closest to $\widehat{j}$ minimize the potential $V_{j}(X_1,X_2,\widehat{\lambda})$ with $j=0,\ldots,M$.

The saddle point equations \eref{sadd-toy.M3}
are invariant under
\be
\widehat j \to \widehat j+\delta \widehat j, \qquad
\lambda \to \lambda+ (X_2-X_1)\delta \widehat j.
\ee
as is the saddle point equation \eref{toy-j-der} for $\widehat j$.  We therefore must have
\begin{equation}
  \label{toy-j-lambda}
\frac{\rmd\widehat{j}}{\rmd\widehat{\lambda}}=\frac{1}{X_2-X_1}>0.
\end{equation}
Since $\lambda$ is a function of $\widehat j$, the saddle point solutions
are functions of $\widehat j$ only and increasing $\lambda$ will increase
$\widehat j$.
For the discretized version $j=0,\ldots,M$
the solutions $X_1$ and $X_2$ get an explicit dependence on $\widehat{\lambda}$
though it has only a limited impact as $j$ tries to be as close as possible
to $\widehat{j}$.
The convexity of the potential as a function of $\wh j$
also tells us that there can be only phase transitions from the phase with
flavor symmetry $\U(j)\times\U(M-j)$ to the phase with flavor symmetry
$\U(j+1)\times\U(M-j-1)$ or $\U(j-1)\times\U(M-j+1)$ for $j=1,\ldots,M-1$.
We still could have
a phase transition between a solutions with all $e_k$  the same.
As we will see below, in the toy model, this happens when $T>1$. Also in the physical system at finite temperature and/or at finite chemical potential there is a region where the system may experience such a direct transition from all $e_k$ equal and negative to all $e_k$ equal and positive, see sections~\ref{sc:temp},~\ref{sc:mu} and~\ref{sc:fullpd}.

In summary, the system runs through all phases corresponding to $\U(M)\to\U(j)\times\U(M-j)$ from $j=0,\ldots,M$ as the real minimizing set $(\widehat{j},X_1(\widehat{j}),X_2(\widehat{j}))$ will depend continuously on $\widehat{\lambda}$. The kinks and, hence, phase transitions only originate from the discreteness of $j$ rather than the continuous variable $\widehat{j}\to j$ implying $X_{1,2}(\widehat{j})\to X_{1,2}(j,{\lambda})$.

Let us underscore the following. What is locally required for the discussion above is a potential with
two minima and the validity of our assumption that the bipartite Ansatz
is valid.
However, we could not exclude the possibility that the flavor symmetry is broken according to
 $\U(M)\to\U(j)\times\U(M-j-1)\times\U(1)$ when  the potential has a minimum with three different $e_k$, two with $g'(e_k)>0$ and one with  $g'(e_k)< 0$. This possibility has to be investigated case by case.

\subsubsection{No-go statement for second order phase transitions}
\label{sec:nogo}

The question that remains is whether any of these phase transitions are  of second order. The transition from $j$ to $j+1$ for $j \ne 1$ and $j \ne M$
necessarily has to be of
first order, because if $X_1 =X_2$  is at a second order phase transition point, we would have a transition
from a state with $j$ of the $e_k$ at $X_2$ to a state with all of the
$e_k$ at $X_2$ as the positivity condition~\eqref{positivity-condition} does not allow any other possibility. The only exceptions are the transitions from $j=0$ to
$j=1$ and from $j=M-1$ to $j=M$. We now consider the latter, while the
former can be worked out in the same way.

As before, the arguments below apply to a potential $v(e)$ that has locally the form of a double well potential and
which is super-linearly increasing for large and small $e$, this means
its derivative $g(e)=v'(e)$ has the shape of a wiggle. 
For a second order phase transition to occur two solutions of the
saddle-point equations have to coalesce. This can only happen at the point
$e_0$  with $g'(e_0) = 0$.
Since we are considering the solution
$M-1 \to M$, we study the behavior of the potential around $e_k = e_0,$
for $k=1,\cdots,M $. At this point we have
\be
2\gamma_1(Me_0-\lambda) + M v'(e_0)&=&0,\nn \\
v''(e_0) &=& 0.
\ee
Although this point is a solution of the saddle-point equations it
does not have to be a global minimum. Below we will show that for
$M>2$ there is one direction in which the potential decreases. 
We use the Ansatz
\be
(e_1, \cdots, e_M) =(\underbrace{e_0+x_1,\cdots, e_0+x_1}_{M-1}, e_0+x_2).
\ee
Because the potential is homogeneous in the $x_k$ the derivatives of
the potential with respect to $x_1$ and $x_2$ also vanish for this Ansatz.
Since the determinant of the Hessian vanishes at $e_0$, there is at
least one direction
in which the second order fluctuations vanish.
To find a decreasing
direction, we thus have to Taylor expand the potential at least to third order
\be
V= V(e_0,\cdots, e_0) + \gamma_1((M-1) x_1+ x_2)^2 +
\frac 1{3!} g''(e_0) ((M-1) x_1^3+ x_2^3)+\cdots.
\ee
In the direction of vanishing second derivative, given by
\be
x_2 = -(M-1)x_1,
\ee
the third order term behaves as
\be
\frac {(M-1)-(M-1)^3}{3!} g''(e_0) x_1^3
=-\frac {M (M-1)(M-2)}{6} g''(e_0) x_1^3.
\ee
For the assumed shape of the potential we have $g''(e_0) > 0$ (minimum of $g(e)$ at $e_0$) so that
generally the third order term becomes negative. One exception is $M=2$.
In that case the fourth order term of the expansion is positive. So for $M=2$
the point $e_1=e_2=e_0$ can be a global minimum. For $M>2$, we always have a decreasing
direction excluding the possibility of a second order phase transition.

 When  $g''(e_0) = 0$,  we need to expand to higher orders. For a local
double well shape of $v(e)$ or local wiggle shape of its derivative $g(e)$ the
first non-vanishing derivative of $v(e)$ must be even. For $L=3,5,\ldots$, we obtain
\begin{equation}
\begin{split}
&\frac{M-1}{L!}g^{(L-1)}(e_0)x_1^L+\frac{1}{L!}g^{(L-1)}(e_0)x_2^L\\
=\ &(M-1)\left[1-(M-1)^{L-1}\right]\frac{x_1^L}{L!}g^{(L-1)}(e_0)<0
\end{split}
\end{equation}
because it must be $g^{(L-1)}(e_0)>0$ if we consider the transition from $j=M-1$ to $j=M$ as $g(e)$ has to have a minimum at $e_0$.

In summary, we can say that for $M\geq 3$ the system always experiences a cascade of first order phase transitions. The only requirement is that
the potential has locally the shape of a double well or its derivative $g(e)$ has the
shape of a ``wiggle'' meaning a maximum followed by a minimum and then growing again. This is the situation also for the physical system in the main text.
There are surely more complex situations when the potential $v(e)$ has more than two minima in a restricted region so that  the saddle point equation~\eqref{sadd-toy} has more than three real solutions. For instance this happens in the middle temperature regime discussed in section~\ref{sc:temp}. However, generally the mechanism for the phase transition is similar.

\subsection{Phase diagram of the toy model for $M=2$}
\label{sec:toym2}

Let us illustrate the phase diagram for $M=2$  with the toy model~\eqref{eq:def_toy}, where we have
\be
  \label{g-toy}
  V(e_1,e_2) = \gamma_1(e_1+e_2-\lambda)^2 +v_{\rm tm}(e_1)+v_{\rm tm}(e_2)
  \ee
  with
  \be
  v_{\rm tm}(e) = (e^2+T-1)^2.
  \ee
  The saddle point equations are given by
  \ba
-  2\gamma_1(e_1+e_2-\lambda)&=g_{\rm tm}(e_1) \nn
  \\
-  2\gamma_1(e_1+e_2-\lambda)&=g_{\rm tm}(e_2),
\label{saddle-tm}
\ea
where the derivative of the potential is defined as
\be
g_{\rm tm}(e)=v_{\rm tm}'(e)=4e(e^2+T-1).
 \ee
 This function can have only two distinct shapes depending on the temperature $T$. For $T\geq1$, it is monotonously increasing so that all
 ${e}_{k}$ need to be equal, and the flavor symmetry is not
 broken in this case.
 For $T<1$, the function develops a local minimum and maximum. Therefore,
 equation~\eqref{sadd-toy} can exhibit three solutions for a fixed $s$, two
 at  ${e}^{(+)}>0> {e}^{(-)}$ with $g_{\rm tm}({e}^{(\pm)})>0$,
 and one at  ${e}^{(0)}$ with  $g_{\rm tm}'({e}^{(0)})<0$.
In agreement with the asymptotic analysis in subsection~\ref{sec:saddeqasymp},
 for sufficiently large or small ${\lambda}$ only one solution exists
 when all ${e}_{k}$ are the same.

 In the toy model~\eqref{eq:def_toy}, we can have a transition from a broken phase
 with $e_1 \ne e_2$ to a phase with unbroken flavor symmetry
 with  $ e_1= e_2$. At a second order transition curve we therefore
 must have $e_1 =  e_2=e_0$ while the determinant of the Hessian must
 vanish. At this point we also must have that $g_{\rm tm}'(e_0) =0$.
 To determine the possible critical  behavior we calculate the Hessian at
 the saddle point in terms of
 $X=( e_1+ e_2)/2$ and $ \Delta = ( e_1- e_2)/2$
 and simplify it with the difference of the two saddle point equations,
\be
 g_{\rm tm}( e_2)-g_{\rm tm}( e_1)=4( e_1 - e_2)
 (e_1^2+ e_2^2+ e_1+e_2 -1+T)=0.
\label{gdif}
 \ee

 On the branch $e_1=e_2=e$ the Hessian reduces to 
\be
 \det H=(12 e^2+4(T-1))^2+ 4\gamma_1 (12 e^2+4(T-1)),
 \ee
which   vanishes when $g_{\rm tm}'(e_0)=0$, i.e. at
 \be\label{ridge}
 e_0= \pm \sqrt {\frac {1-T}3}.
 \ee  
The value of $\lambda$ at the second order
transition follows from the saddle point equations \eref{saddle-tm} for
$e_{1,2}\to e_0$,
 \be
\lambda=\pm\frac{2}{\gamma_1}\left[\gamma_1-\frac{2}{3}(1-T)\right]\sqrt{\frac{1-T}{3}}. 
\label{lam-eq}
\ee

 The line of second order phase transitions in the $(\lambda,T)$ plane may end
 in a tricritical point. At this point a second zero of the determinant
 of the Hessian vanishes. To determine it, we consider
 the branch $ e_1^2+ e_2^2+ e_1e_2 -1+T=0$ of the difference
 of the saddle point equations, see~\eqref{gdif}, where the determinant of the Hessian
in terms of $\Delta$ is given by
\be
\det H = 32 \Delta^2(8 \Delta^2 +\gamma_1 - 6(1-T)).
\ee
The tricritical point, when $\Delta$ changes from $\Delta=0$ to $\Delta\neq0$, is located at
\be
T= 1 - \frac 16 \gamma_1,
\ee
and $\lambda$ given by \eqref{lam-eq}.

\begin{figure}[tb]
	\centering
	\includegraphics[width=.3\textwidth]{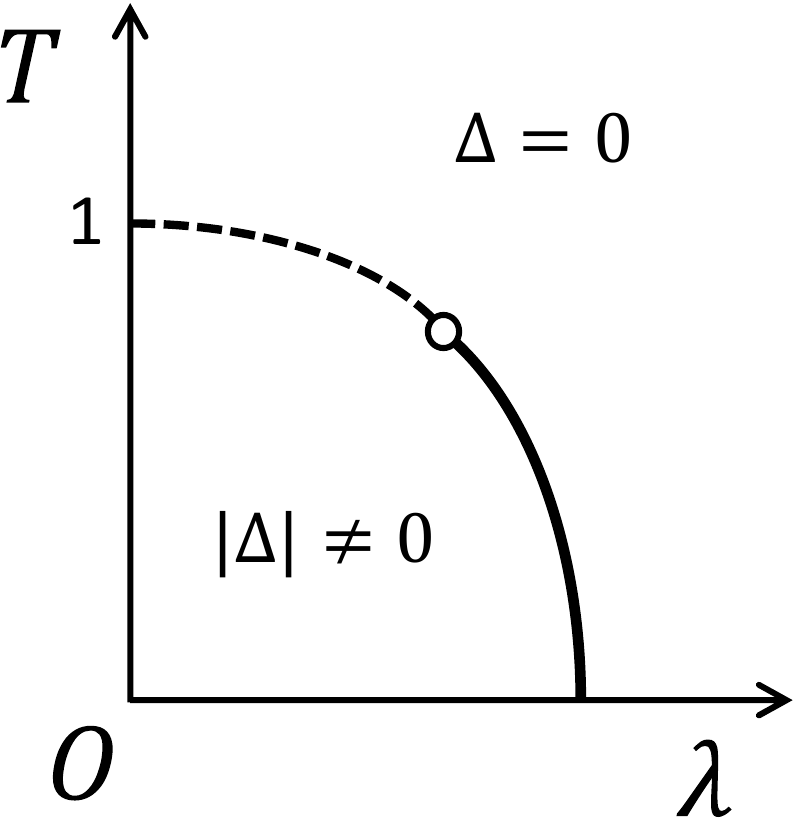}
	\caption{\label{fg:toy_pd_M2}
Sketch of the  phase diagram of the toy model for $M=2$.
The solid curve denotes a first-order phase transition and the dashed curve
 a second-order phase transition. The open circle marks a tri-critical point whose exact location is given by \eqref{eq:toy_crit_p} at $T=1-\gamma_1/6$.}
\end{figure}

If $e_1=e_2=e_0$ is a global minimum, the line of second order phase transitions
has to end at $T= 1-\gamma_1/6$ for $\gamma_1 < 6$. In particular, at zero temperature the phase transition has to be of first order for $\gamma_1 < 6$. The phase
diagram will look like the sketch in figure~\ref{fg:toy_pd_M2}.
There is a region for $1- \gamma_1/6 <T<1$ where a second order phase transition happens. For $T< 1- \gamma_1/6 $ we only have a first order phase transitions.
To determine if  $e_1=e_2=e_0$ is indeed a global minimum, we need explicit
expressions for the solutions and substitute them in the potential which will be done
in the remainder of this subsection.

\begin{figure}[tb]
	\centering
	\includegraphics[width=1\textwidth]{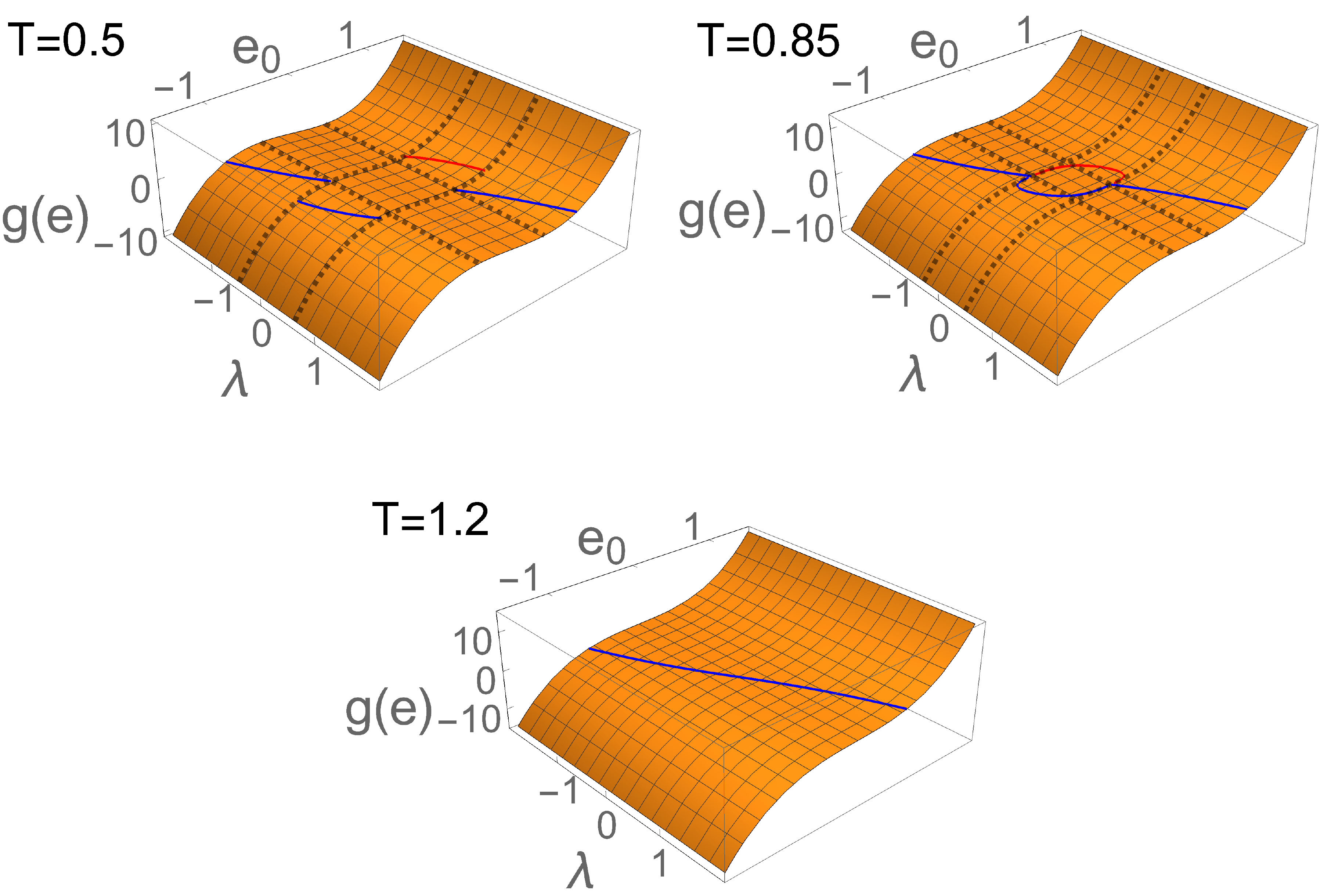}
	\caption{\label{fg:toy_g} The trajectories (red and blue curves) of the two solutions ${e}_1$ and ${e}_2$ as functions of ${\lambda}$ for $M=2$, $\gamma_1=1$ and various $T$. The tri-critical point is at $T=5/6$. To highlight the way how the two solutions ${e}_1={e}_2={e}^{(-)}$
          and ${e}_1={e}_2={e}^{(+)}$
depend on $\lambda$
we also show the function $g_{\rm tm}(e)$.
The dashed curves indicate the  location of the phase transition.
The dashed lines parallel to the $\lambda$ axis are the ``ridges'' of the local minimum and maximum of $g(e)$ given by~\eqref{ridge}, while the dashed curves parallel to $e_0$ are given by~\eqref{lam-eq}.
        }
\end{figure}

We will solve the saddle point equations in terms of $X=(e_1+e_2)/2$ and $\Delta=(e_1-e_2)/2$.
 From the difference of the two saddle point equation
\eref{gdif}
we see that we can have two different cases
\ba\label{Delta}
	\Delta = \begin{cases} 0, & \\ \pm\sqrt{1-T-3X^2}, & \text{only when}\ X^2<\frac{1-T}{3}.\end{cases}
\ea
The solution $\Delta=0$ always exists and is even a solution for general $g(e)$. It corresponds to unbroken flavor symmetry. The non-trivial solution for $\Delta$ obviously represents the case $\U(2)\to\U(1)\times\U(1)$. The potential
in terms of $\Delta$ and $X$ reads as follows
\begin{equation}\label{pot-delt-X}
{V}(X,\Delta, {\lambda}, T) = -16X^4+8(1-T)X^2+\gamma_1\left(2X-{\lambda}\right)^2 + 2(\Delta^2-1+T+3X^2)^2,
\end{equation}
which implies that $\Delta^2=1-T-3X^2$ is the global minimum whenever this solution is allowed.

The resulting saddle point equations for $X$ of the two cases of $\Delta$  differ and are given by
\begin{equation}\label{sad1-toy}
-2\gamma_1(2X_0-{\lambda})=4X_0(X_0^2+T-1)\ \Leftrightarrow\ \gamma_1{\lambda}=2X_0^3+2(\gamma_1+T-1)X_0
\end{equation}
for $\Delta=0$, and
\begin{equation}\label{sad2-toy}
-2\gamma_1(2X_1-{\lambda})=4\left(X_1\pm\sqrt{1-T-3X_1^2}\right)\left[\left(X_1\pm\sqrt{1-T-3X_1^2}\right)^2+T-1\right]
\end{equation}
for $\Delta=\pm\sqrt{1-T-3X_1^2}$. The latter one can be simplified by adding the two equations with signs $\pm$  which gives
\begin{equation}\label{sad2-toy.b}
-4\gamma_1(2X_1-{\lambda})=4[2X_1^3+6X_1(1-T-3X_1^2)+2(T-1)X_1]\ \Leftrightarrow\ \gamma_1{\lambda}=-16X_1^3+2[\gamma_1+2(1-T)]X_1.
\end{equation}
Using eq. \eref{lam-eq} for $\lambda$ at the saddlepoint,
this equation can be factorized as
\be
\left (X_1-\sqrt{\frac{1-T}3} \right)
\left (-16 X_1^2-16 X_1 \sqrt{\frac{1-T}3}-\frac 43 (1-T) +2\gamma_1\right )=0.
\ee
The solutions of the quadratic equation are given by
\be
X_\pm=\frac 12\left( \pm\sqrt{\frac{\gamma_1}2} -\sqrt{(1-T)/3}\right).
  \ee
  One can easily see that the solution $X_+$ has the lower potential. Next compare
  the difference of the potential for $X_1 = \sqrt{(1-T)/3}$ and $X_+$. With some
  work one can show that
\be
  \Delta V = 24 (1-T)^{3/2}(\gamma_1-6(1-T))\left( \sqrt{\frac{\gamma_1}6}-\sqrt{1-T}
  \right ).
  \ee
  This is negative for $T> 1-\gamma_1/6$ which shows that the solutions with
  the second order phase transition is the global minimum. As we have seen
  before from the analysis of the Hessian,  $T= 1-\gamma_1/6$ is the tricritical
  temperature.

  Summarizing, there is a second order phase transition whenever
\begin{equation}\label{eq:toy_crit_p}
(X,\Delta,|{\lambda}|)=\left({\rm sign}({\lambda})\sqrt{\frac{1-T}{3}},0,\frac{2}{\gamma_1}\left[\gamma_1-\frac{2}{3}(1-T)\right]\sqrt{\frac{1-T}{3}}\right)\ {\rm with}\ 1\geq T\geq1-\frac{\gamma_1}{6}.
\end{equation}
The point $T=1-\gamma_1/6$ is a tricritical point where the transition changes into a first order phase transition, we have sketched it in figure~\ref{fg:toy_pd_M2}.

\subsection{Phase diagram of the toy model for $M>2$}
\label{sec:toymgt2}

The situation for larger values of $M$ is more complicated. The saddle point equation have $3^M$ solutions, most of them complex. However, we find a substantial number of real solutions with different values of $\sum_k e_k$. A necessary
condition for a global minimum is that the Hessian is positive definite, but
to uniquely identify the solution we have to substitute it in the potential.
For example, for $M=3$, excluding permutations, we  find three real
solutions with a positive definite Hessian for a significant range of parameters.
We also have a saddle point with all three  $e_k$ different but this had never been
a minimum for the values of the parameters we have analyzed.

However, as we argued in previous sections, the global minimum of the potential
can have at most two different $e_k$. Using this as an Ansatz, this substantially
simplifies the saddle point equations. We still have that a second order phase
transition can only happen at the minimum of $g_{\rm tm}(e)$ which is at
$\sqrt{(1-T)/3}$. The difference of the two saddle-point equation is again given
by \eref{gdif}. The determinant of the Hessian on the branch
$e_1^2+e_2^2+e_1e_2-1+T$ can have only other zeros in addition to $\Delta =0$
when $2j=M$. In that case the location of the tricritical point is given by
\be
T = 1 - \frac{\gamma_1 M}{12}.
\ee
In principle these equation can be solved analytically, and by evaluating the potential
at all minima it is possible to determine whether or not this Ansatz yields
a global minimum. 
However, as was shown in \ref{sec:nogo} 2nd order phase transitions are not
possible for $M>2$ and this Ansatz with $2j=M$ cannot be the true
minimum.

\section{Some computations of  Sec.~\ref{sc:mk}}\label{app:Comp}

In this section we solve the saddle point equations
\be
g(e_k) = -2s, \quad k=1,\cdots, M
\ee
with
\be
g(e) = 2\gamma_2 e +3e(|e|-\sqrt{1+e^2})
\ee
and
\be
s= \gamma_1\mkakko{\sum_{k=1}^M e_k -\lambda}.
\ee
In the first part of this appendix, we calculate the solution with
all $e_k$ equal in which case we can find the exact solution. In the
second part, we compute the solution as a function of $s$ for the
case that there are three different solutions. 

When all $e_k$  take the same value we look for the solution $e_0$ closest
to the origin when
 $s$ and $e_0$ have opposite signs. The saddle point equation  simplifies to
\begin{equation}
2\gamma_1\lambda=2(\gamma_2+M \gamma_1) e_0 +3e_0\left(|e_0|-\sqrt{1+e_0^2}\right)
\end{equation}
which can be rephrased as
\begin{equation}
x_0^3-\frac{2\gamma_1\lambda-3/2}{\gamma_2+M\gamma_1}x_0^2-x_0+\frac{3}{2(\gamma_2+M\gamma_1)}=0\quad {\rm with}\quad e_0=\frac{x_0-x_0^{-1}}{2}.
\end{equation}
Its solution is
\begin{equation}\label{phase1}
\begin{split}
x_0=-a_0+\frac{2^{1/3}(1+3a_0^2)}{\left(\sqrt{b_0^2-108(1+3a_0^2)^3}-b_0\right)^{1/3}}+\frac{\left(\sqrt{b_0^2-108(1+3a_0^2)^3}-b_0\right)^{1/3}}{3\times 2^{1/3}}
\end{split}
\end{equation}
with
\begin{equation}
  a_0=-\frac{1}{2(\gamma_2+M\gamma_1)}
  \left(\frac{4}{3}\gamma_1 \lambda+1\right)\qquad{\rm and}\qquad b_0=27\left(\frac{3}{2(\gamma_2+M\gamma_1)}+a_0+2a_0^3\right).
\end{equation}

For $\gamma_2<3/2$ the function $g(e)$ develops a local minimum and maximum. Thus, the derivative of $g(e)$ with respect to $e$
has to vanish at these points. Its symmetry tells us that there is one zero of $g'(e)$  for $e>0$ and one for $e<0$. Indeed, the equation $g'(e)=0$  can be rewritten as
\begin{equation}
x^3+x-\frac{3}{\gamma_2}=0
\end{equation}
with $|e|=(x-x^{-1})/2$ and $x>1$.
The corresponding solution is given by
\begin{gather}\label{emin}
	\f_{\min}=\max\left\{\frac{x-x^{-1}}{2},0\right\},
	\\
	x=\frac{\left(27/\gamma_2+\sqrt{12+(27/\gamma_2)^2}\right)^{1/3}}{18^{1/3}}
	-\frac{(2/3)^{1/3}}{\left(27/\gamma_2+\sqrt{12+(27/\gamma_2)^2}\right)^{1/3}},
\end{gather}
which is the local minimum of $g(e)$. The local maximum lies at $-e_{\min}$. Hence, for a fixed $s\in[g(\f_{\min})/2,g(-\f_{\min})/2]$ we find three solutions for $g(e)=-2s$.

Two solutions  of $g(e)=-2s$, that we denote by $\f^{(-)}(s)<0<\f^{(+)}(s)$, come with a positive slope $g'(\f^{(\pm)}(s))>0$. The two solutions $\f^{(+)}(s)$ and $\f^{(-)}(s)$ satisfy the relation $\f^{(-)}(s)=-\f^{(+)}(-s)$ due to the symmetry of $g(e)$. Thus, it is enough to state the solution for $\f^{(+)}(s)=(x_+-x_+^{-1})/2$ with
\begin{equation}\label{phase2}
\begin{split}
 x_+=&-a_++\frac{2^{1/3}(1+3a_+^2)}{\left(\sqrt{b_+^2-108(1+3a_+^2)^3}-b_+\right)^{1/3}}+\frac{\left(\sqrt{b_+^2-108(1+3a_+^2)^3}-b_+\right)^{1/3}}{3\times 2^{1/3}}
\end{split}
\end{equation}
and
\begin{equation}\label{aandb}
a_+=\frac{1}{2\gamma_2}\left(\frac{4}{3}s-1\right)\qquad{\rm and}\qquad b_+=27\left(\frac{3}{2\gamma_2}+a_++2a_+^3\right).
\end{equation}
This can be derived by solving the cubic equation
\begin{equation}
\tilde{x}^3+\frac{2s-3/2}{\gamma_2}\tilde x^2-\tilde x+\frac{3}{2\gamma_2}=0
\end{equation}
 in $\tilde{x}>1$ which is equivalent to $g(e)=-2s$ for $e>0$. The correct solution can be selected by the special case $s=0$ which should yield $\tilde{x}=3/(2\gamma_2)$ as can be readily checked for the equation in $g(e)=0$.

 At the third solution $\f^{(0)}(s)\in ]-\f_{\min},\f_{\min}[$, the function $g(e)$ has a negative slope. For $s\in[0,g(-\f_{\min})/2]$, the solution $\f^{(0)}(s)=(x_0-x_0^{-1})/2$ has the form
\begin{equation}\label{phase2.b}
\begin{split}
 x_0=&-a_+-\frac{e^{-i\pi/3}\ 2^{1/3}(1+3a_+^2)}{\left(\sqrt{b_+^2-108(1+3a_+^2)^3}-b_+\right)^{1/3}}-e^{i\pi/3}\frac{\left(\sqrt{b_+^2-108(1+3a_+^2)^3}-b_+\right)^{1/3}}{3\times 2^{1/3}}
\end{split}
\end{equation}
with $a_+$ and $b_+$ as in~\eqref{aandb}, since its limit for $s=0$ should be $x_+=1$. For $s\in[g(\f_{\min})/2,0]$, we can use again the symmetry of $g(e)=-g(-e)$, meaning the solution is then $\f^{(0)}(s)=-\f^{(0)}(-s)$.

\section{\label{ap:comm}Comment on the role of bosonic fluctuations}
In the main text we have explored the pattern of symmetry breaking in the large-$N$ limit. In this limit the fluctuations of the
bosonic fields are completely negligible, but this is no longer the case at finite $N$. Actually the celebrated Coleman-Mermin-Wagner-Hohenberg (CMWH) theorem \cite{Coleman:1973ci,Mermin:1966fe,Hohenberg:1967zz} stipulates that continuous symmetries cannot be spontaneously broken at nonzero temperature in $2+1$ dimensions in the absence of long-range interactions. Thus any symmetry-breaking condensate at $T=0$ must disappear as soon as nonzero temperature is turned on. No massless Nambu-Goldstone modes can appear; in fact they acquire nonzero masses, as demonstrated explicitly for $\O(N)$-invariant models in \cite{PhysRevLett.60.1057,Chakravarty:1989zz,PhysRevB.40.4858,Rosenstein:1989sg,Hasenfratz:1990jw}. In our model, the ground state has to be $E\propto \1_M$ everywhere on the phase diagram at $T>0$. The second-order phase transition line in figure~\ref{fg:pd_mu0_M=2_weak_g2} will be wiped out at finite $N$ and becomes a crossover. As long as $N\gg 1$, the first-order transitions in figures~\ref{fg:pd_mu0_M=2_weak_g2}, \ref{fg:pd_mu0_M=3_weak_g2}, and \ref{fg:mu0_strongphase} may well persist. However, if thermal fluctuations were so strong that the critical point at $\lambda=0$ (present in figures~\ref{fg:pd_mu0_M=3_weak_g2}) is destroyed, then $M$ first-order transition lines emanating from the $T=0$ axis would probably end at $M$ distinct critical points. 

From the viewpoint of the CMWH theorem it may seem that there is no point in talking about symmetry breaking for $T>0$. However this is not the case. Preceding analyses \cite{PhysRevLett.60.1057,Chakravarty:1989zz,PhysRevB.40.4858,Rosenstein:1989sg,Hasenfratz:1990jw} have shown that the mass of the would-be Nambu-Goldstone modes $m_{\rm NG}$ is of order $F^2\exp(-cF^2/T)$, where $c$ is an O(1) pure number and $F^2$ is a square of the ``pion decay constant'', also known as spin stiffness in the literature of quantum magnets. It is well known that $F^2\propto N$ in the large-$N$ limit \cite{Manohar:1998xv}, so we have parametrically $m_{\rm NG}\sim N\exp(-N/T)$ which gets exponentially small at low temperatures or large $N$. In experiments or numerical simulations, the correlation length $\sim m_{\rm NG}^{-1}$ can easily exceed the system size. The system is then virtually indistinguishable from a genuine symmetry-broken phase. The interactions of the would-be Nambu-Goldstone modes get weaker and weaker as the energy scale goes down, but start to increase at the scale $\sim m_{\rm NG}$. As far as physics at length scales $\ll m_{\rm NG}^{-1}$ is concerned, it is perfectly sensible to adopt a description based on spontaneously broken symmetry. A detailed discussion on the consistency between the CMWH theorem and the utility of low-energy effective theories of Nambu-Goldstone modes can be found in \cite{Hofmann:2012uz}. 

To go beyond the mean-field analysis of this paper we must employ nonperturbative methods such as Monte Carlo simulations on a lattice, which seems feasible since the statistical weight \eqref{eq:ZnE} is real and nonnegative for even $N$.

\bibliography{draft-v16.bbl}
\end{document}